\documentclass[a4paper,11pt]{article}
\usepackage[nointlimits]{amsmath}
\usepackage{amsfonts}
\usepackage{amssymb}
\usepackage{amssymb}
\usepackage{amsthm}
\usepackage{float}
\usepackage{graphicx}
\usepackage{epsfig}
\usepackage{layout}
\usepackage[english]{babel}
\usepackage{mathrsfs}
\numberwithin{equation}{section}
\newtheorem{lemma1}     {Lemma}[section]
\newtheorem{teorema1}   [lemma1]{Theorem}
\newtheorem{prop1}      [lemma1]{Proposition}
\newtheorem{coroll1}    [lemma1]{Corollary}
\newtheorem{cong1}      [lemma1]{Conjecture}
\newtheorem{remark1}    [lemma1]{Remark}
\newtheorem{defin1}     [lemma1]{Definition} 
\newtheorem{example1}   [lemma1]{Example} 

\newenvironment{Lemma}[1][]
        {\begin{lemma1}[#1]\begin{samepage}}{\end{samepage}\end{lemma1}}
\newenvironment{Theorem}[1][]
        {\begin{teorema1}[#1]\begin{samepage}}{\end{samepage}\end{teorema1}}
\newenvironment{Proposition}[1][]
        {\begin{prop1}[#1]\begin{samepage}}{\end{samepage}\end{prop1}}
\newenvironment{Corollary}[1][]
        {\begin{coroll1}[#1]\begin{samepage}}{\end{samepage}\end{coroll1}}

\newenvironment{Remark}[1][]
        {\begin{remark1}[#1]\begin{samepage}}{\end{samepage}\end{remark1}}
\newenvironment{Definition}[1][]
        {\begin{defin1}[#1]\begin{samepage}}{\end{samepage}\end{defin1}}
\newenvironment{Example}[1][]
        {\begin{example1}[#1]\begin{samepage}}{\end{samepage}\end{example1}}

\renewcommand{\phi}					{\varphi}
\newcommand{\nada}[1]   {}
\newcommand{\C}         {\mathcal C}
\newcommand{\CcoverB}  {\mathcal C_c\Big(\spacetime\times \overline{\immgrass}\Big)}
\newcommand{\dimension}       {h}
\newcommand{\dimspacetime}       {N}
\newcommand{\energy}       {E}
\newcommand{\eps}       {\varepsilon}
\newcommand{\fq}        {f_{q^{-1}}}
\newcommand{\F}         {{\mathcal F}}
\newcommand{\grad}         {\nabla}
\renewcommand{\H}       {{\sigma}}
\newcommand{\immgrass} {B_{\dimension,\dimspacetime+1}} 
\newcommand{\immmap} {\Sigma} 
\newcommand{\indice}         {\ell}
\newcommand{\indicesuccessione}         {n}
\newcommand{\indicespaziale}         {{\rm a}}
\newcommand{\indispazuno}         {{\rm a}}
\newcommand{\indispazdue}         {{\rm b}}

\newcommand{\lettera}   {\tau}
\newcommand{\lormeas}      {{\sigma^\dimension}}             
\newcommand{\lormeastwo}      {{\sigma^2}}             
\newcommand{\LV}      {\mathcal{LV}_\dimension}             
\newcommand{\LVuno}      {\mathcal{LV}_1}             
\newcommand{\LVdue}      {\mathcal{LV}_2}

\newcommand{\manist}       {\Sigma}
\newcommand{\map}         {\parammap}
\newcommand{\mappa}         {q}
\newcommand{\n}         {{\mathrm n}}

\newcommand{\p}        {\phi}
\newcommand{\parammap}        {X}

\newcommand{\pst}         {z}

\newcommand{\punto} {{\pst}}

\newcommand{\R}         {\ensuremath{\mathbb R}}
\newcommand{\recf}      {f^\infty}
\newcommand{\rect}       {\rectifiable}
\newcommand{\rectifiable}       {\Sigma}
\newcommand{\rightextremum}       {L}
\newcommand{\Rm}        {\ensuremath{\mathbb R^m}}
\newcommand{\Rn}        {\ensuremath{\mathbb R^\dimspacetime}}
\def\rest{\hskip 1pt{\hbox to 10.8pt{\hfill
\vrule height 7pt width 0.4pt depth 0pt\hbox{\vrule height 0.4pt
width 7.6pt depth 0pt}\hfill}}}
 
\newcommand{\spaceimm}       {\gamma}
\newcommand{\spacetime}       {\R^{1+\dimspacetime}} 
 
\newcommand{\sptpt}       {\pst}

\newcommand{\timelikekplanes}       {T_{\dimension,\dimspacetime+1}} 

\newcommand{\vel}       {\mathbb{V}} 
\newcommand{\velnull}       {{\vel_\infty}} 

\newcommand{\vfspacetime}    {\left(\mathcal C_c^1(\spacetime\right))^{N+1}}
\newcommand{\xx}   {{u}}

\begin{document}

\title{Lorentzian varifolds and applications to closed relativistic
strings}

\author{
Giovanni Bellettini\footnote{
Dipartimento di Matematica,
Universit\`a di Roma Tor Vergata,
via della Ricerca Scientifica 1, 00133 Roma, Italy,
and
INFN - Laboratori Nazionali di Frascati (LNF), via E. Fermi 40,
00044 Frascati, Roma,  Italy,
e-mail: Giovanni.Bellettini@lnf.infn.it}
\and
Matteo Novaga\footnote{
Dipartimento di Matematica Pura e Applicata,
Universit\`a di Padova,
via Trieste 63, 35121 Padova, Italy,
e-mail: novaga@math.unipd.it}
\and
Giandomenico Orlandi\footnote{
Dipartimento di Informatica, Universit\`a di
Verona, strada le Grazie 15, 37134
Verona, Italy, email: giandomenico.orlandi@univr.it}
}

\date{}

\maketitle

\begin{abstract}
We develop a suitable generalization of Almgren's theory of varifolds in a
lorentzian setting, focusing on area, first variation,
rectifiability, compactness and closure issues.
Motivated by the asymptotic behaviour of the scaled 
hyperbolic Ginzburg-Landau equations, and by
the presence of singularities in lorentzian
minimal surfaces, 
 we introduce, within the
varifold class, various notions of generalized minimal timelike
submanifolds of arbitrary codimension in flat Minkowski spacetime, which
are global in character and admit conserved quantities, such as relativistic
energy and momentum.
In particular, we show that stationary lorentzian $2$-varifolds
properly include the class of classical 
relativistic and subrelativistic strings. We
also discuss several examples.
\end{abstract}

%\tableofcontents

%\keywords{}
%\subjclass{Primary ; Secondary}

%%%%%%%%%%%%%%%%%%%%%%%%%%%%%%%%%%%%%%%%%%%%%%%%%%%%%%%%%%%%%%%%%%%%
\section{Introduction}
%%%%%%%%%%%%%%%%%%%%%%%%%%%%%%%%%%%%%%%%%%%%%%%%%%%%%%%%%%%%%%%%%%%%
In recent years, a lot of effort 
has been devoted to the 
study of lorentzian stationary (called also minimal) timelike submanifolds 
of arbitrary dimension
$\dimension$ without boundary in the flat Minkowski spacetime,
namely to those $\rect \subset \spacetime$ that, 
whenever sufficiently smooth, satisfy
\begin{equation}\label{eq:minisurfa}
H_\rect=0,
\end{equation}
where $H_\rect$ is the spacetime lorentzian mean curvature of $\rect$,
see for instance \cite{BoIn:34,neu,ViSh:94,Zw:04} and references therein.
A particular case, relevant in physics,
is the one of surfaces, namely when $h=2$.
Under this assumption, minimal surfaces are called  
(classical) closed relativistic strings. 

Equation \eqref{eq:minisurfa} is called the lorentzian minimal surface equation;
differently from the riemannian case, 
it can be regarded as a geometric evolution equation, 
which is hyperbolic in character, due to the signature of the  
lorentzian metric $\eta = {\rm diag}(-1,1,\dots,1)$ considered 
on the ambient space $\spacetime$.
A general short-time existence result of
 smooth solutions to \eqref{eq:minisurfa}
has been obtained
 in \cite{milbredt}. 
However, if one is interested in solutions defined also
for long times, at least one relevant obstruction arises. 
As a matter of fact,
unless the manifold is sufficiently
close to a linear subspace \cite{Br:02}, \cite{Li:04}, 
generically singularities appear \cite{ViSh:94,EH}. 
The onset of singularities is 
one of the main motivations of the present paper: 
indeed, we are interested in weak solutions to the lorentzian minimal surface 
equation,  which are globally defined in time.
As we will see, our approach
is based on the concept of {\it lorentzian varifold}, and is
inspired by the notions of varifold and stationary varifold
introduced by 
Almgren \cite{Alm:65} and developed by 
Allard in the euclidean and riemannian
setting \cite{Al:72,Si:83}. There are
several differences between the riemannian setting and 
the lorentzian one; however, the notion of stationarity 
of a varifold remains the natural concept 
generalizing the zero mean curvature condition \eqref{eq:minisurfa}. Concerning singularities, 
we point out that the  varifolds representation
of the evolving manifolds is parametrization
free, so that in particular changes of topology are allowed.  
A relevant part of the present paper is devoted to the analysis 
 of the concepts of area, first variation and rectifiability 
(according to the proposed definition of weak solution) and to the study of 
compactness and closure
properties for the class of stationary lorentzian varifolds. Several examples
motivate and illustrate our theory.
Notice that the theory we develop here concerns nonspacelike (or causal)
varifolds, as we are interested in generalizations of timelike minimal
surfaces.

\smallskip
Insights on the analysis 
of \eqref{eq:minisurfa} come 
from the study of the asymptotic limits
of solutions to the hyperbolic Ginzburg-Landau (HGL) equation. 
As shown in \cite{neu} by a formal asymptotic expansion argument and for
$h=N=2$, smooth
solutions to \eqref{eq:minisurfa} can be approximated
by solutions to HGL. 
Then, it has been rigorously shown in \cite{J:10} that
solutions to HGL with well-prepared initial data
converge, in a suitable sense, to a smooth lorentzian minimal
submanifold,  provided the latter exists.
 Again, due to the presence of singularities, 
the validity of this convergence result is restricted to
short times.
On the other hand, a preliminary analysis of the 
limit behaviour of HGL
within the varifolds framework has been pursued in \cite{BNO:09},
without restricting to short times, but under
rather strong assumptions on the limit varifold solution. 
In particular, in \cite{BNO:09} it is proposed a first notion of weak 
solution to \eqref{eq:minisurfa} that, in the language of the present paper,
coincides essentially with what we have called here a 
stationary (lorentzian)
rectifiable varifold with {\it no part ``at infinity''}, 
that is no null
(or lightlike) part. 
The assumptions made in \cite{BNO:09} exclude a priori
various examples of weak solutions to \eqref{eq:minisurfa}, such as
singular minimal surfaces where part of the energy is concentrated on
null subsets
of positive measure.
In addition, a limit of a sequence of varifolds considered in \cite{BNO:09} 
is not
necessarily a varifold of the same type even imposing 
uniform lower bounds on the densities.
It is therefore necessary
to relax the definition of rectifiable varifold.
In Definitions \ref{def:timerect} and \ref{def:weaklyrectifiable} we introduce 
the notion of {\it rectifiable} and {\it weakly rectifiable varifold}
respectively,
in the effort of covering 
all relevant examples of limits of minimal surfaces 
at our disposal\footnote{Incidentally, 
in Proposition \ref{proone} we show that for a 
stationary one-dimensional varifold, the concepts of 
rectifiability and weak rectifiability coincide, under a mild
condition on the supports.}, and to
hopely capture limits of 
solutions to HGL. In these two latter
definitions, also null parts are taken into account; in particular,
a rectifiable varifold generalizes the notion of smooth nonspacelike
(or causal) submanifold. 

One of the difficulties in adapting the varifold language
to the hyperbolic case is related to 
 the presence of null parts (possibly with
positive $\dimension$-dimensional measure). In particular, 
troubles arise due to the lack of 
compactness of the embedding 
of the grassmannian of timelike $\dimension$-planes into the space of matrices.
Roughly speaking, in order to keep the property that 
a sequence of varifolds with a uniform bound on the integral
in time of the energy has a converging subsequence, 
we are forced to define the varifolds
on the compactification of the grassmannian. This produces
 a term ``at infinity'' in the varifold
expression.
In this respect, we point out that we often  find to be more natural to describe
a varifold $V$,  splitted as a ``purely timelike''
part $V^0$ and a ``null'' part $V^\infty$ (or part at ``infinity'', which
was excluded in \cite{BNO:09}), 
in two different 
sets of variables: namely, $V^0$ is expressed in the variables
prior to the embedding (and in these variables we denote $V^0$ by $\widetilde V^0$)
while $V^\infty$ is expressed in the compactified variables.

It is worth noticing here that
for a stationary varifold it is still possible to 
define the analog of the notions 
of relativistic energy and momentum, which {\it 
turn out to be conserved quantities}.

\smallskip

\begin{figure}
\begin{center}
\includegraphics[height=4cm]{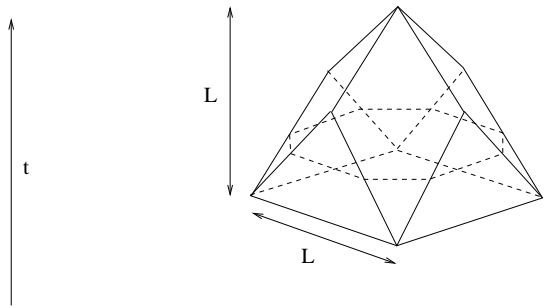}
\smallskip
\caption{\small 
The spacetime evolution of the square in Example \ref{exa:square}: 
this solution must be extended by periodicity}
\label{fig:square3D}
\end{center}
\end{figure}

To have a flavour
of what kind of solutions we can include using the notion 
of weakly rectifiable varifold, take
$N=2$ and $h=2$, and consider 
the string-type solution (introduced in \cite{BHNO}) starting at time
zero from a
square $[-L/2,L/2]^2$, with zero initial velocity. In Figure \ref{fig:square3D}
we show the time-track of the solution (the picture should be 
continued periodically in time for $t>0$ and then reflected for
$t <0$), see 
Example \ref{exa:square} for a detailed discussion.
The interest in this example relies on the fact
that it represents a Lipschitz (actually, a {\it polyhedral}) 
minimal surface containing various null segments,
which in Figure \ref{fig:square3D} are the 
four segments meeting at the upper vertex.  
If one describes parametrically this polyhedral surface as the image of 
a Lipschitz map $(t,u) \in \R \times [0,L) \to 
(t, \gamma(t,u)) \in \R^{1+2}$, where 
$\gamma$ is a weak  solution of the linear wave system
\eqref{linearwave},
it turns out that the set of all $(t,u)$ where ($\gamma_u$ exists and)
$\gamma_u(t,u)=0$ has positive Lebesgue measure. Therefore,
in a parametric language, all these points should be considered as 
singular points. Nevertheless, we can associate to such a solution
a stationary weakly rectifiable varifold admitting the conservation 
of energy.

While a stationary rectifiable varifold generalizes
the concept of lorentzian minimal submanifold,
a stationary weakly rectifiable varifold is rather a limit of stationary
rectifiable
varifolds and its support in $\spacetime$ is {\it not}, in general,
a minimal submanifold, i.e., \eqref{eq:minisurfa} is not necessarily 
satisfied even in regions where it is smooth:
see for instance the cylindrical strings \cite{BHNO} considered 
in Example \ref{exa:sub}.
This phenomenon is due to possible strong oscillations of the tangent spaces
and was, at a formal level, already observed in the paper \cite{neu},
see also \cite{ViSh:94}.
The presence of oscillations of the tangent spaces requires 
to consider the barycenters $\overline P$ (resp.  $\overline Q$) 
of a lorentzian orthogonal
projection on the timelike part (resp. on the null part) 
of 
the varifold 
rather than the orthogonal projections $P$ (resp. $Q$) 
itself: this is
reminiscent of 
the notion of
generalized Young measure \cite{DM:87}, \cite{AB:97}. 

Even if the class of weakly rectifiable stationary varifolds is a rather huge
set of weak solutions to \eqref{eq:minisurfa}, 
it still turns out 
to be  not closed under
varifolds convergence.
As we shall illustrate  in Example \ref{exa:patologico},
limits of stationary weakly rectifiable varifolds
with a uniform energy bound are stationary, but may fail to be weakly rectifiable.
On the other hand,
we expect this closure property to be valid for one-dimensional varifolds.
It can also be of interest to recall that 
the closure of two-dimensional minimal
surfaces (i.e., strings)
has been completely charaterized, at least in a parametric setting, leading
to the concept of subrelativistic string
(see \cite{ViSh:94,neu,Br:05,BHNO} for a detailed discussion):
from the positive side, it turns out that the  stationary varifolds  we are proposing
{\it contain and extend} the notion of subrelativistic string.

Before summarizing the content of the paper, 
another remark is in order. 
We  do not have a  weak-strong uniqueness result
for our generalized solutions. In particular,
assuming that the support of a stationary (weakly) rectifiable 
varifold with multiplicity one
 coincides with a regular solution to \eqref{eq:minisurfa} for 
short times, we do not know whether 
it coincides with such a solution as long as the latter is defined.
We observe that, in view of splitting/collision
Example \ref{exa:collsplit}, the condition that the varifold has multiplicity one is essential. 
We also note that it is not
difficult to check that the subrelativistic strings verify such a
uniqueness property, in view  
of the representation formula \eqref{forsub}.

\smallskip

The plan of the paper is the following. In Section \ref{sec:not}
we introduce some notation and some standard definitions from lorentzian
geometry.
In Section \ref{sub:themap}
we describe our embedding of the set of timelike $\dimension$-planes
in the vector space of all $(\dimspacetime+1)\times (\dimspacetime+1)$ real matrices
(Definition \ref{def:Pr}), and
its compactification $\immgrass$
via the map $q$ (Definition \ref{def:mappa}). 
We also need to describe the embedding of all
null $\dimension$-planes, see formula 
\eqref{eq:pronullo}\footnote{Null  $\dimension$-planes are in the boundary of the grassmannian of
timelike $\dimension$-planes, compare the proof of Lemma \ref{lem:compact}.}.
Some necessary tools of geometric measure theory
are given in Section \ref{sub:GMT}.
On the basis of the definition of lorentzian projection on an 
$h$-dimensional subspace (Definition 
\ref{rapk}),
the lorentzian tangential divergence of a vector field is given in 
formula \eqref{eq:lordivtang}. 
Our first result is Theorem \ref{lem:areanonparametrica}, where 
we find an expression of the lorentzian $\dimension$-dimensional
area element (denoted by $\sigma^\dimension$ in the sequel, to keep
distinct from the euclidean $h$-dimensional Hausdorff
measure $\mathcal H^h$) independent
of parametrizations, 
namely
\begin{equation}\label{arba}
\sigma^h(B) = \int_B \sqrt{-\nu_t^2 + \sum_{i=1}^N 
(\nu_{x^i})^2} ~d\mathcal H^h.
\end{equation}
Here the covector field $\nu$ with time-space components
$(\nu_t, \nu_x)$ is defined as $\nu 
:= \frac{\eta\n_1}{
\vert \n_1\vert_{\rm e}}$, where 
$\n_1$ is a {\it distinguished} vector in the normal space
(see Definition \ref{def:Pr}), and $\vert \cdot\vert_{\rm e}$
is the euclidean norm. 
We stress that \eqref{arba} is valid in arbitrary codimension $N+1-h$.
In Corollary \ref{cor:area} we express $\sigma^h(B)$ using
the horizontal velocity vector $\vel$ defined in \eqref{eq:horvel},
possibly also integrating first in time and then 
on the time-section. 
We conclude Section \ref{sub:GMT} 
with Theorems \ref{teo:GG} and \ref{no} (first variation of area) which, together with formula 
\eqref{divPH}, represent a first link between geometric measure
theory and classical lorentzian differential geometry. 
In Section \ref{sec:timelike} we introduce the class of test
functions $\mathcal F$ (Definition \ref{deftest}): the definition is given
in such a way that the recession function $f^\infty$ (Definition 
\ref{def:recession}) 
is well defined for any
 $f \in
\mathcal F$. The class
$\LV$ of lorentzian varifolds is introduced in Definition 
\ref{defvar}, in duality with the class $\mathcal F$. 
Stationarity is introduced in Definition \ref{defsta}:
it is based on formula \eqref{eqfirst},
and on the notion of lorentzian tangential divergence.
The definition 
is formally the same as in the riemannian case,
and (as in that case) it acquires a clear meaning 
looking at the first variation of area, in this case 
the lorentzian $h$-dimensional
area $\sigma^h$.  The splitting of 
a varifold into its timelike part $V^0$ and its null part $V^\infty$ is given
in Definition \ref{def:VzeroVinfinito}, and the (already
mentioned) measure $\widetilde V^0$ is defined 
in \eqref{eq:Vzerotilde}. The disintegrations of these
three measures are given in \eqref{decodeco} of Section \ref{subsec:disi},
through which we can write the action of a varifold on a test
function in a more useful way (Lemma \ref{lem:azione}). 
In the same section we introduce the barycenters
$\overline P$ and 
$\overline Q$.
These are concepts involving only the part of $V$ on the grassmannian
(see Definition \ref{defbar}) and are crucial in the study of
weakly rectifiable varifolds, which are often obtained as weak
(and strongly oscillating) limits of smooth timelike
lorentzian minimal surfaces.   Proper, rectifiable
and weakly rectifiable varifolds are introduced in Section \ref{sec:firstproperties},
together with some preliminary properties. It is interesting
to observe that, under suitable circumstances, 
it is still possible to derive a distributional 
stationarity 
equation in the weakly rectifiable case: this is accomplished in
Section \ref{unce}. 
Section \ref{seccons} has a central role: here we prove
various conservation
laws for stationary varifolds. To properly introduce the notion
of relativistic energy and momentum (Definition \ref{def:enemome})
we need to disintegrate the projected part on $\spacetime$ 
of the varifold
using the Lebesgue measure on the time-axis:
see \eqref{mipre}. The analog quantities in case the 
varifold reduces to a smooth timelike submanifold are the usual
relativistic energy and momentum, see Remark \ref{rem:consrett}.
In Section \ref{sub:relstri} we prove one of the main results of 
the paper,
namely that {\it it is always  possible to 
associate with a relativistic 
string a stationary rectifiable varifold} (Theorem \ref{senzalabel}), {\it and 
with a subrelativistic string a stationary
weakly rectifiable varifold} (Theorem \ref{senzanome}).
We believe these two results to
be an encouraging indication for the validity of our notion
of generalized solution to the lorentzian minimal surface equation 
\eqref{eq:minisurfa}.
Examples of varifolds associated with relativistic and subrelativistic
strings are given in Section \ref{subkink}, in particular 
cylindrical strings and the already mentioned polyhedral string. 
Section \ref{secexa} treats two other examples, 
the second one being rather interesting. 
\begin{figure}
\begin{center}
\includegraphics[height=4cm]{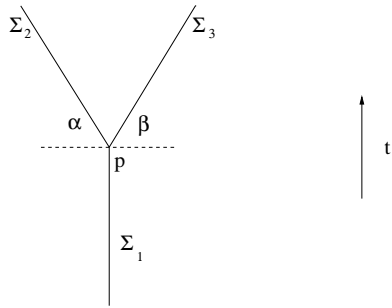}
\smallskip
\caption{\small A stationary triple junction. 
}\label{fig:creation}
\end{center}
\end{figure}
In the first Example \ref{exa:subspace} we show that 
our theory allows to rigorously prove that a null $h$-plane
is minimal, despite the fact that normal vectors in this case 
are not well defined. The second Example \ref{exa:collsplit} describes
a one-dimensional varifold associated with a splitting (or, using
time reversal, a collision). We consider an incoming half-line,
for instance a vertical half-line $\Sigma_1$, with a real positive multiplicity
$\theta_1$ on it; next we make the half-line split, at a triple
junction $p \in \R^{1+1}$, 
into two timelike half-lines $\Sigma_2$, $\Sigma_3$: see 
Figure \ref{fig:creation}. We then focus on the following problem:
which conditions one must impose on the splitting angles $\alpha$, $\beta$
and on the multiplicity $\theta_i$ on $\Sigma_i$ ($i=2,3$) 
in order Figure \ref{fig:creation} to represent a stationary
$1$-varifold in $\R^{1+1}$? This problem has a (nonunique) solution,
which can be obtained inspecting the weak notion of stationarity
around the triple junction. It turns out that solutions
are obtained imposing a sort of 
weighted balance condition at $p$ involving the three lorentzian normal 
vectors to the three half-lines: see equation \eqref{tripunti}
and Figure  
 \ref{fig:vectors}. 
Interestingly, the problem can be equivalently
solved imposing the conservation laws, see equations
\eqref{eq:conscons}. We also discuss the case when $\alpha$ and 
$\beta$ tend to the null directions.
In Section \ref{sec:pathological} 
we present an elementary example concerning 
the limit of a zig-zag piecewise affine 
curve having only null directions (Example \ref{exa:zigzag}), 
and two rather pathological examples.
In Example \ref{exa:patella} 
we show a not rectifiable stationary
purely singular varifold, obtained as a limit of a sequence
of rectifiable stationary varifolds; in Example \ref{exa:patologico}
we show a not rectifiable stationary
purely diffuse varifold, obtained as a limit of a sequence
of rectifiable stationary varifolds. These two examples 
illustrate the difficulty of characterizing the closure of the 
class of (weakly) rectifiable varifolds.
In the Appendix (Section \ref{sec:mea}) we recall various concepts
from measure theory, in particular the generalized Radon-Nikod\'ym
theorem (Theorem \ref{th:gRN}), and the disintegration of a measure
(Theorem \ref{th:disint})
needed throughout the paper.

 %%%%%%%%%%%%%%%%%%%%%%%%%%%%%%%%%%%%%%%%%%%%%%%%%%%%%
\section{Notation}\label{sec:not}
 %%%%%%%%%%%%%%%%%%%%%%%%%%%%%%%%%%%%%%%%%%%%%%%%%%%%%
Let $\dimspacetime\geq 1$. 
A point in the Minkowski spacetime $\spacetime$
will be usually denoted by 
$z = (t,x) \in \R \times \Rn$.
We indicate by 
$$
|\cdot|_{\rm e} \qquad {\rm and} \qquad
 (\cdot, \cdot )_{\rm e}
$$
the euclidean norm and scalar product in $\spacetime$, respectively. 
 We adopt the same notation for the euclidean norm and scalar product in $\Rn$.
We use the greek letters $\alpha,\beta, \gamma, \rho$ to denote 
indices ranging from $0$ to $N$, while we use the roman letters 
$\indispazuno, \indispazdue$  to denote spatial indices 
ranging from $1$ to $N$. 
Unless otherwise specified, we usually adopt the Einstein's convention of summation over spacetime
repeated 
indices or over repeated space
indices.

We denote by 
$\{e_0,\dots,e_\dimspacetime\}$ the 
canonical euclidean orthonormal basis of $\spacetime$.
We indicate by 
$\{e^0,\dots,e^\dimspacetime\}$ 
the dual
basis of $\{e_0,\dots,e_\dimspacetime\}$, i.e.,
$\langle e^\alpha, e_\beta\rangle=  \delta^\alpha_\beta$,
where
$\langle \cdot, \cdot \rangle$ denotes duality between 
covectors and vectors and $(\delta^\alpha_\beta) = {\rm Id}$ is the identity 
matrix.
Vectors have components labelled with upper
indices, 
while covectors 
have components labelled with lower
indices.

We set $\mathbb S^{N} := \{v \in \spacetime : \vert v\vert_{\rm e}=1\}$,
$\mathbb S^{\dimspacetime-1} := \{v \in \Rn : \vert v\vert_{\rm e}=1\}$.
Given $\pst \in \spacetime$ and $\rho>0$ we set $B_\rho(\pst) := \{\zeta \in 
\spacetime: \vert \zeta-\pst\vert_{\rm e} < \rho\}$.  

We denote by $\eta$ 
the lorentzian metric tensor in $\spacetime$,
$$
\eta ={\rm diag}(-1,1,\dots,1) = (\eta_{\alpha \beta}),
$$
and by 
$\eta^{-1} = (\eta^{\alpha\beta})$ the inverse of $\eta$.
Note that $\eta e_0 = - e^0$ and $\eta e_{\indicespaziale}
= e^{\indicespaziale}$ for $\indicespaziale \in \{1,\dots,N\}$.

Given $v = (v^0,\ldots,v^N)$ and  $w = 
(w^0,\ldots,w^N)$ vectors in $\spacetime$, we let 
\[
(v, w) := 
\eta_{\alpha\beta}^{}
~ v^\alpha 
w^\beta = v_\beta w^\beta
\]
be the {\it lorentzian} scalar product between $v$ and $w$, 
where  $(v_0,\dots,v_N)$ are the components of  the covector corresponding to $v$.

We recall that a vector $v \in \spacetime\setminus \{0\}$ is called 
\begin{itemize}
\item[-] 
spacelike 
if $(v,v) >0$;
\item[-] timelike if $(v,v) <0$;
\item[-] null if $(v,v) =0$;
\item[-] nonspacelike if $(v,v) \leq 0$, namely if it is either timelike
or null.
\end{itemize}

If $v\in \spacetime$ is either spacelike or null, we let
$$
\vert v\vert := \sqrt{(v,v)} \geq 0.
$$
We adopt a similar notation for covectors.

Occasionally, the time component $v^0$ of the vector $v
\in \spacetime$ is denoted by $v_t
\in \R$,
and the space component $(v^1,\dots,v^N)$ of $v$ by $v_x\in \Rn$.

%%%%%%%%%%%%%%%%%%%%%%%%%%%%%%%%%%%%%%%%%%%%%%%%%%%%%%%%%%%%%%%%%%%%%%% 
\section{The set $T_{h,\dimspacetime+1}$, the
map $\mappa$ and the set $\immgrass$}\label{sub:themap}
%%%%%%%%%%%%%%%%%%%%%%%%%%%%%%%%%%%%%%%%%%%%%%%%%%%%%%%%%%%%%%%%%%%%%%%
Let $h \in \{1,\dots,\dimspacetime\}$. 
Given an unoriented $\dimension$-dimensional vector space
$\Pi\subset \spacetime$ 
(an  $\dimension$-plane for short), 
we let
$$ 
\Pi^{\bot} := \Big\{\zeta \in 
\spacetime : (\zeta, \xi) =0 ~ ~\forall \xi \in \Pi\Big\}
$$ 
be the subspace orthogonal to $\Pi$ in the {\it lorentzian} sense.

We recall \cite{On:} that 
$\Pi$ is called
\begin{itemize}
\item[-] 
 timelike if 
${\mathrm n}$ is spacelike
for all ${\mathrm n}\in \Pi^{\bot}$, and in this case
${\rm dim}(\Pi^{\bot}) = \dimspacetime+1-\dimension$;
\item[-] null if $\n$ is spacelike or null for all $\n \in \Pi^\perp$, 
and at least one $\n \in \Pi^\perp\setminus \{0\}$ is null;
%$\Pi$ contains a nonzero null vector and if it does not contain
%any timelike vector;
\item[-] nonspacelike (or causal) if $\Pi$ is either timelike or null.
\end{itemize}

The set of all timelike $\dimension$-planes is open, and 
we embed it
%the set of timelike $\dimension$-planes
into the vector space of $(\dimspacetime+1)\times (\dimspacetime+1)$-real 
matrices $M_{\dimspacetime+1}
\simeq \R^{(\dimspacetime+1)^2}$ 
as follows: we associate with a timelike $\dimension$-plane $\Pi$ the matrix $P_\Pi$ 
corresponding to the lorentzian orthogonal projection $\spacetime \to \spacetime$ onto $\Pi$ 
(see also \cite{peter}).
More precisely we give the following
\begin{Definition}[{\bf 
The matrix $P_\Pi$ and the vectors $\n_1,\dots,\n_{\dimspacetime+1-\dimension}$}]\label{def:Pr}
Let $\Pi$ be a timelike $\dimension$-plane. We define
\begin{equation}\label{rapk}
P_\Pi := {\rm Id} - \sum_{j=1}^{\dimspacetime+1-\dimension} \n_j\otimes \eta \n_j,
\end{equation}
where 
\begin{itemize}
\item[-]  $\n_1,\dots \n_{\dimspacetime+1-\dimension} \in \Pi^{\perp}$ 
are spacelike vectors, 
\item[-] the time component  of $\n_1$ is nonnegative,
\item[-] $\n_2,\dots \n_{\dimspacetime+1-\dimension}$ have
vanishing time component,  
\item[-] for $i,j\in \{1,\ldots,\dimspacetime+1-\dimension\}$
\begin{equation}\label{eq:normalizzni}
(\n_i,
\n_j)= \begin{cases}
1 & {\rm if}~ i=j,
\\
0 & {\rm if}~ i \neq j.
\end{cases}
\end{equation}
\end{itemize}
\end{Definition}

Sometimes we write $P$ or also
$P(\n_1,\dots,\n_{\dimspacetime+1-\dimension})$
in place of $P_\Pi$. Despite the index $j$ is repeated, 
we prefer not to drop the symbol of summation in \eqref{rapk}. 

\smallskip

The $(1,1)$-tensor $P$ in \eqref{rapk}, if applied to a vector (resp.
to a covector) gives a vector (resp. a covector).
Equation \eqref{rapk} written in components reads as
$$
P(e^\alpha, e_\beta)=
\langle e^\alpha, P(e_\beta)\rangle=P_\beta^\alpha = \delta_\beta^\alpha - \eta_{\beta\gamma} \n_j^\alpha \n_j^\gamma,
\qquad \alpha, \beta \in \{0,\dots, \dimspacetime\}.
$$
\begin{Remark}\label{rem:parecchioutile}\rm 
Let $P$ be as in \eqref{rapk}. 
Then:
\begin{itemize}
\item[(i)] $P$ is not necessarily 
symmetric, while $\eta P$ is symmetric;
\item[(ii)] 
the restriction of $P$ to $\{0\} \times \Rn$ is symmetric, since $\eta$ acts 
as the
identity on $\{0\}
\times \Rn$;
\item[(iii)] $P_\alpha^\alpha =
\dimspacetime+1 - \sum_{j=1}^{\dimspacetime+1-h} (\n_j, \n_j) =
 \dimension$;
\item[(iv)] given a Lorentz transformation $L$, the lorentzian
orthogonal projection onto $L(\Pi)$ is given by $LPL^{-1}$.
\end{itemize}
\end{Remark}

\begin{Remark}\label{rem:indepchoice}\rm
$P$ does not depend on the choice of 
the $(\dimspacetime+1-\dimension)$-tuple of vectors 
$\n_1,\dots,\n_{\dimspacetime+1-\dimension}$ satisfying the properties
listed in Definition \ref{def:Pr}. Indeed:
\begin{itemize}
\item[-] case 1: $e_0 \in P$. Then $\n_1$ has vanishing time component, and  
the restriction of $P$ to 
$\Pi \cap (\{0\}\times \Rn)$
equals (letting ${\rm Id}_\dimspacetime$ the identity matrix
of $\Rn$)
$$
{\rm Id}_\dimspacetime - \sum_{j=1}^{\dimspacetime+1-\dimension} \n_j\otimes \eta \n_j 
= 
{\rm Id}_\dimspacetime - \sum_{j=1}^{\dimspacetime+1-\dimension} \n_j\otimes \n_j,
$$ 
which is an orthogonal euclidean projection, hence independent
of the choice of the $(N+1-\dimension)$-tuple of vectors 
$\n_1,\dots,\n_{\dimspacetime+1-\dimension}$.
\item[-] case 2: $e_0 \notin P$.
The restriction of $P$ to 
$\Pi \cap (\{0\}\times \Rn)$
equals 
$$
{\rm Id}_\dimspacetime - \sum_{j=2}^{\dimspacetime+1-\dimension} \n_j\otimes \eta \n_j 
= 
{\rm Id}_\dimspacetime - \sum_{j=2}^{\dimspacetime+1-\dimension} \n_j\otimes \n_j,
$$ 
which is an orthogonal euclidean projection, hence independent
of the choice of the $(N-\dimension)$-tuple of vectors 
$\n_2,\dots,\n_{\dimspacetime+1-\dimension}$;
\item[-] $\n_1 \in \Pi^\perp \cap  
({\rm span}\{\n_2, \dots,
\n_{\dimspacetime+1-h}\})^\perp$, and 
$({\rm span}\{\n_2, \dots,
\n_{\dimspacetime+1-h}\})^\perp = {\rm span}\{\Pi, e_0\}$. Hence
\begin{equation}\label{unid}
{\rm span}\{\n_1\} 
= \Pi^\perp \cap {\rm span}\{\Pi, e_0\}.
\end{equation}
In particular 
$\n_1$ is uniquely determined, since the right hand side of \eqref{unid} is 
one-dimensional, $(\n_1,\n_1)=1$, and
by assumption $\n_1^0 \geq 0$.
\end{itemize}
Notice that if $e_0 \in P$  the vector $\n_1$ is not uniquely determined.
\end{Remark}

\begin{Definition}[{\bf The set $\timelikekplanes$}]
We denote by $\timelikekplanes \subset M_{\dimspacetime+1}$ the set 
of all $(\dimspacetime+1)\times (\dimspacetime+1)$-real matrices $P=P_\Pi$ corresponding
to timelike $\dimension$-planes $\Pi$ 
in the sense of Definition \ref{def:Pr}.
\end{Definition}

As it is well known, as $\n \in \Pi^{\perp}$ approaches the light cone,
it tends to become parallel to $\Pi$, and its euclidean
norm $\vert \n \vert_{\rm e}$ tends to $+\infty$. Therefore the set
$\timelikekplanes$
is not bounded.
For our purposes, the closure of $\timelikekplanes$ needs to be compactified.
We choose a way to compactify $\timelikekplanes$
which consists in dividing
$P$ by its ${}^0_0$ component. Let us be more precise.

Given $P \in \timelikekplanes$ 
we have
\begin{equation}\label{eq:P00}
P_{0}^0 
= 1 + \sum_{j=1}^{\dimspacetime+1-\dimension} (\n_j^0)^2 =  1 + (\n_1^0)^2 \geq 1,
\end{equation}
and 
\begin{equation}\label{eq:n0lambda}
P_{0}^{\indicespaziale} = -\sum_{j=1}^{\dimspacetime+1-\dimension} 
\eta_{0\beta}
 \n_j^\beta 
\n_j^{\indicespaziale} 
 =\n_1^0 
 \n_1^{\indicespaziale}, \qquad \indicespaziale \in \{1,\dots,\dimspacetime\}.
\end{equation}

\begin{Remark}\label{rem:controllo}\rm The set $\{v \in \spacetime : v ~{\rm spacelike}, ~
\vert v\vert^2=1\}$ is unbounded. However, if $v = (v_t, v_x)$, we have
$$
\vert v\vert^2=1 ~ \Rightarrow ~ \frac{\vert v_x\vert_{\rm e}^2}{1 + 2 (v_t)^2}
 \leq 1. 
$$ 
Therefore, a bound 
on the time component 
of a spacelike vector
of $\{v \in \spacetime : 
\vert v\vert^2=1\}$ gives a bound
on the euclidean norm of its spatial component.
\end{Remark}

We are now in a position to give the following
\begin{Definition}[{\bf The map $q$ and the set $\immgrass$}]\label{def:mappa}
We define the 
 map $\mappa : \timelikekplanes \to M_{\dimspacetime+1}$ as
\begin{equation}\label{eq:themapq}
\mappa(P) := \frac{P}{P_0^0} = \frac{P}{1 + (\n_1^0)^2}, \qquad P
= P(\n_1,\dots,\n_{\dimspacetime+1-\dimension}) \in \timelikekplanes, 
\end{equation}
and we set
$$
\immgrass:= \mappa(\timelikekplanes).
$$
\end{Definition}
\begin{Remark}\label{rem:compattificazione}\rm
The set $\immgrass \subset M_{\dimspacetime+1}$ is open and bounded 
(recall Remark
\ref{rem:controllo}), in particular
its closure $\overline{\immgrass}$ is compact. 
\end{Remark}

%%%%%%%%%%%%%%%%%%%%%%%%%%%%%%%%%%%%%%%%%%%%%%%%%%%%%%%
\subsection{Projections on null $\dimension$-planes}
%%%%%%%%%%%%%%%%%%%%%%%%%%%%%%%%%%%%%%%%%%%%%%%%%%%%%%

The following lemma gives some insight on the geometry
of  $\overline{\immgrass}$.

\begin{Lemma}[{\bf Boundary of $\immgrass$}]\label{lem:compact}
The boundary of $\immgrass$ has the following representation\footnote{
When we write $\eta (1,\velnull)$ we implicitely consider 
$(1, \velnull)$ as a column.}:
\begin{equation}\label{eq:bdrySn}
\partial \immgrass = \left\{-(1,\velnull) \otimes \eta(1,\velnull) 
: \velnull \in \mathbb S^{\dimspacetime-1}
\right\}.
\end{equation}
In particular, $\partial \immgrass$ is independent of 
the integer $\dimension \in \{1,\dots, \dimspacetime\}$.
\end{Lemma}
\begin{proof}
Let $\{P_\indice\}\subset \timelikekplanes$ be a sequence of matrices,
define 
$Q_\indice := \mappa(P_\indice)$,  and assume that 
the sequence $\{Q_\indice\}$
converges
to some $Q \in \partial \immgrass$ as $\indice\to +\infty$.
Following the notation in Definition \ref{def:Pr}, 
for any $\indice \in \mathbb N$ we can  
write $P_\indice = {\rm Id} - \sum_{j=1}^{
\dimspacetime
+1-\dimension} 
 \n_j^{(\indice)} \otimes 
\eta \n_j^{(\indice)}$.
Let us indicate for notational 
 simplicity by 
$\lettera_\indice$ the square of the time component of 
$\n_1^{(\indice)}$, i.e., $\lettera_\indice := 
(\n_1^{(\indice)~\! 0})^2$, so that in particular 
$\displaystyle \lim_{\indice \to +\infty} \tau_\ell = +\infty$. 
Then
$$
Q_\indice 
=
\frac{{\rm Id} - \displaystyle \sum_{j=2}^{\dimspacetime
+1-\dimension}
\n_j^{(\indice)}\otimes
\eta \n_j^{(\indice)}}{
1 + \lettera_\indice 
}-
 \frac{\n_1^{(\indice)} \otimes 
\eta \n_1^{(\indice)}}{
1 + \lettera_\indice 
}.
$$
Hence
\begin{equation}\label{eq:paramsfere}
\lim_{\indice \to +\infty}
Q_\indice 
=-
\lim_{\indice \to +\infty}
\frac{ 
\n_1^{(\indice)}  \otimes 
\eta\n_1^{(\indice)}
}{
1 + \lettera_\indice 
} 
=
-
\lim_{\indice \to +\infty}
\frac{
\n_1^{(\indice)} }{
\sqrt{\lettera_\indice}
} \otimes 
\frac{
\eta\n_1^{(\indice)}
}{
\sqrt{\lettera_\indice}
}.
\end{equation}
Since by assumption 
$\n_1^{{(\indice)}~\!0}\geq 0$, it follows 
$\displaystyle \lim_{\indice \to +\infty} 
\frac{\n_1^{(\indice) ~\!0}}{\sqrt{\lettera_\indice}}
=1$. Taking also into account that 
$\{Q_\indice\}$ is a converging sequence, we can define
$$
\velnull := \lim_{\indice \to +\infty} 
\frac{(\n_1^{(\indice)})_{x}}{\sqrt{\lettera_\indice}} ~ \in \Rn.
$$
It then follows
from \eqref{eq:paramsfere} that
$$
-
\lim_{\indice \to +\infty}
\frac{
\n_1^{(\indice)} }{
\sqrt{\lettera_\indice}
} \otimes 
\frac{
\eta \n_1^{(\indice)}
}{
\sqrt{\lettera_\indice}
} 
=
- (1,\velnull) \otimes \eta (1,\velnull) 
= (1,\velnull) \otimes (1,-\velnull).
$$
It remains to show that $\vert \velnull\vert_{\rm e}=1$, and this follows passing to 
the limit as $\ell \to +\infty$ in 
the 
equality
$$
\sum_{\mathrm a=1}^{\dimspacetime+1}
\frac{\left(\n_1^{(\indice) ~\!\mathrm a}\right)^2}{\tau_\ell} =
\frac{1
+\left(\n_1^{(\indice) ~\!0}\right)^2}{\tau_\ell},
$$
since the left hand side converges to $\vert \velnull\vert^2_{\rm e}$,
while the right hand side converges to $1$.
\end{proof}

\begin{Remark}\label{rem:nullo}\rm
Note that $(1,\velnull)$ is a null vector.
Note also that 
from \eqref{eq:bdrySn} it follows that $\partial \immgrass$
is diffeomorphic to $\mathbb S^{\dimspacetime-1}$ (which is independent
of $\dimension$).
\end{Remark}

The intersection of a null  
$\dimension$-plane $\Pi$ with the positive light cone
is  a half-line, since $\Pi$ is tangent to the half-cone.
The euclidean orthogonal projection 
on $\Rn$ of such a half-line
is (a half-line) identified with a vector $\velnull \in \mathbb S^{\dimspacetime-1}$.
If $\dimspacetime + 1 - \dimension>1$, there are several null $\dimension$-planes $\Pi$
having the same $\velnull$.

\begin{Remark}\label{rem:rigido}\rm 
Let
$v_1 \neq v_2$ be vectors of $\Rn$ such that 
$\vert v_1\vert_{\rm e} = \vert v_2\vert_{\rm e}=1$.
The convex combination of 
$-(1,v_1) \otimes \eta(1,v_1)$ and $-(1,v_2) \otimes \eta(1,v_2)$
 is not 
of the form $-(1,v_3) \otimes \eta(1,v_3)$ 
for some $v_3 \in \mathbb S^{\dimspacetime-1}$.
Indeed, the image of 
$- (1, v_1)\otimes \eta (1,v_1)$ 
(resp. of $- (1,v_2) \otimes \eta (1,v_2)$) 
is generated by $(1,v_1)$ (resp. by $(1,v_2$))
so that the image of the mean value $\frac{1}{2}
\left(
- (1,v_1) \otimes \eta (1,v_1)
-
(1,v_2) \otimes \eta (1,v_2)
\right)$ is the timelike $2$-plane generated by $(1,v_1)$ and $(1,v_2)$.
\end{Remark}

\begin{Remark}\label{rem:buttazzo}\rm
The restriction  to $\{0\} \times\Rn$ 
of the matrix 
$-(1,\velnull) \otimes \eta(1,\velnull)$,
given in the proof of Lemma \ref{lem:compact},
is symmetric.
\end{Remark} 

We can now describe how to associate
with any null
$\dimension$-plane a  projection. 
Let $\Pi$ be a null $\dimension$-plane; we 
can uniquely choose 
$\velnull
\in \mathbb S^{\dimspacetime-1}$ satisfying  the condition $(1,
\velnull) \in \Pi$.
Then, we uniquely
associate with $\Pi$ the map 
\begin{equation}\label{eq:pronullo}
Q_\Pi := -(1,\velnull) \otimes \eta (1,\velnull) \in \partial \immgrass.
\end{equation}
Observe that the image of $Q_\Pi$ is contained in the span of $(1, \velnull)$.

%%%%%%%%%%%%%%%%%%%%%%%%%%%%%%%%%%%%%%%%%%%%%%%%%%%%%
\section{Geometric measure theory}\label{sub:GMT}
%%%%%%%%%%%%%%%%%%%%%%%%%%%%%%%%%%%%%%%%%%%%%%%%%%%%%
We denote by 
$\mathcal H^k$ the euclidean $k$-dimensional Hausdorff measure in $\spacetime$, for $k\in \{0,1,\dots,h\}$.

Let $\rect \subset \spacetime$ be an $\mathcal H^\dimension$-measurable
set. We say that $\rect$ is countably $\dimension$-rectifiable ($h$-rectifiable
for short) if $\mathcal H^\dimension$-almost all of the set
$\rect$ can be covered by a countable union of Lipschitz graphs,
see \cite{AmFuPa:00}. Therefore, an $h$-rectifiable set admits
tangent space $\mathcal H^\dimension$-almost evereywhere.

We let $T_\pst \manist$ be the tangent space to $\manist$ at 
$\punto$ (where it is defined).
$\manist$ is called timelike (resp. null) if 
$T_\punto\manist$ is timelike (resp. 
$T_\punto\manist$ is null) 
for all $z\in\manist$ where $T_z\Sigma$ exists 
(in particular $\mathcal H^\dimension$-almost
everywhere on $\manist$). 
$\manist$ is called {\it nonspacelike} (or causal) if 
$T_\punto\manist$ is either timelike or
null.

Let $\manist$ be nonspacelike and 
let $\pst \in \manist$ be such that $T_z\Sigma$ exists. 
We introduce the following notation: 
\begin{itemize}
\item[-] if $T_z \manist$ is timelike, we set 
$$
P_\manist(\pst): \spacetime \to \spacetime
$$ 
the lorentzian orthogonal projection onto $T_\pst \manist$,  that has the 
expression 
\begin{equation}\label{strap}
P_\manist (\pst) 
= {\rm Id} - \sum_{j=1}^{\dimspacetime
+1-\dimension} \n_j(\pst)\otimes \eta \n_j(\pst),
\end{equation}
where $\n_1(\pst) = {\n_1}_\Sigma(\pst),\dots,
\n_{\dimspacetime+1-h}(\pst)= 
{\n_{\dimspacetime+1-h}}_\Sigma(\pst)$ are required to satisfy the 
properties listed in Definition \ref{def:Pr}, provided $\Pi$ 
is replaced by $T_\pst \rect$;
\item[-] if $T_z \manist$ is null, we set 
$$
Q_\manist(z) = -(1, \velnull(z)) \otimes \eta (1, \velnull(z)),
$$
where $\velnull(z)$ is required to satisfy the properties listed at the end
of Section \ref{sub:themap}, provided the null $\dimension$-plane $\Pi$ 
is replaced by $T_\pst \rect$.
\end{itemize}

%%%%%%%%%%%%%%%%%%%%%%%%%%%%%%%%%%%%%%%%%%%%%%%%%%%%%%%%%%%%%%%%%%%
\subsection{Lorentzian tangential operators}\label{sub:lortan}
%%%%%%%%%%%%%%%%%%%%%%%%%%%%%%%%%%%%%%%%%%%%%%%%%%%%%%%%%%%%%%%%%%%
Assume that $\rect$ is timelike.
Let $\psi\in {\rm Lip}_c(\manist)$ (that is, $\psi$ is Lipschitz
on $\rect$ and with compact support). Suppose that  
there exists an extension $\Psi$ of $\psi$
with\footnote{Assuming $\Psi$ only Lipschitz (with compact
support) on $\spacetime$ 
does not guarantee
that $\Psi$ is differentiable on $\rect$. On the other hand,
in some examples 
we need to consider $\rect$ to be $h$-rectifiable, and not necessarily
of class $\mathcal C^1$, and therefore we cannot assume $\psi$ of class $\mathcal C^1$.}
 $\Psi\in \C^1_c(\spacetime)$.
We 
denote by $d_\tau \psi$ 
the lorentzian tangential differential 
of $\psi$ on $\manist$,
defined as
\begin{equation}\label{dopomoltedisc}
d_\tau\psi := P_{\manist}^*~ d \Psi \qquad
{\rm on}~  \manist,
\end{equation}
where 
$d\Psi$ is the differential of $\Psi$. 
In equation \eqref{dopomoltedisc}, $P_\Sigma^*$ is nothing else
but $P_\Sigma$, whenever considered as acting on the covector field
$d\Psi$, namely
\begin{equation}\label{strapstar}
P^*_\manist (\pst) d\Psi(\pst)  
= \left({\rm Id} - \sum_{j=1}^{\dimspacetime
+1-\dimension} \eta \n_j(\pst)\otimes \n_j(\pst)\right)
d \Psi(\pst).
\end{equation}
Note that 
$$
P^*_\Sigma(z)(e^\alpha, e_\beta)
= 
\langle P_\Sigma^*(z)(e^\alpha), e_\beta\rangle
= 
{P_{\manist}^*(z)}^\alpha_\beta= 
\langle e^\alpha,  P_\Sigma(z)(e_\beta)\rangle
=
{P_{\manist}(z)}^\alpha_\beta.
$$
Therefore, in the following we will identify $P_\Sigma$ with $P_\Sigma^*$,
and we will omit the ${}^*$ in \eqref{dopomoltedisc}.

Notice that the tangential differential of $\psi$ is independent of
the extension $\Psi$\footnote{If $\psi$ is zero
on $\rect$, the tangent space to $\Sigma$ is in the kernel 
of $d\Psi$. Hence, if $v$ is a vector, $\langle P^*_\Sigma d\Psi, v\rangle=
\langle d\Psi, P_\rect v\rangle =0$, since $P_\rect v$ is a tangent vector.}. 

Let $Y\in ({\rm Lip}_c(\rect))^{N+1}$. Assume that 
there exists an extension
$\mathcal Y \in (\mathcal C^1_c(\spacetime))^{N+1}$ of $Y$.  
We define the lorentzian tangential divergence ${\rm div}_\tau Y$ 
on $\manist$ as follows:
\begin{eqnarray*}
{\rm div}_\tau Y &:=& 
d\mathcal Y^\alpha_\alpha
- d\mathcal Y^\alpha_\beta
~\n_j^\beta~ \eta_{\gamma\alpha} ~\n_j^\gamma 
\qquad {\rm on}~  \manist,
\end{eqnarray*}
where $d \mathcal Y$ is the differential of $\mathcal Y$.
Such a tangential divergence is independent of the extension $\mathcal Y$.

Notice\footnote{
Given a tensor $T = T^\beta_\alpha$ of type $(1,1)$, we set
${\rm tr}(T) := T_\alpha^\alpha$.} that 
\begin{equation}\label{eq:lordivtang}
{\rm div}_\tau Y
= 
{\rm tr}\left(P_\manist ~d \mathcal Y\right)
\qquad {\rm on}~ \manist.
\end{equation}
Indeed, 
$$
{\rm tr}(P_\Sigma d \mathcal Y) 
= {\rm tr}\left( \big(
{\rm Id} - \sum_{j=1}^{\dimspacetime+1-\dimension}
\eta \n_j \otimes \n_j\big) d\mathcal Y\right) = 
d\mathcal Y^\alpha_\alpha
- d\mathcal Y^\alpha_\beta
~\n_j^\beta~ \eta_{\gamma\alpha}~ \n_j^\gamma
\qquad {\rm on}~
\manist.
$$
Note also that 
\begin{equation}\label{divproduct}
{\rm div}_\tau (\psi Y) = \psi\, {\rm div}_\tau Y + \langle d_\tau \psi,Y\rangle.
\end{equation}
Let
$T\in 
\mathcal {\rm Lip}_c(\manist;\R^{(\dimspacetime
+1)^2})$ be a 
 $(1,1)$-tensor field. Assume that there exists an extension 
$\mathcal T \in \mathcal C^1_c(\spacetime; \R^{(\dimspacetime
+1)^2})$ of $T$. 
We define the lorentzian
tangential divergence ${\rm div}_\tau T$ of $T$ as
\begin{equation}\label{eq:divtangtensor}
{{\rm div}_\tau T}_{~\!\!\alpha} := 
d \mathcal T_{\alpha\beta}^\beta - 
d\mathcal T_{\alpha\beta}^\rho
~\n_j^\beta~ \eta_{\gamma\rho}
~\n_j^\gamma \qquad {\rm on}~ \manist, \ \alpha \in \{0,\dots,\dimspacetime\},
\end{equation}
or equivalently
\begin{eqnarray*}
{\rm div}_\tau T(\pst) 
&:=& {\rm tr}\left(P_\manist(\pst) d \mathcal T(\pst)\right), 
\qquad \pst \in \manist.
\end{eqnarray*}
Finally, if $\rect$ is timelike and in addition
is of class $\mathcal C^2$, we let 
\begin{equation}\label{eq:defH}
H_\rect := \sum_{j=1}^{\dimspacetime
+1-\dimension}{\rm div}_\tau \n_j ~ \n_j
\qquad {\rm on}~ \rect,
\end{equation}
be the {\it lorentzian}
mean curvature vector of $\rect$. Observe that 
\begin{equation}\label{divPH}
{\rm div}_\tau P_\rect = 
-\eta\,H_\rect,
\end{equation}
since, using $\n_j^\beta {\n_j}_\beta=1$ for any
$j=1,\dots,N+1-h$, it follows
 ${{\rm div}_\tau P_\rect}_{~\!\alpha} 
= 
-\sum_{j=1}^{N+1-h} (d\n_j^\beta)_\beta {\n_j}_\alpha
-\n_j^\beta
(d{\n_j}_\alpha)_\beta = -\sum_{j=1}^{N+1-h} (d\n_j^\beta)_\beta {\n_j}_\alpha$.

%%%%%%%%%%%%%%%%%%%%%%%%%%%%%%%%%%%%%%%%%%%%%%%%%%%%%%%%%%%%%%%%%%%%%%
\subsection{The lorentzian $N+1-h$-codimensional area: parametrization free expression}\label{sub:lor}
%%%%%%%%%%%%%%%%%%%%%%%%%%%%%%%%%%%%%%%%%%%%%%%%%%%%%%%%%%%%%%%%%%%%%%
We recall \cite{Zw:04} that the 
$h$-dimensional lorentzian area $\mathcal S_h(\rect)$ of a timelike 
$\dimension$-dimensional rectifiable 
set $\immmap = X(\Omega) \subset \spacetime$, where 
$\Omega \subset \R^\dimension$ is an open set, and 
$X :
\Omega  
\to \spacetime$ is a Lipschitz
embedding, is given by 
\begin{equation}\label{eq:azioneS}
\mathcal S_\dimension(\immmap) = \int_{\Omega} \sqrt{- {\rm det} g}~ du_1 \dots du_\dimension,
\end{equation}
where $g$ is the 
matrix with components 
$$
g_{ij} := \left(X_{u_i}, X_{u_j}\right),
$$
that are almost everywhere defined in $\Omega$.

We are interested in 
representing $\mathcal S_h(\rect)$
using only the image of the map $X$. 
It is useful to 
introduce the following notation, valid for any $\dimension \in \{1,\dots,
\dimspacetime\}$.

\begin{Definition}[{\bf The covector field $\nu$}]\label{def:nu}
Let $\rect\subset \spacetime$ be a nonspacelike $h$-rectifiable set. 
We define $\mathcal H^h$-almost everywhere on 
$\Sigma$ the covector field 
$\nu = \nu_\Sigma$ as
\begin{equation}\label{def:nufi}
\nu(z) := 
\begin{cases}
\displaystyle \frac{\eta \n_1(z)}{\vert \eta \n_1(z)\vert_{\rm e}} 
& 
{\rm if}~ T_z\rect
~{\rm is~timelike},
\\
\\
 \displaystyle -\frac{1}{\sqrt{2}}
\eta (1,\velnull(z)) &
 {\rm if}~T_z\rect
~{\rm is~null}.
\end{cases}
\end{equation}
\end{Definition}
%%%%%%%%%%%%%%%%%%%%%%%%%%%%$\n_1$ per me e'  $n_\phi$
%%%%%%%%%%%%%%%%%%%%%%%%%%%%%%%%%%%%%%%%%%%%%%%%%
The covector field $\nu$ has unit euclidean 
norm, and its definition
does not involve the remaining normal vectors
$\n_2, \dots, \n_{\dimspacetime + 1 - \dimension}$. 
If $\dimspacetime+1-\dimension>1$ the covector field $\nu$ 
is nothing else but a suitable unit (in euclidean sense) 
covector normal to $\rect$.

Observe that at points $z \in \rect$ where $T_z\rect$ is timelike we have
\begin{equation}\label{mattex}
\n_1(z) = \frac{\eta^{-1} \nu(z)}{\vert \eta^{-1} \nu(z)\vert},
\end{equation}
and $\eta^{-1}\nu(z) = \displaystyle \frac{\n_1(z)}{\vert \n_1(z)\vert_{\rm e}}$.

The following result gives the expression of the $h$-dimensional
area of a nonspacelike
manifold $\immmap$ in terms of 
$\nu$,  
in arbitrary codimension $N+1-h$. We write $\nu$ in components
as $\nu = (\nu_t, \nu_x) \in \R \times \R^\dimspacetime$.

\begin{Theorem}[{\bf $\dimension$-dimensional
area}]\label{lem:areanonparametrica}
Let $\rect\subset \spacetime$ be a nonspacelike $h$-rectifiable set.
For any $\dimension \in \{1,\dots,\dimspacetime\}$ we have 
\begin{equation}\label{sigman}
\mathcal S_\dimension(\immmap) 
=\int_{\immmap} |\nu| ~d\mathcal H^\dimension(\sptpt)
= \int_{\immmap} \sqrt{-\nu_t^2 + \vert \nu_x\vert_{\rm e}^2}
~d \mathcal H^\dimension(\sptpt).
\end{equation}
\end{Theorem}
\begin{proof}
The integral in \eqref{sigman} is 
restricted to the points $z$ of $\rect$ where $T_z\rect$ is timelike,
since otherwise the integrand vanishes.
Therefore, it is not restrictive to assume 
that $\rect$ is timelike.

Using \eqref{eq:azioneS} and the euclidean area formula \cite{Fe:68} we have 
$$
\mathcal S_\dimension(\immmap) = 
\int_{\immmap} \frac{\sqrt{- {\rm det} g}}{\sqrt{\vert {\rm det} G}\vert} 
~ d\mathcal H^\dimension(\sptpt),
$$
where $G$ is the matrix with components
$$
G_{ij} = (X_{u_i}, X_{u_j})_{\rm e}, \qquad i,j \in \{1,\dots,h\},
$$
and 
${\rm det} g$ and ${\rm det} G$ are calculated at $(u_1,\dots,u_h) = X^{-1}(\sptpt)$.
We choose a local parametrization $X$ around
a point $\overline u = (\overline u_1,\dots,\overline u_h) \in \Omega$
so that the time component of 
$X_{u_i}(\overline u)$ is zero for any $i=2,\dots,\dimension$, 
and moreover
$$
g(\overline u) = {\rm diag}\Big(
\left(X_{u_1}(\overline u), 
X_{u_1}(\overline u)\right), 
1,\dots,1\Big).
$$
Observe that 
$X_{u_1}(\overline u)$ is timelike, i.e., 
$\left(X_{u_1}(\overline u), X_{u_1}(\overline u)\right)<0$.
In this way we have 
${\rm det} g(\overline u)
= 
\left(X_{u_1}(\overline u), X_{u_1}(\overline u)\right)
$, and 
${\rm det} G(\overline u)
= \left(X_{u_1}(\overline u), X_{u_1}(\overline u)\right)_{\rm e}$.

Therefore, to prove \eqref{sigman} we have to show that 
$$
-\overline 
\nu_t^2 + \vert 
\overline \nu_x
\vert_{\rm e}^2 = 
-\frac{\left(X_{u_1}(\overline u), X_{u_1}(\overline u)\right)}
{\left(X_{u_1}(\overline u), X_{u_1}(\overline u)\right)_{\rm e}}, 
$$
where 
$\overline \nu_t := \nu_t(\overline \sptpt)$,
$\overline \nu_x :=  \nu_x(\overline \sptpt)$, 
$\overline \nu = (\overline \nu_t,\overline \nu_x)$,
and 
$\overline \sptpt = \map(\overline u)$.

By construction\footnote{Note that $e_0 \notin N_{\overline z}\Sigma$,
and $X_{u_1}(\overline u) \notin N_{\overline z} \rect$. Therefore
the inclusion 
$X_{u_1}(\overline u) \in 
{\rm span}\{N_{\overline 
z} \rect, e_0\}$ is equivalent to the inclusion 
$e_0
\in 
{\rm span}\{N_{\overline 
z} \rect, X_{u_1}(\overline u)\}$. 
} we have 
$$
X_{u_1}(\overline u) \in 
T_{\overline z} \immmap \cap {\rm span}\{N_{\overline 
z} \rect, e_0\}.
$$
In addition 
$$
T_{\overline z} \immmap \cap {\rm span}\{N_{\overline 
z} \rect, e_0\}=
T_{\overline z} \rect \cap {\rm span}\{\overline \nu, e_0\},
$$
since obviously 
$T_{\overline z} \immmap \cap {\rm span}\{N_{\overline 
z} \rect, e_0\}\subseteq
T_{\overline z} \rect \cap {\rm span}\{\overline \nu, e_0\}$,
and moreover
$T_{\overline z} \immmap \cap {\rm span}\{N_{\overline 
z} \rect, e_0\}$ is one-dimensional, and 
$T_{\overline z} \rect \cap {\rm span}\{\overline \nu, e_0\}\ne \emptyset$
by \eqref{def:nufi}. Hence 
$$
X_{u_1}(\overline u) \in 
T_{\overline z} \rect \cap {\rm span}\{\overline \nu, e_0\}.
$$
We now observe that
$(\vert \overline \nu_x\vert_{\rm e}^2, - \overline \nu_t \overline \nu_x)$
is orthogonal, in euclidean sense, to $\overline \nu$. In addition
$(\vert \overline \nu_x\vert_{\rm e}^2, - \overline \nu_t \overline \nu_x)
\in {\rm span}\{\overline\nu, e_0\}$, since, recalling also
that $\vert \overline \nu \vert_{\rm e}^2=1$,
$$
e_0 - \overline \nu_t 
\overline \nu = (1-\nu_t^2,-\overline \nu_t\overline\nu_x) 
= (|\overline \nu_x|_{\rm e}^2,-\overline \nu_t\overline \nu_x).
$$
It follows that 
$X_{u_1}(\overline u)$
is parallel to 
$(\vert \nu_x\vert_{\rm e}^2, - \nu_t \nu_x)$, 
and therefore there exists a constant
$\lambda \in \R\setminus \{0\}$ such that 
$$
X_{u_1}(\overline u) = \lambda
(\vert \nu_x(\overline \sptpt)\vert_{\rm e}^2, - \nu_t(\overline \sptpt) \nu_x
(\overline \sptpt)).
$$
Hence, since $\overline \nu_x\neq 0$,
$$
\frac{(X_{u_1}(\overline u), X_{u_1}(\overline u))}
{(X_{u_1}(\overline u), X_{u_1}(\overline u))_{\rm e}}=
\frac{
-\vert \overline \nu_x\vert_{\rm e}^4 + \overline \nu_t^2 \vert \overline \nu_x\vert^2_{\rm e}
}{
\vert \overline \nu_x\vert_{\rm e}^4 + \overline \nu_t^2 \vert \overline \nu_x\vert_{\rm e}^2
} = 
\frac{
-\vert \overline \nu_x\vert_{\rm e}^2 + \overline \nu_t^2 
}{
\vert \overline \nu_x\vert_{\rm e}^2+ \overline \nu_t^2  
} = 
-\vert \overline \nu_x\vert_{\rm e}^2 + \overline \nu_t^2.
$$
\end{proof}

\begin{Definition}[{\bf Lorentzian $\dimension$-dimensional
area}]\label{def:misuradilorentz}
Let $h \in \{1,\dots,N\}$.
Let $\rect \subset \spacetime$ be a nonspacelike $h$-rectifiable set. 
Given a Borel set $B \subseteq \rect$ we define
\begin{equation}\label{eq:sigmah}
\lormeas(B) := \int_B \sqrt{- \nu_t^2 + \vert \nu_x\vert^2_{\rm e}}~d\mathcal H^\dimension.
\end{equation}
\end{Definition}

\begin{Definition}[{\bf Horizontal velocity}]\label{def:velovecto}
Let $\rect\subset\spacetime$ be a nonspacelike $h$-rectifiable set. 
We
 define the horizontal normal velocity vector field $\vel$ at a 
differentiability point
$z$ of $\rect$ as
\begin{equation}\label{eq:horvel}
\vel(z) := 
\begin{cases}
\displaystyle \frac{{\n_1}^0}{\vert {\n_1}_x\vert_{\rm e}} ~\frac{{\n_1}_x}{\vert
{\n_1}_x\vert_{\rm e}} &  {\rm if~} T_z\rect {\rm ~is ~timelike},
\\
\\
\velnull(z) & {\rm if}~ T_z\rect {\rm ~is ~null}. 
\end{cases}
\end{equation}
\end{Definition}
The vector field $\vel(t,\cdot)$ 
 represents the 
normal velocity of the time slice
$$
\rect(t) := \rect \cap \{\sptpt^0 = t \}
$$
of $\rect$.
Notice that 
$$
\sptpt \in \rect \Rightarrow e_0 + \vel(\sptpt) \in T_\sptpt(\rect),
$$
since one checks directly that $(\n_i, e_0+\vel)=0$ for any
$i\in \{1,\dots,N+1-h\}$.

\begin{Remark}\label{rem:utile}\rm
At timelike points of $\rect$ we have,
using the definition of $\vel$ and $- (\n_1^0)^2
+\vert {\n_1}_x\vert^2_{\rm e} 
=1$, 
\begin{equation}\label{eq:n10v}
(\n_1^0)^2 = \frac{\vert \vel\vert_{\rm e}^2}{1 - \vert \vel\vert_{\rm e}^2}.
\end{equation}
Hence, from \eqref{eq:P00}, 
\begin{equation}\label{eq:poo}
P^0_0 = \frac{1}{1 - \vert \vel\vert_{\rm e}^2},
\end{equation}
and from \eqref{eq:n0lambda}
\begin{equation}\label{eq:pooo}
P_{0}^{\mathrm a} 
= \n_1^0 \n_1^{\mathrm a}  = 
 (1+ (\n_1^0)^2) \vel^{\mathrm a}
=  \frac{\vel^{\mathrm a}}{1-\vert \vel\vert_{\rm e}^2}, \qquad
\mathrm a \in \{1,\dots,N\}. 
\end{equation}
Note also that 
$$
\vert \vel\vert_{\rm e}^2 = \frac{(\n_1^0)^2}{\vert {\n_1}_x
 \vert^2_{\rm e}}
=
\frac{(\nu_t)^2}{\vert \nu_x\vert_{\rm e}^2}
= 
\frac{1-\vert \nu_x\vert_{\rm e}^2}{\vert \nu_x\vert_{\rm e}^2}.
$$
\end{Remark}

\begin{Corollary}\label{cor:area}
Let $\rect\subset \spacetime$ be a nonspacelike $h$-rectifiable set. 
For any Borel set
$B \subset \immmap$ we have 
\begin{equation}\label{eq:areavel}
\int_{B} d \lormeas = 
\int_{B} \sqrt{\frac{1 - \vert \vel \vert_{\rm e}^2}{1+\vert
\vel\vert_{\rm e}^2}}~ d\mathcal H^\dimension 
=
\int_{\R} \int_{B(t)} 
\sqrt{1 - \vert \vel \vert_{\rm e}^2}~ d\mathcal H^{\dimension 
-1} dt,
\end{equation}
where $B(t) := B \cap \{x^0=t\}$.
\end{Corollary}
\begin{proof} 
The first equality follows from \eqref{sigman} and 
$$
- (\nu_t)^2 + \vert \nu_x\vert_{\rm e}^2 = 
\left(- \frac{(\nu_t)^2}{\vert \nu_x\vert_{\rm e}^2} + 1\right) \vert \nu_x\vert_{\rm e}^2
=
 \left(- \vert \vel\vert_{\rm e}^2+ 1\right) \vert \nu_x\vert_{\rm e}^2
 = \frac{1-\vert \vel\vert_{\rm e}^2}{1 + \vert \vel\vert_{\rm e}^2}.
 $$
To prove the second equality in \eqref{eq:areavel} 
we recall the coarea formula on an $h$-rectifiable set \cite{Fe:68}: 
\begin{equation}\label{eq:trap}
\int_\immmap f ~d\mathcal H^h
= \int_{\R} \int_{\immmap(t)} \frac{f}{\vert \grad_\immmap p\vert_{\rm e}} ~d\mathcal H^{h-1} dt,
\end{equation}
where $p : \spacetime \to \R$ 
is defined as 
\begin{equation}\label{def:p}
p(t,x) := t, \qquad (t,x) \in \spacetime,
\end{equation}
 and $\grad_{\immmap}$
denotes the euclidean tangential gradient to $\rect$.
The assertion then follows taking 
$f = \displaystyle \frac{1-\vert \vel\vert^2_{\rm e}}{
1+\vert \vel\vert^2_{\rm e}}$ in \eqref{eq:trap}, and observing\footnote{
If $X_{u_1}$ is as in the proof 
of Theorem \ref{lem:areanonparametrica}, we have that $\grad_\Sigma p = 
\left(\grad p, \frac{X_{u_1}}{\vert X_{u_1}\vert_{\rm e}}\right)_{\rm e}
X_{u_1}
$. Since $X_{u_1}$ is parallel to $(1,-\frac{\nu_t\nu_x}{\vert \nu_x\vert^2_{\rm e}}) = (1,\vel)$, formula
\eqref{achi} follows.} that 
\begin{equation}\label{achi}
\vert \grad_\rect p \vert_{\rm e}^2 = \frac{1}{1  + \vert \vel \vert_{\rm e}^2}.
\end{equation}
\end{proof}

In the lorentzian setting we have 
the following integration by parts formula, which is at the 
core of the definition of stationary varifold. 

\begin{Theorem}[{\bf Gauss-Green Formula}]\label{teo:GG}
Let $\manist
\subset\spacetime$ be an $\dimension$-dimensional timelike 
embedded oriented submanifold without boundary  of class ${\mathcal C}^2$. Let 
$Y\in 
(\mathcal C^1_c(\manist))^{N+1}$ 
 be a vector field which 
is tangential to $\manist$. Then
\begin{equation}\label{eq:stazio}
\int_\manist {\rm div}_\tau Y~ d\lormeas =0.
\end{equation}
Therefore for any $\psi\in \mathcal 
C^1_c(\spacetime)$ and any  $Z \in 
(\mathcal C^1_c(\manist))^{N+1}$ 
\begin{equation}\label{GG}
\int_\manist \psi\, {\rm div}_\tau Z  \,d\H^\dimension = 
\int_\manist \psi~ (H_\rect,Z) \,d\H^\dimension-
\int_\manist \langle d_\tau\psi,Z\rangle \,d\H^\dimension.
\end{equation}
\end{Theorem}
\begin{proof}
Let $\mathcal Y\in \C^2(\spacetime)^{N+1}$ be a smooth extension 
of $Y$.
Let us  observe that 
$D Y = P_\manist~ d {\mathcal Y}$, where 
$D$ is the covariant derivative. To show this, we 
observe that $d\eta=0$, and in particular $P_\manist d \eta=0$. Hence, it is enough to 
prove \cite[Theorem 3.3.1]{Wald} that 
$P_\manist d$ is torsion free, and this can be proven 
as in \cite[pag. 11]{KoNo}. Then \eqref{eq:stazio} follows
from \cite[Theorem B.2.1, (B.2.26)]{Wald}.

We now set
$$
Z^\top := P_\rect Z, \qquad
Z^{\perp} := Z- Z^\top = 
\sum_{i=1}^{N+1-h} 
(Z,\n_i) \n_i.
$$
Then, using also \eqref{divproduct},
\begin{equation}\label{perpm}
{\rm div}_\tau Z^{\perp} = 
\sum_{i=1}^{N+1-h} 
{\rm div}_\tau \Big((Z, \n_i) \n_i\Big)= 
\sum_{i=1}^{N+1-h} 
(Z, \n_i) 
{\rm div}_\tau 
\n_i= 
(H_\rect, Z^{\perp}) = (H_\rect, Z).
\end{equation}
Assertion \eqref{GG} follows,
using \eqref{eq:stazio} (with $\psi Z^\top$ replacing $Y$) 
and \eqref{perpm}.
\end{proof}

\begin{Theorem}[{\bf First variation}]\label{no}
Let $\rect\subset \spacetime$ be an $h$-dimensional timelike
embedded submanifold withouth boundary of class $\mathcal C^1$. 
Let $Y
\in (\mathcal C^1_c(\spacetime))^{N+1}$, 
and let $\Omega \subset \spacetime$ be a bounded open set containing 
the support of $Y$.
For any $s \in \R$
and $z \in \spacetime$ define $\Phi_s(z) := z + s Y(z)$. 
Then 
\begin{equation}\label{eq:firstvariation}
\frac{d}{ds} \mathcal S_h(\Omega \cap \Phi_s(\Sigma))_{\big\vert s=0}
= \int_\Sigma {\rm div}_\tau Y ~d\sigma^h.
\end{equation}
\end{Theorem}
\begin{proof}
It follows arguing as  
in \cite{Al:72}, \cite{Si:83bis}\footnote{
In the case $h=N$ formula
\eqref{eq:firstvariation} follows arguing for instance as
in the proof of \cite[Theorem 5.1]{BePa:96} with the choice $\p^o(\xi^*) =
\sqrt{-(\xi_t^*)^2 + \vert \xi_x^*\vert_{\rm e}^2}$ (that part of 
the proof 
holds without assuming 
the convexity of $\phi^o$), $n_\phi=\n_1$ and 
$\nu_\phi= \eta \n_1$.}.
%\cite[Section 9]{Si:83}.
\end{proof}

%%%%%%%%%%%%%%%%%%%%%%%%%%%%%%%%%%%%%%%%%%%%%%%%%%%%%%%
\section{Lorentzian $\dimension$-varifolds}\label{sec:timelike}
%%%%%%%%%%%%%%%%%%%%%%%%%%%%%%%%%%%%%%%%%%%%%%%%%%%%%%%
The generalized manifolds we are interested in in this paper are the 
{\em lorentzian} $\dimension$-{\em varifolds} 
which, as we shall see in 
Definition \ref{def:VzeroVinfinito}, have a timelike
part and  a null part. Besides nonsmoothness, 
also the presence of a null
part is source of various difficulties.
As we shall see, the notion of lorentzian $\dimension$-varifold
is reminiscent of the 
generalized Young measures \cite{DM:87}, \cite{AB:97}.

Denote by $q^{-1} : \immgrass = q(\timelikekplanes) \to \timelikekplanes$ 
the inverse of the map $q$ introduced in Definition \ref{def:mappa},
namely
\begin{equation}\label{inversa}
q^{-1}(Q) = \Big(1 + (\n_1^0)^2\Big) Q,
\qquad 
Q= q\left(P(\n_1,\dots,\n_{\dimspacetime+1-\dimension})\right) \in \immgrass.
\end{equation}
Given $f \in \mathcal C(\spacetime\times \timelikekplanes)$ 
we define the composition
$$
\fq : \spacetime \times \immgrass \to \R
$$
of $f$ via the inverse of $({\rm Id}_{\spacetime}, q)$, 
furtherly divided by a positive factor,
as follows. 
\begin{Definition}[{\bf The map $\fq$}]\label{def:themapqf}
Given any pair $(\sptpt,Q)\in \spacetime \times {\immgrass}$,
where $Q= q(P(\n_1,\dots,\n_{\dimspacetime+1-\dimension})) \in \immgrass$,
we set
\begin{equation}\label{eq:qf}
\fq
(\sptpt,Q):= 
\frac{f(\sptpt,\mappa^{-1}(Q))}{\mappa^{-1}(Q)_{0}^0} = 
\frac{f(z, \left(1+(\n_1^0)^2) Q\right)}{1 + (\n_1^0)^2}.
\end{equation}
\end{Definition}

In the next definition we specify a class of admissible test
functions.

\begin{Definition}[{\bf The space $\F$}]\label{deftest}
We let $\F$ be the vector space of all  
functions $f\in 
\mathcal C(\spacetime\times \timelikekplanes)$ such that 
$\fq$ 
 can be continuously extended 
to $\spacetime \times \overline{\immgrass}$, and such an extension
(still denoted by $\fq$) has compact support. 
\end{Definition}
{}From Definition \ref{deftest} we have that 
a necessary condition satisfied by the elements of $\F$ is the following:
given $f\in \F$ 
there exists $\Lambda 
\in [0,+\infty)$
such that 
\begin{equation}\label{eqgrowth}
|f(\sptpt,P)|\le \Lambda \,P_{0}^0, \qquad (\sptpt,P) \in \spacetime \times 
\timelikekplanes.
\end{equation}
Some sufficient conditions will be given in Lemma \ref{lemmaab} below. 

We endow $\F$  with the following convergence: a sequence 
$\{f_n\}\subset \F$ converges to $f \in \F$ if there exists
a compact set $K\subset \spacetime$ containing the supports
of all $f_n$, and 
\[
\lim_{n \to +\infty}\sup_{(\sptpt,P)\in K \times \timelikekplanes}
\frac{\vert f_n(\sptpt,P)-f(\sptpt,P)\vert}{P_{0}^0} = 0.
\]

The following observation will be useful when considering 
the action on $\F$ of an element of the dual of $\F$,
see Lemma \ref{lem:azione} below.

\begin{Remark}[{\bf The isomorphism $i$}]\label{rem:iso}\rm 
The space $\F$ is isomorphic to 
the space $\CcoverB$,
via
the linear isomorphism $i : \F \to \CcoverB$ defined by 
$$
i(f) := \fq, \qquad 
f \in \F.
$$
\end{Remark}

\begin{Definition}[{\bf Recession function}]\label{def:recession}
Given any $f\in \F$ 
we define the recession function $\recf \in 
\mathcal C_c(\spacetime\times \partial \immgrass)$ of $f$ as
\[
\recf(\sptpt,Q) := 
~ \lim_{P \in \timelikekplanes, ~\!\mappa(P) \to Q}
~ \frac{f(\sptpt,P)}{P_{0}^0},
\qquad (\sptpt,Q)\in \spacetime\times \partial \immgrass.
\]
\end{Definition}

The following example, as well as the next lemma,
will be useful in the sequel, since they show that 
functions linear in $P_0^0$ are admissible.
\begin{Example}\label{exa:variabsep}\rm
Let $\phi \in \mathcal C_c(\spacetime)$. The function 
$f$ defined by 
\begin{equation}\label{piedi}
f(\pst,P) = 
\phi(\pst)
P_0^0,
\qquad (\pst,P) \in \spacetime
\times \timelikekplanes,
\end{equation}
belongs to $\F$, and we have $f^\infty = \varphi$.
Note that the choice $\phi \equiv 1$ on the whole of $\spacetime$
is not allowed.
\end{Example}
Other examples of functions belonging to $\F$ are given by 
the following result (see \cite[Lemma 2.2]{AB:97}
for a proof that can be adapted to our setting). Recall that 
$M_{N+1}$ denotes the space of all $(N+1)\times (N+1)$-symmetric
matrices.
\begin{Lemma}\label{lemmaab}
Let $f\in 
\mathcal 
C(\spacetime\times M_{\dimspacetime+1})$ satisfy \eqref{eqgrowth} and have support in 
$K\times \timelikekplanes$, for some compact $K\subset \spacetime$. 
Then $f_{\vert \spacetime\times \timelikekplanes}$ belongs to $\F$ 
in one of the following two cases:
\begin{itemize}
\item[-] $f$ is bounded. 
In this case we have $\recf=0$.
\item[-] $f(z,\cdot)$ is positively one-homogeneous, that is
\[
f(\pst,\lambda P)= \lambda\,f(\pst,P), \qquad 
(\pst,P)\in \spacetime\times M_{\dimspacetime+1},\ \lambda\ge 0.
\] 
\end{itemize}
\end{Lemma}

Taking into account also Remark \ref{rem:iso}, we are finally
in a position to define a lorentzian varifold. 

\begin{Definition}[{\bf Lorentzian $\dimension$-varifolds}]\label{defvar}
We say that $V$ is a lorentzian $\dimension$-varifold, and 
we write 
$$
V \in 
\LV,
$$
if $V$ 
is a positive Radon measure  on $\spacetime \times 
\overline{\immgrass}$.
\end{Definition}

\begin{Remark}\label{rem:deli}\rm
Any element of $\LV$ belongs 
to the dual $\CcoverB'$ of the locally
convex space $\CcoverB$. In addition
$\CcoverB'$ is isomorphic to the dual $\F'$ of $\F$ via 
the map $i' : \CcoverB' \to \F'$, 
$$
i'(V)(f) := 
 V(i(f)), \quad f \in \F.
$$
Hence to any $V\in \LV$ we can uniquely associate
$i'(V) \in \F'$.

{\it Warning}: when we write 
$V(f)$, for a given function $f \in \F$ and a measure
$V \in \LV$, we will always mean $i'(V) (f)$.  
\end{Remark}

\smallskip

Making use of Remark \ref{rem:deli}, 
the action of a varifold on a test function 
will be better specified below, at the end of Section \ref{subsec:disi}.

The notion of convergence for varifolds reads as follows.

\begin{Definition}[{\bf Varifolds convergence}]\label{def:variconv}
Let $V\in \LV$ 
 and $\{V_j\}\subset
\LV$.  We write
$$
V_j\rightharpoonup V
$$
 if
\begin{equation}\label{eqtest}
\lim_{j \to +\infty} V_j(f) = V(f), \qquad f \in \F.
\end{equation}
\end{Definition}
%%%%%%%%%%%%%%%%%%%%%%%%%%%%%%%%%%%%%%%%%%%%%%%%%%%%%%%%%%%
\subsection{First variation and stationarity}\label{sec:firstvariation}
%%%%%%%%%%%%%%%%%%%%%%%%%%%%%%%%%%%%%%%%%%%%%%%%%%%%%%%%%%%
Thanks to Lemma \ref{lemmaab}, if $Y \in (\mathcal C^1_c(\spacetime))^{N+1}$,
\begin{equation}\label{eq:thefunction}
{\rm the~ function~}
(z, P) \in \spacetime \times T_{h,\dimspacetime
+1} \to {\rm tr}\left(P dY(z)
\right)
~ {\rm belongs ~to~} \F.
\end{equation}
Therefore, taking into account Theorem \ref{teo:GG}, 
similarly to the riemannian case \cite{Si:83}
we can give the following definition.
\begin{Definition}[{\bf First variation}]\label{def:firstvariazzion}
Let 
$V\in\LV$. 
The first variation of $V$ is the 
vector distribution $\delta V$ in $\spacetime$
defined as follows:
\begin{equation}\label{eqfirst}
\begin{aligned}
& \delta  V (Y) := V(\textup{tr}(P dY)), \qquad 
Y \in \vfspacetime.
\end{aligned}
\end{equation}
\end{Definition}

Also the following definition is the same as in the riemannian case.

\begin{Definition}[{\bf Stationarity}]\label{defsta}
Let  $V\in \LV$. 
We say that  $V$ is stationary\footnote{ 
More generally, we say that
$V$ has bounded first variation if $\delta V$ 
is a Radon measure on $\spacetime$, that is, if
there exists $C>0$ such that 
$\delta V(Y) \le C\,\max_{\spacetime}
|Y|_{\rm e}$ for any $Y\in \vfspacetime$.
}
if 
$$
\delta V(Y)=0,  \qquad  Y \in \vfspacetime.
$$
\end{Definition}

\begin{Remark}\rm
When $\manist$ is a smooth null manifold, there are not
smooth compactly supported variations $Y$ normal to $\manist$,
guaranteeing  that the varied manifold
remains either null, or
partly null and partly timelike. Therefore, we do not
have any formula similar to \eqref{eq:firstvariation} for null
smooth manifolds. Despite this fact, 
Definition \ref{def:firstvariazzion}  seems one of the simplest extensions of the first
variation concept to null manifolds.
Definition \ref{def:firstvariazzion}  guarantees
that the limit of stationary varifolds
is still stationary, as shown in the next observation. 
\end{Remark}

\begin{Remark}\label{remfirstvar}\rm 
Let $\{V_j\} \subset \LV$ be a sequence converging to $V \in \LV$, 
and assume that each $V_j$  is 
stationary.
Then\footnote{Similarly, if each $V_j$  
has bounded first variation, then $V$ 
has bounded first variation.}
$V$ is 
stationary. 
Indeed, 
by Definition \ref{def:variconv} and \eqref{eq:thefunction}, if $V_j\rightharpoonup V$ then
$$
\delta V_j (Y)
 \to \delta V(Y), 
\qquad Y\in \vfspacetime.
$$
\end{Remark}
%%%%%%%%%%%%%%%%%%%%%%%%%%%%%%%%%%%%%%%%%%%%%%%%%%%%%%%%%%%%%%%%%%%%%%%%
\subsection{Splitting of $V$ into $V^0$ and $V^\infty$}
%%%%%%%%%%%%%%%%%%%%%%%%%%%%%%%%%%%%%%%%%%%%%%%%%%%%%%%%%%%%%%%%%%%%%%%%
A first decomposition of a varifold consists in taking its ``timelike part''
and its ``null part''. We denote by 
$\rest$ the restriction of a measure.

\begin{Definition}[{The measures \bf $V^0$ and $V^\infty$}]\label{def:VzeroVinfinito}
Let $V 
\in \LV$. We define 
\begin{itemize}
\item[] 
$V^0:= V\rest  
\left(\spacetime\times\immgrass\right)$, 
\item[] $V^\infty:= 
V \rest \left( 
\spacetime\times \partial {\immgrass}\right)$.
\end{itemize}
\end{Definition}

A  
lorentzian $\dimension$-varifold $V\in \LV$
can be uniquely decomposed as 
\begin{equation}\label{eq:decV}
V= V^0 + V^\infty.
\end{equation}
We will see that in certain cases
 the measure $V^\infty$ 
is the part of the varifold which, roughly speaking, takes into account the 
set of all points of the associated
generalized manifold where the tangent space is null.

In the expression of the action 
of a varifold on a test function, 
it is  convenient to introduce another measure $\widetilde V^0$
in the space $\spacetime \times \timelikekplanes$.
To this purpose, recall that the map $q$ is defined in \eqref{eq:themapq},
and recall Definition \ref{def:themapqf} of $q^{-1}$.

\begin{Definition}[{\bf The measure $\widetilde V^0$}]\label{def:Vzerotilde}
Let $V \in \LV$. 
We define the Radon measure $\widetilde V^0$
on $\spacetime \times \timelikekplanes$ as follows: for any
$f \in \F$ 
\begin{equation}\label{eq:Vzerotilde}
\widetilde V^0 (f) = 
\int_{\spacetime \times \timelikekplanes}
f(z,P) ~d \widetilde V^0(z,P) := 
\int_{\spacetime \times \immgrass} \frac{f(z, q^{-1}(Q))}{q^{-1}(Q)^0_0}
~ dV^0(z,Q).
\end{equation} 
\end{Definition}

The measure $\widetilde V^0$ is therefore the image of 
the measure $V^0$ through the map
$({\rm id}_{\spacetime}, q^{-1})$, furtherly divided
by a positive factor.

%%%%%%%%%%%%%%%%%%%%%%%%%%%%%%%%%%%%%%%%%%%%%%%%%%%%%%%%%%%%%%%%%%%%%%%%
\section{Disintegrations, barycenter and decompositions}\label{subsec:disi}
%%%%%%%%%%%%%%%%%%%%%%%%%%%%%%%%%%%%%%%%%%%%%%%%%%%%%%%%%%%%%%%%%%%%%%%%
Let $ V \in \LV$. 
In what follows we need to suitably disintegrate the measures $V^0, V^\infty$ and 
$\widetilde V^0$. In order to do this,
we denote by 
$$
\pi: \spacetime \times {\overline \immgrass} \to \spacetime
$$
the projection 
on the first factor.

The fact that $V$ is a Radon measure and
the compactness
of $\overline{\immgrass}$ imply that 
$$
V\left(K \times M_{N+1}\right) = V\Big(
K \times \overline{\immgrass}\Big) < +\infty
\qquad {\rm for~any~compact~set~} K \subset \spacetime.
$$
Hence we can apply  
Theorem \ref{th:disint} in the Appendix, so that there exists a disintegration 
of $V$, namely
$$
V = \mu^{}_V\otimes V_\sptpt,
$$
where 
\begin{itemize}
\item[] $\mu^{}_V := \pi_\# V$ is a positive Radon measure on $\spacetime$,  
\item[]
  $V_\sptpt$ is a probability measure on $\overline{\immgrass}$
defined for $\mu^{}_V$-almost every $\sptpt \in \spacetime$.
\end{itemize}

Similarly, there are disintegrations
\begin{equation}\label{decodeco}
\begin{aligned}
V^0 = & \mu_{V^0} \otimes V^0_\sptpt, 
\qquad
\ \ 
\mu_{V^0} := \pi_\# V^0, 
\\
\\
V^\infty = & \mu_{V^\infty} \otimes V^\infty_\sptpt,
\qquad \mu_{V^\infty} := \pi_\# V^\infty,
\\
\\
\widetilde V^0 = &\mu_{\widetilde V^0}\otimes \widetilde V^0_z, 
\qquad 
\ \ \mu_{\widetilde V^0} := \widetilde \pi_\# \widetilde V^0,
\end{aligned}
\end{equation}
where $\widetilde \pi: \spacetime \times \timelikekplanes \to
\spacetime$ is the projection on the first factor.

The measures $\mu^{}_V$, $\mu_{V^\infty}$ and
$\mu_{\widetilde V^0}$ 
will be splitted in \eqref{diffuse} below. Moreover, they will 
be
furtherly disintegrated in Section \ref{seccons}, see 
in particular formula \eqref{mipre}, in connection with conservation laws.
The measure 
$\mu_{\widetilde V^0}$ is the generalization of the area $\sigma^h$
 to the varifold setting.

\begin{Remark}\rm 
Despite the decomposition in \eqref{eq:decV}, $V_\sptpt$ cannot
be equal to 
$V_\sptpt^0 + V_\sptpt^\infty$, since 
$V_z$, $V_\sptpt^0$ and $V_\sptpt^\infty$ are
probability measures.
 Notice however that 
projecting the equality $\mu_V \otimes V_z = 
\mu_{V^0} \otimes V_z^0+ \mu_{V^\infty} \otimes V_z^\infty$ 
on $\spacetime$ via the map $\pi$, and using the fact that $V_z^0$ and $V_z^\infty$
are probability measures, gives
\begin{equation}\label{anc}
\mu_V = \mu_{V^0} + \mu_{V^\infty}.
\end{equation}
\end{Remark}

Taking into account the above definitions, 
we can represent the action of $V\in \LV$ on $\F$ as follows: 

\begin{Lemma}[{\bf Action of a varifold}]\label{lem:azione}
Let $V \in \LV$ and 
$f \in \F$. Then  
\begin{equation}\label{eq:azione}
\begin{aligned}
V(f) 
= & 
\int_{\spacetime\times \timelikekplanes}
f(\pst,P)
\,d\widetilde V^0 (\pst,P)
+ 
\int_{\spacetime\times \partial {\immgrass}} \recf(\pst,Q)
\,dV^\infty(\pst,Q)
\\
= & 
\int_{\spacetime}
\left(
\int_{\timelikekplanes}
f(\pst,P)\,d{\widetilde V^0}_\pst(P)\right) d\mu_{\widetilde V^0}(\pst)
\\
&+ 
\int_{\spacetime}
\left(
\int_{\partial \immgrass} \recf(\pst,Q)
\,dV^\infty_\pst(Q)\right) d\mu_{V^\infty}(\pst)\,.
\end{aligned}
\end{equation}
\end{Lemma}
\begin{proof}
Using \eqref{eq:decV} and 
recalling Remark \ref{rem:deli} and Definition \ref{def:themapqf}, 
we have 
\begin{equation*}
\begin{aligned}
V(f) =& V(i(f)) = V^0(i(f)) + V^\infty(i(f)) 
\\
= & 
\int_{\spacetime\times \immgrass}
\frac{f(\pst,q^{-1}(Q))}{q^{-1}(Q)^0_0}
~ dV^0 (\pst,Q)
\\
& + 
\int_{\spacetime\times \partial {\immgrass}} \recf(\pst,Q)
\,dV^\infty(\pst, Q).
\end{aligned}
\end{equation*}
Hence, using Definition \ref{def:Vzerotilde}, the first 
equality in \eqref{eq:azione} follows.
The second equality is a direct consequence of the
disintegrations \eqref{decodeco}.
\end{proof}

\begin{Remark}\rm
Note carefully that the first addendum on the right hand
side of \eqref{eq:azione} is an integral over $\spacetime
\times \timelikekplanes$, while the second addendum
is an integral over $\spacetime \times \partial \immgrass$.
\end{Remark}

Notice that $V_j\rightharpoonup V$ does not  imply 
$V^0_j\rightharpoonup V^0$ or $V^\infty_j\rightharpoonup V^\infty$, and does
not imply that separately the projections converge: this can be
seen by examples, such as Example \ref{exa:subspace}, where $V^\infty_j=0$,
while $V = V^\infty\neq 0$.
\begin{Remark}\label{rem:passano}\rm
We have 
$$
V_j\rightharpoonup V \Rightarrow
\mu_{V_j} \rightharpoonup \mu^{}_V.
$$
Indeed, let $\phi\in \C_c(\spacetime)$, and 
take $f$ as in \eqref{piedi}.
Then 
\begin{equation*}
\begin{aligned}
V_j(f) = &
\int_{\spacetime\times \timelikekplanes}
\phi(\pst) 
P_0^0 ~
d 
{\widetilde V_j^0}
{}_z
~
d\widetilde \mu_{V_j^0}(\pst)
+ 
\int_{\spacetime\times \partial {\immgrass}} \phi(\pst)
\,d {V_j^\infty}_{\!\!\!\punto}(Q)\, d\mu_{V^\infty_j}(\pst)
\\
= & \int_{\spacetime} \phi ~d\mu_{V^0_j} 
+ \int_{\spacetime} \phi~ d\mu_{V_j^\infty} = \int_{\spacetime}
\phi~ d\mu_{V_j},
\end{aligned}
\end{equation*}
where in the last equality we have used \eqref{anc}. 
\end{Remark}

The following result will be used in Remark \ref{rem:equi} and 
 Theorem \ref{senzanome}. 

\begin{Proposition}[{\bf Compactness}]\label{procida}
Let $\{V_j\}\subset \LV$ 
be a sequence of lorentzian $\dimension$-varifolds such that 
\begin{equation}\label{eqmass}
\sup_j 
\mu_{V_j}(K)
<+\infty,
\qquad  K\subset\spacetime\ {\rm compact.}
\end{equation}
Then there exist $V\in\LV$ and a subsequence
$\{V_{j_k}\}$ of $\{V_j\}$ such that 
$V_{j_k}\rightharpoonup V$ as $k \to +\infty$.
\end{Proposition}

\begin{proof}
Since ${V_{j}}_z$ are probability measures, we have 
$$
V_j\Big(K \times \overline{\immgrass}\Big) = \mu_{V_j}(K).
$$
The assertion then follows from \eqref{eqmass}, 
and from De La Vall\'ee Poussin Compactness Theorem (see \cite[Cor. 1.60]{AmFuPa:00}).
\end{proof}

Recalling the disintegration of $\widetilde V^0$ in \eqref{decodeco}, we 
can now
give the following definition, which will allow to take
into account the oscillations 
of the tangent spaces.

\begin{Definition}[{\bf Barycenter}]\label{defbar}
Let  $V\in 
\LV$. We set
$$
\overline P(\pst):= 
\int_{\timelikekplanes}
P~ d\widetilde V^0_\pst(P)
\qquad {\rm for}~ \mu_{\widetilde V^0}-{\rm a.e.~} \pst\in \spacetime.
$$ 
Similarly, we set 
$$
\overline Q(\pst):= 
\int_{\partial \immgrass}
Q~ dV^\infty_\pst(Q)
\qquad {\rm for}~ \mu_{V^\infty}-{\rm a.e.~} \pst\in \spacetime.
$$ 
\end{Definition}

\begin{Remark}\label{rem:sbaglia}\rm  For $\mu_{\widetilde V^0}$-almost
every $\pst\in \spacetime$ we have the following assertions:
\begin{itemize}
\item[-] 
the matrix $\overline P(z)$  is  well defined 
by Lemma \ref{lemmaab}, since linear functions of the projections
can be integrated with respect to $\widetilde V_\pst^0$ and $\pst$ is fixed;
\item[-]
$\overline P(z)$ is not necessarily symmetric, while 
$\eta \overline P(z)$ is symmetric;
\item[-]
in general
$\overline P(\pst)\not\in \timelikekplanes$,
since 
$\timelikekplanes$ is not a convex set.
\end{itemize}
Similar properties (with obvious modifications) hold for $\overline Q$.
 \end{Remark}
For a lorentzian $\dimension$-varifold, $\overline P(z)$ is 
not necessarily a projection on a timelike or null
$\dimension$-plane
in the sense described in Section \ref{sub:themap}. However, still its
trace equals $\dimension$.
More interestingly, 
if $\overline P(z)$ is a projection matrix, then the measure
$\widetilde V^0_z$ is a Dirac delta. Precisely, we have 
the following result.

\begin{Proposition}[{\bf Properties of $\overline P$}]\label{lembarp}
Let $V \in \LV$. Then 
\begin{equation}\label{eqtr}
\overline P(\pst)^\alpha_\alpha = \dimension
\qquad {\rm for}~ \mu_{\widetilde V^0}-{\rm a.e.~} \pst\in \spacetime.
\end{equation}
Moreover
\begin{equation}\label{ogni}
 \overline P(\pst) 
\in \timelikekplanes \ \Rightarrow \ \widetilde V^0_\pst=\delta_{\overline
P(\pst)}.
\end{equation}
\end{Proposition}

\begin{proof}
Assertion \eqref{eqtr} follows from the fact that the trace is a linear
operator, and $P_\alpha^\alpha=h$ 
for all $P\in \timelikekplanes$.

Let us prove \eqref{ogni}.
Being $z$ fixed, we write for simplicity $\overline P  = \overline P(z)$.
Since $\overline P\in \timelikekplanes$,  we can find a Lorentz transformation 
$L$ such that $L^{-1} \overline P L$ takes the form 
$$
L^{-1} \overline  P L = {\rm diag}(1,\ldots,1,0,\ldots,0), 
$$
where $1$ appears $\dimension$-times.
Recalling  \eqref{eq:P00},
we have
\begin{equation}\label{eqproj}
1 \leq (L^{-1} P L)_{0}^0, 
\end{equation}
with equality if and only if 
\begin{equation}\label{eqproq}
L^{-1} P L = 
\left(
\begin{array}{cc}
1 & (0,\ldots,0) \\
(0,\ldots,0)^T & R 
\end{array}
\right),
\end{equation}
where $(0,\dots,0) \in \Rn$, and 
$R:\R^N\to\R^N$ 
is a euclidean orthogonal projection onto an $(\dimension-1)$-plane. 
Without loss of generality, we can assume that this $(h-1)$-plane
is spanned by $\{e_1,\dots,e_{h-1}\}$.
Integrating \eqref{eqproj} on $\timelikekplanes$ with respect to the 
probability measure $\widetilde V^0_\pst$,
we get
\begin{equation}\label{eqpronto}
%\begin{aligned}
1 \le  \int_{\timelikekplanes} (L^{-1} P L)_{0}^0 \,d\widetilde V^0_\pst(P)= 
(L^{-1} \overline  P L)_{0}^0 = 1.
\end{equation}
This implies that the measure $\widetilde V^0_\pst$ is concentrated 
on the set $S$ of matrices of the form \eqref{eqproq},
namely $\widetilde V^0_z(\timelikekplanes \setminus S)=0$.
In particular, for such a matrix $L^{-1} P L\in S$ there holds 
\begin{equation}\label{eqprolambda}
1 \geq 
(L^{-1} P L)_{\indicespaziale}^{\indicespaziale}
= R^\indicespaziale_\indicespaziale, \qquad 1\le\indicespaziale\le N,
\end{equation}
where we do not sum over $\indicespaziale$.
Integrating now \eqref{eqprolambda} on $\timelikekplanes$
with respect to $\widetilde V^0_z$ and using \eqref{eqproj},  we obtain
\begin{equation}\label{eqprontobis}
1 \ge \int_{\timelikekplanes}(L^{-1}  P L)_{\indicespaziale}^{\indicespaziale} \,d\widetilde V^0_\pst(P)= 
(L^{-1} \overline  P L)_{\indicespaziale}^{\indicespaziale} = 1,
\qquad\ 1\le\indicespaziale \leq \dimension -1,
\end{equation}
where again we do not sum over $\indicespaziale$.
This and \eqref{eqpronto} imply $P=\overline  P$, hence 
$\widetilde V^0_\pst = \delta_{\overline P(\pst)}$.
\end{proof}

We will see in Example \ref{exa:sub} an interesting case of a varifold  
for which
$\overline P$ is not the projection on the tangent space.

Taking into account
Lemma \ref{lem:azione} and Definition \ref{defbar} of $\overline P$, 
we can write the first variation of $V$ in \eqref{eqfirst} as 
\begin{equation*}
\begin{aligned}
\delta V(Y) =&
\int_{\spacetime \times \timelikekplanes}
\textup{tr}\left(P dY\right)\,d{\widetilde V^0}(z,P)
+ \int_{\spacetime\times \partial \immgrass}
\textup{tr}\left(Q dY\right)\,
d{V^\infty}(z,Q)
\\ 
=& \int_{\spacetime}\textup{tr}\left(
\overline P d Y\right)\,d\mu_{\widetilde V^0}(z)
+ \int_{\spacetime} 
\textup{tr}\left(\overline Q dY\right)\,
d\mu_{V^\infty}(z).
\end{aligned}
\end{equation*}
%

%%%%%%%%%%%%%%%%%%%%%%%%%%%%%%%%%%%%%%%%%%%%%%%%
\subsubsection{Radon-Nikod\'ym decompositions}\label{sub:dec}
%%%%%%%%%%%%%%%%%%%%%%%%%%%%%%%%%%%%%%%%%%%%%%%%
Using the generalized Radon-Nikod\'ym theorem (Theorem \ref{th:gRN}
in the Appendix; recall that $\mathcal H^h$
is not $\sigma$-finite) we can decompose the measures 
$\mu_{V^0}$ and $\mu_{V^\infty}$ in 
\eqref{decodeco} into their absolutely continuous, singular
and diffuse parts respectively: 
\begin{equation}\label{diffuse}
\begin{aligned}
\mu_{V}= &
\mu_{V}^{ac}
+\mu_{V}^s + \mu_{V}^d,
\\
\mu_{V^0}= &
\mu_{V^0}^{ac}
+\mu_{V^0}^s + \mu_{V^0}^d,
\\
\mu_{V^\infty}= &
\mu_{V^\infty}^{ac}
+\mu_{V^\infty}^s + \mu_{V^\infty}^d,
\\
\mu_{\widetilde V^0}= &
\mu_{\widetilde V^0}^{ac}
+\mu_{\widetilde V^0}^s + \mu_{\widetilde V^0}^d,
\end{aligned}
\end{equation}
where 
$$
\mu_{V}^{ac}<< \mathcal H^\dimension, \quad 
\mu_{V^0}^{ac}<< \mathcal H^\dimension, \quad
\mu_{V^\infty}^{ac}<< \mathcal H^\dimension, \quad
\mu_{\widetilde V^0}^{ac}<< \mathcal H^\dimension.
$$
Being 
$\mu_V$ 
a Radon measure, it follows that 
$\mu_{V}^{ac}$, 
$\mu_{V}^s$ and  $\mu_{V}^d$ are mutually singular. 
The same property holds for the decompositions of 
$\mu_{V^0}$,  
$\mu_{V^\infty}$, and  
$\mu_{\widetilde V^0}$. 

%%%%%%%%%%%%%%%%%%%%%%%%%%%%%%%%%%%%%%%%%%%%%%%%%%%%%%%%%%%%%%%%%%%%%%%
\section{Proper, rectifiable and weakly 
rectifiable  varifolds}\label{sec:firstproperties}
%%%%%%%%%%%%%%%%%%%%%%%%%%%%%%%%%%%%%%%%%%%%%%%%%%%%%%%%%%%%%%%%%%%%%%%
We now introduce the notions of proper, rectifiable
and weakly rectifiable  varifold. 
Proper rectifiable varifolds 
consist of timelike $\dimension$-varifolds
without singular or diffuse part, and 
essentially have been considered in \cite{BNO:09}. 
\begin{Definition}[{\bf Timelike and proper varifolds}]\label{def:prop}
Let  $V\in \LV$. We say that $V$ is timelike if 
\begin{equation*}
V^\infty= 0.
\end{equation*}
If in addition 
\begin{equation*}
\mu^s_{V^0}=0,
\qquad
\mu_{V^0}^d=0,
\end{equation*}
then we say that $V$ is proper.
\end{Definition}
As we shall see, even subrelativistic strings are not, in general,
proper varifolds, and therefore Definition \ref{def:prop} must be weakened. 
However, 
it may be useful
to find sufficient conditions
ensuring that the limit of a sequence of proper varifolds
is proper. For instance, the following observation (that will be used  in 
Example \ref{exa:patella}) holds.

\begin{Remark}[{\bf Criterion for being proper}]\label{rem:equi}\rm
Let $\{V_j\} \subset \LV$ be a sequence 
satisfying the bound \eqref{eqmass} on $\mu_{V_j}$. 
Assume in addition that
\begin{equation}\label{eq:cosc}
\begin{aligned}
& {\rm for\ all\ } \eps>0 
{\rm\ and\ all\ compact~} K\subset\spacetime \ {\rm there\ exists\ a\
compact\ }
\mathcal T \subset \timelikekplanes
\\
& {\rm \ such\ that}
\ \sup_j 
\left(
\int_{K\times\left(\timelikekplanes\setminus 
\mathcal T\right)} 
P_{0}^0\,d\widetilde V^0_j
+ 
\mu_{V^\infty_j}(K)\right) \le\eps.
\end{aligned}
\end{equation}
Then (using for instance the lower semicontinuity
inequality on open sets as in \cite[(1.9)]{AmFuPa:00})
the limit varifold $V$ 
given by Proposition \ref{procida} 
is proper. Condition \eqref{eq:cosc} is verified for instance 
if the varifolds $V_j$ are proper and there exists $p >1$ 
such that 
\begin{equation}\label{eqLp}
\forall~ {\rm compact~} K \subset \spacetime
\ \ \exists
C >0:
\quad \sup_j \int_{K\times\timelikekplanes} 
\left(P_{0}^0\right)^p\,d\widetilde V^0_j 
\le C.
\end{equation}
\end{Remark}

%%%%%%%%%%%%%%%%%%%%%%%%%%%%%%%%%%%%%%%%%%%%%%%%
\subsection{Rectifiable and weakly rectifiable varifolds}\label{sub:rectweak}
%%%%%%%%%%%%%%%%%%%%%%%%%%%%%%%%%%%%%%%%%%%%%%%%
As already discussed in the Introduction, in this work 
an important role is played
by the lorentzian varifolds that we will call weakly rectifiable.

To understand the next definitions, 
it is useful to keep in mind that, with any 
$\dimension$-rectifiable timelike set $\rect \subset \spacetime$,
we can associate in a natural way a varifold  defined
by 
$$
 \left(\lormeas \rest \rect\right) \otimes \delta_{P_\rect},
$$
where we recall that $\lormeas$ is defined in \eqref{eq:sigmah}. 

\begin{Definition}[{\bf Rectifiable varifolds}]\label{def:timerect}
Let $V \in \LV$. We say that $V$ is rectifiable if the following properties hold:
\begin{itemize}
\item[1.] there exist a timelike 
$\dimension$-rectifiable set $\manist^0\subset \spacetime$ 
and a positive multiplicity
function $\theta^0\in L^1_{\rm loc}(\manist^0,\H^\dimension)$
such that
\begin{equation}\label{eq:tildeVzero}
\widetilde V^0 
=\theta^0~\left(
\H^\dimension\rest \,\manist^0 \right)\otimes \delta_{P_{\manist^0}},
\end{equation}
\item[2.] there exist a null $\dimension$-rectifiable set $\manist^\infty\subset \spacetime$ 
and a positive multiplicity
function $\theta^\infty\in L^1_{\rm loc}(\manist^\infty,\mathcal H^\dimension)$
such that
\begin{equation}\label{eq:Vinfinito}
V^\infty =\theta^\infty~ \left(\mathcal H^\dimension\rest \,\manist^\infty\right) \otimes \delta_{Q_{\manist^\infty}}.
\end{equation}
\end{itemize}
\end{Definition}

A rectifiable varifold\footnote{
When the multiplicity take values in the
positive integer numbers, $V$ is called integer, and the 
same for Definition \ref{def:weaklyrectifiable}.
We will not deepen the properties of integer varifolds
in the present paper.} is described by two $h$-rectifiable sets, 
one timelike and the other null, each one equipped with a multiplicity function.
The measure  part of the varifold on the grassmannian is concentrated on the 
orthogonal projection onto the tangent space to its support.
A rectifiable varifold is therefore a generalization of what we could
call nonspacelike $h$-rectifiable set, possibly endowed with a real
positive multiplicity function.

\begin{Remark}\label{rem:ovv}\rm 
Notice that in Definition \ref{def:timerect}, item 1, we use 
$\lormeas$, while in item 2 we use $\mathcal H^h$. This is due to the fact that 
$\lormeas$ vanishes on null sets\footnote{Recall that 
for a rectifiable set, null means lightlike.}.
\end{Remark}

We now define a class of varifolds which contains all the 
relevant examples of relativistic strings considered in Section 
\ref{sub:relstri}.

\begin{Definition}[{\bf Weakly rectifiable varifolds}]
\label{def:weaklyrectifiable}
Let $V \in \LV$. We say that $V$ is weakly rectifiable if 
the following properties hold:
\begin{itemize}
\item[1.]
there exist a timelike $\dimension$-rectifiable set $\manist^0$ and
a positive multiplicity function 
$\theta^0\in L^1_{\rm loc}(\manist^0,\H^\dimension)$ 
such that: 
\begin{itemize}
\item[1a.]  $\quad \mu_{\widetilde V^0}^{ac}
=\theta^0\H^\dimension\rest \,\manist^0$, 
\item[1b.] $\quad \mu_{\widetilde V^0}^d =0$,
\item[1c.]
\begin{equation}\label{eqTP}
{\rm Range} \Big(\overline 
P_{\manist^0}(\pst)\Big)\subseteq T_\pst\manist^0 \qquad {\rm for~ } \H^\dimension-{\rm a.e.}~ 
\pst \in \manist^0,
\end{equation}
\end{itemize}
\item[2.] 
there exist 
a null $\dimension$-rectifiable set $\manist^\infty \subset \spacetime$ and 
a positive multiplicity function
$\theta^\infty\in L^1_{\rm loc}(\manist^\infty,\mathcal H^\dimension)$, 
such that:
\begin{itemize}
\item[2a.] $\quad \mu_{V^\infty}^{ac}
=\theta^\infty\mathcal H^\dimension\rest \,\manist^\infty$,
\item[2b.] $\quad \mu_{V^\infty}^d=0$,
\item[2c.] $\quad\!\!$ for $\mathcal H^\dimension$--almost
every $\pst \in \manist^\infty$ we have 
$V^\infty_\pst = \delta_{Q_{\manist^\infty}(\pst)}$.
\end{itemize}
\end{itemize}
\end{Definition}

Notice that
$$
\mu^{ac}_{V^0}
\perp 
\mu^{ac}_{V^\infty}.
$$
Note also that no conditions on the singular parts are imposed
for a weakly rectifiable varifold.

\begin{Remark}\rm
The difference between conditions 2a, 2b, 2c of Definition 
\ref{def:weaklyrectifiable}
and condition 2 of Definition \ref{def:timerect} is that in 
Definition \ref{def:weaklyrectifiable} we allow the presence of a singular 
part in $\mu_{V^\infty}$. Example \ref{exa:square} shows that, in general,
such a part does not vanish. 

Condition 1c is reminiscent of the fact that the measure $\widetilde V_z^0$ 
has barycenter in the lorentzian projection onto $T_z\rect^0$, even if it is 
not necessarily concentrated on it.
On the other hand, condition 2c requires the measure to be concentrated on
the lorentzian 
projection onto $T_z\rect^\infty$.
\end{Remark}

\begin{Remark}\label{rem:rem}\rm
In Proposition \ref{proone} we show essentially that stationary 
weakly rectifiable 1-varifolds are necessarily rectifiable.
In Example \ref{exa:sub} we exhibit a weakly rectifiable $2$-varifold which is not
rectifiable. 
\end{Remark}

The following result is a motivation for 
introducing conditions 1c and 2c in Definition \ref{def:weaklyrectifiable}. 

\begin{Proposition}\label{rem:blowup}
Let $V \in \LV$ 
be a stationary varifold.
Then 
conditions 1a, 1b, 2a and 2b of Definition \ref{def:weaklyrectifiable} imply conditions 1c and 2c.
\end{Proposition}
\begin{proof}
If $Y \in \vfspacetime$, from the stationarity of $V$ it follows
$$
\begin{aligned}
& \int_{\manist^0} \theta^0~ 
{\rm tr}\left(\overline P dY\right) ~d\H^\dimension
+
\int_{\spacetime} {\rm tr}\left(\overline P dY\right)~d\mu^s_{\widetilde V^0}
\\
& + \int_{\manist^\infty} \theta^\infty~ {\rm tr}\left(\overline Q  dY\right)~ d
\mathcal H^\dimension
+
\int_{\spacetime} {\rm tr}\left(\overline Q dY\right)~d\mu^s_{V^\infty}
=0.
\end{aligned}
$$
Choosing $Y = \phi e_i$ for $\phi \in \mathcal C^1_c(\spacetime)$
and using the arbitrariness of $i\in \{0,\dots,\dimspacetime\}$, it follows the 
vector equality
\begin{equation}\label{eq:stazphi}
\int_{\manist^0} \theta^0 \overline P  d\phi ~d\H^\dimension+
\int_{\spacetime} \overline P d\varphi~ d\mu^s_{\widetilde V^0}
+
\int_{\manist^\infty} \theta^\infty \overline Q d\phi ~d\mathcal H^\dimension
+ \int_{\spacetime} \overline Q d\varphi~d\mu^s_{V^\infty}=0.
\end{equation}
Let us now show that condition 1c holds.
We follow the blow-up argument in  
\cite[Theorem 3.8]{AmSo:98}. We take a point $z \in \manist^0 \setminus
\manist^\infty$ 
where there exists $T_z\rect^0$, and 
such that the density of $\mu^s_V$ with respect
to $\mathcal H^h$ is zero. 
Assume also that 
$\pst\in \manist^0$ is a 
Lebesgue point both for $\theta^0$ and $\overline P$.
Then a rescaling argument in \eqref{eqfirst}   gives
\[
\overline P(\pst) \int_{T_{\pst}\manist^0} d\phi ~d \mathcal H^\dimension=0\,, 
\qquad
\phi\in \mathcal C^1_c(\spacetime). 
\]
Taking $\phi$ constant on $T_z \rect^0$,
multiplied by a suitable cut-off function with support invading 
$T_z\rect^0$,
  implies that $\overline P$ annihilates 
the normal covectors. Therefore, the image of $\overline P(z)$
(considered now as an operator taking vectors into vectors) is contained 
in $T_z(\Sigma)$, and 
condition 1c holds.

A similar proof gives that for $\mathcal H^h$-almost every 
$z \in \manist^\infty$ we have 
\begin{equation}\label{pagdrepio}
\overline Q(\pst) 
\int_{T_{\pst}\manist^\infty} d\phi ~d \mathcal H^\dimension=0\,, 
\qquad
\phi\in \mathcal C^1_c(\spacetime).
\end{equation}
To show that 
condition 2c holds, we have  to prove that $V^\infty_{z}$
is a Dirac delta.
{}From \eqref{pagdrepio} and arguing as above it follows that 
\begin{equation}\label{telepadrepio}
{\rm Range}\left(\overline Q(z)\right) \subseteq T_{z}\rect^\infty.
\end{equation}
In particular, since any vector in $T_z \rect^\infty$ is not timelike, we deduce
\begin{equation}\label{eq:nottime}
\overline Q(z) e_0 {\rm ~ is~ not~ timelike}.
\end{equation}
On the other hand, 
considering the 
map $m$ which associates with $Q\in \partial \immgrass$
the element $\velnull \in  \mathbb S^{\dimspacetime-1}$ given 
by \eqref{eq:bdrySn},
and setting 
$$
\lambda := m_{\#} V^\infty_{z},
$$
 we have\footnote{$Q e_0 = - (1, \velnull)\otimes \eta (1, \velnull)e_0
= (1, \velnull) \otimes (1, -\velnull) e_0 = (1,\velnull)$.}
\begin{equation}\label{piopadr}
\begin{aligned}
\overline Q(z) e_0 = & \int_{\partial 
\immgrass} Q(z) e_0 ~ dV^\infty_{z}(Q)
\\
=&  
\int_{\partial \immgrass}
(1, m(Q))~ dV^\infty_{z}(Q)
= \left(1, 
\int_{\theta \in \mathbb S^{\dimspacetime-1}}
\theta~ d\lambda(\theta)
\right).
\end{aligned}
\end{equation}
Since $\vert \int_{\theta \in \mathbb S^{\dimspacetime-1}}
\theta~ d\lambda(\theta)\vert_{\rm e} \leq 1$, we deduce that 
the vector $(1, \int_{\theta \in \mathbb S^{\dimspacetime-1}}
\theta~ d\lambda(\theta))$ is either timelike or null.
We conclude, using 
\eqref{eq:nottime}, that 
$\overline Q(z)(e_0)$ is null, so that 
$\vert \int_{\theta \in \mathbb S^{\dimspacetime-1}}
\theta~ d\lambda(\theta)\vert_{\rm e} =1$.
{}From this it follows that $\lambda$ is a Dirac
delta. This conclusion implies that also $V^\infty_{z}$ is a Dirac delta.
Finally, 
 remembering \eqref{telepadrepio}, we have\footnote{Recall
that there is only one null line contained in $T_z(\Sigma^\infty)$,
and this  is the one
generated by $(1, \velnull(z))$.} 
that  $V^\infty_z$ in concentrated on $Q_{\manist^\infty}(z)$.
\end{proof} 

%%%%%%%%%%%%%%%%%%%%%%%%%%%%%%%%%%%%%%%%%%%%%%
\subsection{Stationarity conditions for weakly rectifiable varifolds}\label{unce}
%%%%%%%%%%%%%%%%%%%%%%%%%%%%%%%%%%%%%%%%%%%%%%%
Recall that if $\manist$ is smooth and timelike, and if $T$ is a $(1,1)$-tensor
field defined on $\manist$, the tangential
divergence ${\rm div}_\tau T$ is defined in \eqref{eq:divtangtensor}.
Under smoothness assumptions, we can still derive the necessary stationarity
conditions for a stationary weakly rectifiable varifold.

\begin{Proposition}[{\bf Stationarity condition, I}]\label{progen}
Let $\theta^0
\left(\H^\dimension\rest \,\manist^0\right)\otimes V_\punto^0
\in \LV$ be a proper stationary weakly rectifiable varifold,
and 
assume that $\manist^0\subset \spacetime$ is an $h$-dimensional embedded manifold of class $\C^2$
without boundary. 
Then
\begin{equation}\label{eqHgen}
\overline P_{\manist^0}~ d_\tau\theta^0 + \theta^0
\,\textup{div}_\tau
\overline P_{\manist^0} =0
\qquad {\rm on\ }\manist^0
\end{equation}
in the sense of distributions.
\end{Proposition}

\begin{proof} 
Let $Y \in \vfspacetime$.
{}From the stationarity assumption we have 
$$
0=\delta V(Y)
=\int_{\manist^0}\theta^0\,\textup{tr}
\left(
{\overline P}_{\manist^0} d Y
\right)\, d\H^\dimension.
$$
Observe now that  
$$
\begin{aligned}
{\rm div}_\tau (\overline P_{\Sigma^0} Y) =& 
{\rm tr}(P_{\Sigma_0} d\overline P_{\Sigma^0} Y) + 
{\rm tr}(P_{\Sigma^0} \overline P_{\Sigma^0} dY)
\\
= & \langle {\rm div}_\tau \overline P_{\Sigma^0}, Y\rangle
+{\rm tr}(\overline P_{\Sigma^0} dY),
\end{aligned}
$$
where we used the fact that 
\eqref{eqTP} implies $P_{\Sigma^0} \overline P_{\Sigma^0}
= \overline P_{\Sigma^0}$. 
Integrating by parts \eqref{eqfirst} and
using \eqref{GG}, we obtain
\begin{equation*}
\begin{aligned}
0
&=\int_{\manist^0}\theta^0\,\Big( 
\textup{div}_\tau\left(\overline P_{\manist^0} Y\right)-
\langle \textup{div}_\tau \overline P_{\manist^0},Y\rangle\Big)
  \, d\H^\dimension
\\
&= - \int_{\manist^0}\langle~
{\overline P}_{\manist^0} d_\tau \theta^0 + \theta^0\textup{div}_\tau
\overline P_{\manist^0}, Y~\rangle~ d\H^\dimension\,,
\end{aligned}
\end{equation*}
which gives \eqref{eqHgen}.
\end{proof}
The stationarity condition \eqref{eqHgen} can be written,
in an equivalent and  more readable way,  in terms
of the mean curvature of $\Sigma^0$ and of the ``defect
$P_{\Sigma^0}-\overline P_{\Sigma^0}$
of being a projection'' 
 as follows.

\begin{Remark}[{\bf Stationarity condition, II}]\label{rem:difettt}\rm
Adding and subtracting ${\rm div}_\tau P_{\Sigma^0}$ to \eqref{eqHgen},
and 
recalling the relation \eqref{divPH} between the projection 
and the mean curvature, we have 
$$
\begin{aligned}
0 =&
\theta^0 
 \textup{div}_\tau
(P_{\manist^0} 
- \overline P_{\manist^0})
- \theta^0 
 \textup{div}_\tau
P_{\manist^0}
- \overline P_{\Sigma^0} d_\tau \theta^0 
\\
=& 
\theta^0 
 \textup{div}_\tau
(P_{\manist^0} 
- \overline P_{\manist^0}) 
+ \theta^0 \eta H_{\Sigma^0}
- \overline P_{\Sigma^0} d_\tau \theta^0.
\end{aligned}
$$
Splitting this equation into
its normal and tangential components gives
\begin{equation}\label{eqCweak}
\left\{
\begin{aligned}
& \theta^0\left(H_{\manist^0} +
\left(\eta^{-1} \textup{div}_\tau(
P_{\manist^0} - \overline P_{\manist^0} 
)\right)^\bot\right) =0
\\
& \eta^{-1}
\overline P_{\manist^0} ~d_\tau\theta^0 - \theta^0 \Big(
\eta^{-1}\textup{div}_\tau(
P_{\manist^0} - \overline P_{\manist^0} 
)\Big)^\top = 0
\end{aligned}\qquad {\rm on}~ \rect^0,
\right.
\end{equation}
\end{Remark}
in the sense of distributions.

The following result immediately follows from \eqref{eqCweak}.

\begin{Proposition}\label{prosta}
Let $\theta^0\left(\H^\dimension\rest \,\manist^0\right)
\otimes \delta_{P_{\manist^0}} \in \LV$ be a proper stationary rectifiable varifold,
and  
assume that $\manist^0\subset \spacetime$ is an $h$-dimensional 
embedded oriented manifold of class $\C^2$
without boundary. Suppose also that 
 $\theta^0>0$.
Then $\theta^0$ is constant on $\manist^0$, and $\manist^0$ is a timelike 
minimal surface, that is, it is a smooth solution to \eqref{eq:minisurfa}.
\end{Proposition}
\begin{proof}
The hypotheses imply that $P_{\Sigma^0} = \overline P_{\Sigma^0}$. 
Since by assumption $\theta^0>0$, the first equation in \eqref{eqCweak}
implies $H_{\Sigma^0}=0$.
\end{proof}

%%%%%%%%%%%%%%%%%%%%%%%%%%%%%%%%%%%%%%%%%%%%%%%%%%%%%%%%%%%%%%%%%
\subsection{The one-dimensional case}
%%%%%%%%%%%%%%%%%%%%%%%%%%%%%%%%%%%%%%%%%%%%%%%%%%%%%%%%%%%%%%%%%

In the one-dimensional case, for a weakly 
rectifiable varifold, $\overline P_{\manist^0}$
is a projection, as we show in the next proposition.

\begin{Proposition}\label{prouno}
Let $V \in \LVuno$ be weakly rectifiable, and assume that 
\begin{equation}\label{franza}
\mu^s_{V}=0.
\end{equation}
Then $V$ is rectifiable. 
\end{Proposition}
\begin{proof}
Fix $\pst\in\manist^0$ such that \eqref{eqTP} holds, and choose a 
Lorentz transformation $L$ such that 
${\rm Range}(L^{-1}\overline P(z) L) = \R e_0$.
{}From \eqref{eqtr} applied with $\dimension=1$ it follows that 
$(L^{-1} \overline P(z) L)^0_0=1$. Hence, reasoning as in the proof of Lemma \ref{lembarp}
it follows $L^{-1} \overline P(z) L={\rm diag}(1,0,\dots,0)$, 
which proves that $\overline P$ is a projection, hence
$$
\overline P(z)= P_{\manist^0}(z).
$$
This assertion, together with hypothesis \eqref{franza}, imply that $V$ is rectifiable.
\end{proof}

We point out that, due to the 
examples in \cite{neu}, \cite{BHNO}, the thesis of Proposition \ref{prouno} 
is generally not true for $\dimension >1$, see for instance
Example \ref{exa:sub}. It would be interesting to 
investigate which conditions are implied on the singular part $\mu_V^s$ from 
the stationarity condition. 
In particular, we now prove
that $\mu_V^s=0$ for stationary weakly rectifiable $1$-varifolds, thus implying the rectifiability. 

\begin{Proposition}\label{proone}
Let $V \in \LVuno$ be a stationary and weakly rectifiable varifold
such that 
\begin{equation}\label{eqkappabis}
\forall I\subseteq \R {\rm ~interval~} \exists
~ K \subset \Rn {\rm ~compact~such~that~}
 {\rm spt}(\mu^{}_{V}) \cap (I \times \Rn) \subseteq I 
\times K.
\end{equation}
Then 
$V$ is rectifiable.
\end{Proposition}

\begin{proof}
In view of Proposition \ref{prouno} it is enough to prove that $\mu^s_V=0$.
By Proposition \ref{proconsuno} below we have 
$$
p_\#\mu_V<<\mathcal L^1,
$$
where we recall that 
$p : \spacetime\to \R$ is defined in \eqref{def:p},
and where $\mathcal L^1$
is the Lebesgue measure.
This implies in particular that 
$p_\#\mu^s_V<<\mathcal L^1$. 
Hence $p_\#\mu^s_V=0$, which in turn implies $\mu^s_V=0$.
\end{proof}

%%%%%%%%%%%%%%%%%%%%%%%%%%%%%%%%%%%%%%%%%%%%%%%%%%%%%%%%%%%%%%%%%%%
\section{Conserved quantities}\label{seccons}
%%%%%%%%%%%%%%%%%%%%%%%%%%%%%%%%%%%%%%%%%%%%%%%%%%%%%%%%%%%%%%%%%%%%
In this section we show that for stationary varifolds, 
despite their nonsmoothness, 
we can still speak about various conserved quantities. These conservation laws
will be useful in the nonsmooth examples considered in Sections 
\ref{sub:relstri} and \ref{secexa}: see in particular Examples \ref{exa:square}
and \ref{exa:collsplit}. 

We need the following result, which is 
given in a time-localized form. 

\begin{Proposition}[{\bf Absolute continuity of 
$p_\# \mu_V$}]\label{proconsuno}
Let $V\in  \LV$ be a stationary varifold. Assume that 
\begin{equation}\label{eqkappa}
\exists I\subseteq \R {\rm ~interval~and~} \exists K 
\subset \Rn {\rm ~compact:}~ \quad
 {\rm spt}(\mu^{}_{V}) \cap (I \times \Rn) \subseteq I 
\times K.
\end{equation}
Then 
\begin{equation}\label{eq:ipocons}
p_\#\mu^{}_{V}
 \rest I << \mathcal L^1.
\end{equation} 
\end{Proposition}

\begin{proof}
Assumption \eqref{eqkappa} allows to apply 
Theorem \ref{th:disint} of the Appendix
(with the choices $d=1$, $m=N$, $\pi = p$ and $\nu = \mu_V$).
Therefore we can disintegrate 
$\mu^{}_V$ as 
$$
\mu^{}_V = 
p_\#\mu^{}_{V}
\otimes \lambda_t.
$$
Observe now that $\mu^{}_{\widetilde V^0}$
and $\mu^{}_{V^\infty}$ are both absolutely continuous with respect to 
$\mu^{}_V$. Therefore, also $p_\# \mu^{}_{V^0}$ and $p_\# \mu^{}_{V^\infty}$ 
are absolutely continuous with respect to $
p_\#\mu^{}_{V}$. Hence 
we can write
\begin{equation}\label{eq:splitdoppio}
\mu^{}_{\widetilde V^0} = p_\#\mu^{}_{V}
 \otimes \widehat\mu^{}_{\widetilde V^0_t},\qquad 
\mu^{}_{V^\infty} = p_\#\mu^{}_{V} \otimes \widehat\mu^{}_{V^\infty_t}.
\end{equation}
We now use the hypothesis that $V$ is stationary. Take
$$
Y(t,x)=\phi(t) \psi(x) e_0,
$$
 with 
$\phi \in \mathcal C^1_c(I)$ and $\psi \in \mathcal C^1_c(\Rn)$, with 
$\psi \equiv 1$ in a neighbourhood of $K$. Then 
the stationarity condition $\delta V(Y)=0$ implies
\begin{equation}\label{eq:zz}
0 = 
\int_{I\times\R^N} 
{\overline P}_{0}^{0}(t,x)\,\phi'(t)\,
d\mu^{}_{\widetilde V^0} + \int_{I\times\R^N}
{\overline Q}_0^0(t,x) \phi'(t) ~d\mu^{}_{V^\infty}.
\end{equation}
Notice that ${Q}^0_0=1$, so that passing to the mean value
also 
\begin{equation}\label
{eq:Q00uguale1}
{\overline Q}_{0}^{0}= 1,
\end{equation}
and therefore \eqref{eq:zz} becomes
\begin{equation}\label{eq:zzz}
0 = 
\int_{I\times\R^N} 
{\overline P}_{0}^{0}(t,x)\,\phi'(t)\,
d\mu^{}_{\widetilde V^0} + \int_{I\times\R^N}
\phi'(t) ~d\mu^{}_{V^\infty}.
\end{equation}

Using \eqref{eq:splitdoppio} we can write 
\eqref{eq:zzz} as
\begin{equation}\label{viet}
0=\int_{I} \phi'(t)f(t) ~dp_\#\mu^{}_{V}(t),
\end{equation}
where 
$$
f(t) := \int_{\Rn}
{\overline P}_{0}^{0}(t,x) d\widehat \mu^{}_{\widetilde V^0_t}(x)+
\widehat \mu^{}_{V^\infty_t}(\Rn), 
\qquad p_\# \mu_V-{\rm a.e.}~ t \in I.
$$
Observe that $f \geq 1$, since 
${\overline P}_{0}^{0}\ge 1$ and \eqref{eq:Q00uguale1} holds.
It follows
$$
f(t) = \int_{\Rn}
{\overline P}_{0}^{0}(t,x) d\widehat \mu^{}_{\widetilde V^0_t}(x)+
\widehat \mu^{}_{V^\infty_t}(\Rn) \geq 1,
\qquad p_\# \mu_V-{\rm a.e.}~ t \in I.
$$
{}From \eqref{viet}  and Lemma \ref{lem:lemmetto} below, it follows that 
there exists a constant $C \geq 1$ such that 
\begin{equation}\label{calfa}
f ~ p_\#\mu^{}_{V}
~\rest I  = C \mathcal L^1.
\end{equation}
Being $f \geq 1$, from \eqref{calfa} we deduce 
$p_\#\mu^{}_{V}\rest I<< \mathcal L^1$.
\end{proof}

\begin{Lemma}\label{lem:lemmetto}
Let $\mu$ be a positive Radon measure
on 
$\R$ such that 
$$
\int_\R \phi'~ d\mu = 0, \qquad \phi\in \mathcal C^1_c(\R).
$$
Then there exists a constant $C \geq 0$ such that
$\mu = C \mathcal L^1$.
 In particular $\mu<< \mathcal L^1$.
 \end{Lemma}
\begin{proof}
It is sufficient to prove 
that $\mu$ is translation invariant on open intervals. Therefore,
let $a,b \in \R$, $a<b$, 
  and $\tau \in \R$. 
We have to show that 
$$
\mu((a,b)) = \mu((a+\tau, b+\tau)).
$$
Assume that $b < a+\tau$, the other cases being similar. Let $\phi \in {\rm Lip}_c(\R)$ be defined as 
follows:
$\phi :=0$ in $(-\infty,a]\cup 
[b+\tau, +\infty)$, $\phi(t):=t-a$ in $(a,b)$, $\phi(t) := b-a$ 
in $(b, a+\tau)$, and $\phi(t) := -t + b + \tau$ in 
$(a+\tau, b+\tau)$. For any positive $\eps$ sufficiently small let $\phi_{\eps}\in \mathcal C^1_c(\R)$ 
be a function such that  $\phi_\eps = \phi$ in 
$(-\infty, a] \cup (a+\eps, b-\eps)\cup (b, a+\tau) \cup (a+\tau+\eps, b+\tau-\eps)\cup
[b+\tau,+\infty)$. We can in addition suppose 
$\phi_\eps$ to be  equi-Lipschitz with respect to $\eps$.
Then 
$$
\begin{aligned}
0=& \int_{(a,a+\eps)} \phi_\eps' ~d\mu 
+\int_{(b-\eps,b)} \phi_\eps' ~d\mu 
+
\int_{(a+\tau,a+\tau+\eps)} \phi_\eps' ~d\mu
+\int_{(b+\tau-\eps,b+\tau)} \phi_\eps' ~d\mu
\\
& + 
\mu([a+\eps, b-\eps]) - \mu([a+\tau+\eps, b+\tau-\eps]).
\end{aligned}
$$
Letting $\eps \to 0^+$ and using the equi-lipschitzianity of $\varphi_\eps$
it follows that the first four terms on the right hand side
converge to zero as $\eps \to 0^+$. Hence
$$
0 = \lim_{\eps \to 0^+} 
\Big(
\mu([a+\eps, b-\eps]) - \mu([a+\tau+\eps, b+\tau-\eps])  
\Big),
$$
and the assertion follows
from the inner regularity \cite{AmFuPa:00} of $\mu$.
\end{proof}

Let $V\in  \LV$ be a stationary varifold and assume that \eqref{eqkappa}
holds. Then using Proposition \ref{prosta} 
we can disintegrate $\mu^{}_{\widetilde V^0}$ and $\mu^{}_{V^\infty}$ with respect to $\mathcal L^1$,
 so that we can give the following
\begin{Definition}[{\bf The measures $\mu_{\widetilde V^0_t}$, $\mu_{V_t^\infty}$}]\label{def:ppo}
We set
\begin{equation}\label{mipre}
\mu^{}_{\widetilde V^0} = \mathcal L^1 \otimes {\mu^{}_{\widetilde V^0_t}},
\qquad
\mu^{}_{V^\infty} = \mathcal L^1 \otimes \mu^{}_{V^\infty_t}.
\end{equation}
\end{Definition}

We are now in a position to 
give the following definitions.

\begin{Definition}[{\bf Energy-momenta}]\label{def:enemome}
Let $V\in\LV$ and 
assume that \eqref{eqkappa} holds.
Then, for $\mathcal L^1$-almost every $t \in I$,
we define 
\begin{itemize}
\item[-] 
the energy momentum vector of $V$ at time $t$ as 
\begin{equation}\label{eq:enemome}
\begin{aligned}
\mathcal E(t) &:= 
\int_{\R^N}{\overline 
P}_{0}^{0}(t,x)\,d\mu^{}_{\widetilde V_t^0}(x)
+
\mu^{}_{V^\infty_t}(\R^N),
\\
\mathcal P^\indispazuno(t)&:= \int_{\R^N}{\overline 
P}_{0}^{\indicespaziale}(t,x)\,d\mu^{}_{\widetilde V_t^0}(x)+
\int_{\R^N} 
{\overline Q}_{0}^{\indicespaziale}(t,x)~d \mu^{}_{V^\infty_t}(x), 
\qquad \indicespaziale \in \{1,\dots,\dimspacetime\};
\end{aligned}
\end{equation}
\item[-]
the angular momentum of $V$ at time $t$ as 
\begin{equation}\label{eq:enemomebis}
\begin{aligned}
\Omega^{\alpha\beta}(t) :=
& \int_{\R^N}
\left(x^\alpha {\overline 
P}_0^\beta(t,x)-x^\beta {\overline P}_0^\alpha (t,x) 
\right)\,d \mu^{}_{\widetilde V_t^0}(x)
\\
& +
\int_{\R^N} 
\left(x^\alpha {\overline Q}_{0}^\beta(t,x) - 
x^\beta {\overline Q}_0^\alpha(t,x)\right)~d \mu^{}_{V^\infty_t}(x),
\end{aligned}
\qquad
\alpha, \beta \in \{0,\dots,N\}.
\end{equation}
\end{itemize}
\end{Definition}

\begin{Remark}\rm
{}From \eqref{anc} and \eqref{mipre} it follows that for $t_1, t_2 \in I$, $t_1<t_2$,
\begin{equation}\label{ambro}
\int_{t_1}^{t_2} \mathcal E(t)~dt = 
\mu_V\left((t_1,t_2) \times \Rn
\right).
\end{equation}
Indeed, 
$$
\begin{aligned}
\int_{t_1}^{t_2} 
\mathcal E(t)~dt = &
\int_{t_1}^{t_2} \int_{\R^N}{\overline 
P}_{0}^{0}(t,x)\,d\mu^{}_{\widetilde V_t^0}(x)~dt +
\int_{t_1}^{t_2} 
\mu^{}_{V^\infty_t}(\R^N)
~dt
\\
=& 
\int_{(t_1,t_2)\times \Rn} {\overline 
P}_{0}^{0}~ d\mu^{}_{\widetilde V^0} 
+
\mu^{}_{V^\infty}((t_1,t_2) \times \Rn)
\\
=& 
\int_{(t_1,t_2)\times \Rn} \int_{\timelikekplanes}
P_{0}^{0}(z)~ 
d\widetilde V^0_z (P)~d\mu^{}_{\widetilde V^0}(z) 
+
\mu^{}_{V^\infty}((t_1,t_2) \times \Rn)
\\
=& 
\mu_{V^0}((t_1,t_2) \times \Rn)
+\mu_{V^\infty}((t_1,t_2) \times \Rn)
=  
\mu_V((t_1,t_2) \times \Rn),
\end{aligned}
$$
where in the first equality of the last line we are 
using \eqref{eqkappa}, namely we are localized in a compact
subset of $\spacetime$.
\end{Remark}

For a rectifiable varifold, Definition \ref{def:enemome} gives
the usual definitions of relativistic energy and momentum, as shown in the next
observation.

\begin{Remark}[{\bf $\mathcal E$, $\mathcal P^{\mathrm a}$, $\Omega^{\alpha \beta}$ for a rectifiable varifold}]\label{rem:consrett}\rm
Let $V \in \LV$ be a rectifiable varifold satisfying 
\eqref{eqkappa}, and let 
$\rect^0, \theta^0, \rect^\infty$ and $\theta^\infty$
be as in Definition \ref{def:timerect}. 
Set
$$
\rect^0(t) := \rect^0 \cap \{x^0=t\},
\qquad
\Sigma^\infty(t) := \Sigma^\infty \cap \{x^0=t\}.
$$
Then, from \eqref{eq:tildeVzero} and  \eqref{eq:areavel} we have
\begin{equation}\label{musc}
\mu^{}_{\widetilde V^0} = \theta^0 \sigma^h \rest \Sigma^0  = 
\theta^0 
~\mathcal L^1 \otimes 
\left( \sqrt{1 - \vert \vel(t,\cdot)\vert^2_{\rm e}} 
~\mathcal 
H^{h-1} \rest \Sigma^0(t)\right), 
\end{equation}
and from \eqref{eq:Vinfinito}, 
\begin{equation}\label{olare}
\mu^{}_{V^\infty} = \theta^\infty \mathcal H^h \rest \Sigma^\infty
= \theta^\infty \sqrt{2} 
~\mathcal L^1\otimes
\left(\mathcal H^{h-1} \rest \Sigma^\infty(t)\right).
\end{equation}
Moreover, from \eqref{eq:poo} and \eqref{eq:pooo} we have
$$
{\overline P}^0_0 = {P}_0^0 = \frac{1}{1-\vert \vel \vert^2_{\rm e}},
\qquad
{\overline P}^\indispazuno_0 = {P}^\indispazuno_0 = \frac{\vel^\indispazuno}{1-\vert
\vel \vert^2_{\rm e}}, \quad \indispazuno \in \{1,\dots,N\},
$$
and from \eqref{eq:pronullo}
$$
{\overline Q}^0_0 = {Q}^0_0 = 1, 
\qquad
{\overline Q}^\indispazuno_0 = {Q}^\indispazuno_0 =   \vel_\infty^{\indispazuno}, \qquad
\indispazuno \in \{1,\dots,N\}.
$$
Hence, using \eqref{musc} and \eqref{olare},
for $\mathcal L^1$-almost every $t\in I$ 
and $\indispazuno, 
\indispazdue \in \{1,\dots,N\}$, we have
\begin{eqnarray}\label{eq:conse}
\mathcal E(t) &=& \int_{\manist^0(t)} \frac{
\theta^0}{\sqrt{1-\vert \vel\vert_{\rm e}^2}}~d \mathcal H^{\dimension-1}
+  \sqrt{2} \int_{\manist^\infty(t)} \theta^\infty ~d \mathcal H^{\dimension-1},
\\
\label{eq:consedavv}
\mathcal P^\indispazuno(t) &=&  
\int_{\manist^0(t)} \frac{\vel^\indispazuno}{\sqrt{1-\vert \vel\vert_{\rm e}^2}} ~\theta^0~d \mathcal H^{\dimension-1}
+ \sqrt{2} \int_{\manist^\infty(t)} 
\vel_\infty^{\indispazuno}~\theta^\infty ~d \mathcal H^{\dimension-1},
\end{eqnarray}
\begin{equation}\label{eq:angumome}
\begin{aligned}
\Omega^{{\rm ab}}(t) =& 
\int_{\manist^0(t)} \left(\frac{x^\indispazuno \vel^\indispazdue-
x^\indispazdue \vel^\indispazuno}{\sqrt{1 - \vert \vel\vert^2_{\rm e}}}\right)  ~\theta^0~d\mathcal H^{h-1}
\\
& +  \sqrt{2} \int_{\manist^\infty(t)} 
\left( x^\indispazuno \vel_\infty^{{\rm b}}-
x^\indispazdue \vel_\infty^{\indispazuno}\right) \theta^\infty 
~d\mathcal H^{h-1},
\end{aligned}
\end{equation}
and
\begin{equation}\label{palle}
\begin{aligned}
\Omega^{0\indispazuno}(t) = & 
\int_{\manist^0(t)} \left(\frac{t \vel^\indispazuno -
x^\indispazuno}{\sqrt{1 - \vert \vel\vert^2_{\rm e}}}\right)  ~\theta^0~d\mathcal H^{h-1}
\\
& +  \sqrt{2} \int_{\manist^\infty(t)} 
\left(t \vel_\infty^{\indispazuno}-x^\indispazuno\right) \theta^\infty 
~d\mathcal H^{h-1}.
\end{aligned}
\end{equation}
\end{Remark}

We now show that these quantities are conserved in time 
in the case of a stationary varifold.

\begin{Theorem}[{\bf Conserved quantities}]\label{procons}
Let $V\in  \LV$ be a stationary varifold and assume that 
\eqref{eqkappa} holds.
Then $\mathcal E$, $\mathcal P^\indispazuno$, and $\Omega^{\alpha\beta}$
do not depend on $t$.
\end{Theorem}
\begin{proof}
The constancy  of $\mathcal E$
follows from \eqref{calfa} by noticing that 
$$
\mathcal E=f\frac{{\rm d}p_\#\mu_V}{\rm d \mathcal L^1}=C, \qquad \mathcal 
L^1-{\rm a.e.~in~} I,
$$
where $\frac{\rm d}{\rm d \mathcal L^1}$ denotes the 
Radon-Nikod\'ym  derivative with respect to $\mathcal L^1$.

The constancy of $\mathcal P^\indispazuno$ 
can be proven arguing as in the proof 
of Proposition \ref{proconsuno}, 
 by testing the stationarity condition $\delta(V)=0$  with 
$Y(t,x)=\phi(t) \psi(x) e_\indispazuno$.

The constancy of $\Omega^{\mathrm a \mathrm b}$
can be proven by testing 
with $Y(t,x)=\phi(t)\psi(x) \left(x^\indispazuno e_\indispazdue-x^\indispazdue e_\indispazuno\right)$: 
in this case \eqref{eqfirst} gives 
$$
\begin{aligned}
0  =& 
\int_I \int_{\Rn} \varphi(t) {\rm tr}\Big({\overline P} (e^\indispazuno \otimes e_\indispazdue - e^\indispazdue \otimes
e_\indispazuno)\Big)  ~d\mu^{}_{\widetilde V^0_t}~dt
\\
& +
\int_I \int_{\Rn} \varphi(t) {\rm tr}\Big({\overline Q} (e^\indispazuno \otimes e_\indispazdue - e^\indispazdue \otimes
e_\indispazuno)\Big)  ~d\mu^{}_{V^\infty_t}~dt
\\
 &+ 
\int_I \int_{\Rn} \varphi'(t) {\rm tr}\Big({\overline P}e^0\otimes(x^\indispazuno
 e_\indispazdue - x^\indispazdue e_\indispazuno)\Big)
~d\mu^{}_{\widetilde V^0_t} dt
\\
&+
\int_I \int_{\Rn} \varphi'(t) {\rm tr}\Big({\overline Q}e^0\otimes(x^\indispazuno
 e_\indispazdue - x^\indispazdue e_\indispazuno)\Big)
~d\mu^{}_{V^\infty_t} dt
\\
= &
-\int_I \int_{\Rn} \varphi'(t) \Big(x^\indispazuno{{\overline P}}_0^\indispazdue - x^\indispazdue {{\overline P}}_0^\indispazuno\Big)
~d\mu^{}_{\widetilde V^0_t} dt
\\
& -
\int_I \int_{\Rn} \varphi'(t) \Big(x^\indispazuno{{\overline Q}}^0_\indispazdue- x^\indispazdue {{\overline P}}^0_\indispazuno\Big)
~d\mu^{}_{V^\infty_t} dt,
\end{aligned}
$$
where we used that 
${P}_\indispazuno^\indispazdue={P}_\indispazdue^\indispazuno$, ${P}_0^\indispazuno=-{P}_\indispazuno^0$, 
${Q}_\indispazuno^\indispazdue={Q}_\indispazdue^\indispazuno$ 
and ${Q}_0^\indispazuno=-{Q}_\indispazuno^0$. 
Then the assertion follows again from Lemma \ref{lem:lemmetto}. 

Let us now consider $\Omega^{0\mathrm a}$:
testing \eqref{eqfirst} 
with $Y(t,x)=\phi(t)\psi(x) \left(t e_\indispazuno+x^\indispazuno e_0\right)$ 
we obtain
$$
\begin{aligned}
0  =& \int_I \int_{\Rn} \varphi(t) {\rm tr}\Big({\overline P} (e^0 \otimes e_\indispazuno +
 e^\indispazuno \otimes e_0)\Big) ~d\mu^{}_{\widetilde V^0_t}~dt
\\
& +
\int_I \int_{\Rn} \varphi(t) {\rm tr}\Big({\overline Q} (e^0 \otimes e_\indispazuno +
 e^\indispazuno \otimes e_0)\Big) ~d\mu^{}_{V^\infty_t}~dt
\\
 &+ 
\int_I \int_{\Rn} \varphi'(t) {\rm tr}\Big({\overline P}(t e_\indispazuno + x^\indispazuno e_0)\Big)
~d\mu^{}_{\widetilde V^0_t}~dt
\\
 &+ 
\int_I \int_{\Rn} \varphi'(t) {\rm tr}\Big({\overline Q}(t e_\indispazuno + x^\indispazuno e_0)\Big)
~d\mu^{}_{V^\infty_t}~dt
\\
= & \int_I \int_{\Rn} 
\phi'(t) {\rm tr}\Big({\overline P} e^0 (t e_\indispazuno + x^\indispazuno e_0)\Big)~d\mu^{}_{\widetilde V^0_t}~dt
\\
 &+ 
\int_I \int_{\Rn} 
\phi'(t) {\rm tr}\Big({\overline Q} e^0 (t e_\indispazuno + x^\indispazuno e_0)\Big)~d\mu^{}_{V^\infty_t}~dt
\\  
= &\int_I \int_{\Rn} 
  \varphi'(t) \Big(t{{\overline P}}^0_\indispazuno + x^\indispazuno 
\Big)~d\mu^{}_{\widetilde V^0_t}~dt
+ 
\int_I \int_{\Rn} 
\varphi'(t) \Big(t{{\overline Q}}^0_\indispazuno + x^\indispazuno 
\Big)
~d\mu^{}_{V^\infty_t}~dt
\\
= & - \int_I \int_{\Rn}  \varphi'(t) \Big(t {{\overline P}}^\indispazuno_0 - x^\indispazuno {{\overline P}}_0^0\Big)
~d\mu^{}_{\widetilde V^0_t}~dt
-
\int_I \int_{\Rn}\varphi'(t) \Big(t {{\overline Q}}^\indispazuno_0 - x^\indispazuno \Big)
~d\mu^{}_{V^\infty_t}~dt,
\end{aligned}
$$
and the assertion follows as above from Lemma \ref{lem:lemmetto}. 
\end{proof}
We conclude this section by pointing out that condition \eqref{eqkappa}
is fulfilled in all examples considered in the present paper.

%%%%%%%%%%%%%%%%%%%%%%%%%%%%%%%%%%%%%%%%%%%%%%%%%%%%%%%%%%%
\section{Closed relativistic and subrelativistic strings}\label{sub:relstri}
%%%%%%%%%%%%%%%%%%%%%%%%%%%%%%%%%%%%%%%%%%%%%%%%%%%%%%%%%%%
We start now by considering the relevant examples which motivate
our theory.
 
Let $L>0$ and let $a,b\in \mathcal C^2(\R;\Rn)$ be two
$L$-periodic maps\footnote{Note that, for instance,
$b=0$ is not allowed, while it will be allowed in the definition 
of subrelativistic string.}  such that 
$$
|a'|_{\rm e}=|b'|_{\rm e}=1 \qquad {\rm in}~ \R.
$$
Define
\begin{equation}\label{gammaab}
\gamma(t,\xx):=\frac{a(\xx+t)+b(\xx-t)}{2}, \qquad (t,\xx) \in \R^2.
\end{equation}
Then $\gamma\in \mathcal C^2(\R^2; \Rn)$
is $L$-periodic both in $t$ and in $\xx$, and
 satisfies the linear wave system 
\begin{equation}\label{linearwave}
\gamma_{tt} = \gamma_{\xx\xx} \qquad {\rm in}~ \R^2,
\end{equation}
and the constraints
\begin{equation}\label{para:armo}
(\gamma_t, \gamma_\xx)_{\rm e}=0, \qquad 
\vert \gamma_t\vert^2_{\rm e} + \vert \gamma_\xx\vert^2_{\rm e} =1 \qquad{\rm in~}
\R^2.
\end{equation}
In particular, when $\gamma_\xx(t,\xx)=0$ we have $\vert \gamma_t(t,\xx)\vert^2_{\rm e}=1$, 
and hence $\gamma_t(t,\xx)$ is a null vector.

Define
the $\mathcal C^2$ map 
$\Phi_\gamma: \R^2 \to
\spacetime$ as
\begin{equation}\label{figamma}
\Phi_\gamma(t,\xx) := (t,\gamma(t,\xx)), \qquad (t,\xx) \in \R^2.
\end{equation}
Set also 
$$
S_\gamma := \{(t,\xx) \in \R\times [0,L): \gamma_\xx(t,\xx)=0\}, \qquad \qquad C_\gamma:=
\Phi_\gamma(S_\gamma),
$$
and note that, using also the periodicity assumption, it follows that 
$C_\gamma$ is closed.

If we assume that 
\begin{equation}\label{ifweassume}
\mathcal L^2(S_\gamma)=0,
\end{equation}
then 
$$
\mathcal H^2(C_\gamma)=0,
$$
and $\Phi_\gamma(\R^2)
\setminus C_\gamma$
is a lorentzian
minimal surface 
(see for instance \cite{ViSh:94}), 
namely, 
in the relatively open set $\Phi_\gamma(\R^2) \setminus C_\gamma$, its
lorentzian mean curvature vanishes. 

\begin{Remark}\label{nonmichiamo}\rm
In general, $C_\gamma$ may be nonempty, see for instance 
Example \ref{exa:kink}. 
\end{Remark}

\begin{Definition}[{\bf Closed relativistic string}]
A map  $\gamma \in \mathcal C^2(\R^2; \Rn)$ as in \eqref{gammaab}
and satisfying \eqref{ifweassume} is  called $L$-periodic (or also closed)
relativistic string.
\end{Definition}

\begin{Definition}[{\bf Varifold 
associated with a relativistic string}]\label{anchelui}
Let $\gamma$ be an $L$-periodic relativistic string, and set
$$
\Sigma_\gamma^0:=\Phi_\gamma\left(\R \times [0,L)\right),
$$
and
\begin{equation}\label{defteta}
\theta_\gamma^0(t,x) := \#\{ \xx\in [0,L):\ \gamma(t,\xx)=x\},
\qquad (t,x)\in\Sigma^0_\gamma.
\end{equation}
We define the rectifiable varifold $V_\gamma \in \LVdue$ associated
with $\gamma$ as
$V_\gamma := V_\gamma^0 + V_\gamma^\infty$, where
\begin{equation}\label{aglidounnome}
\widetilde V^0_\gamma:= 
\theta_\gamma^0 \left(\lormeastwo \rest \Sigma^0_\gamma\right) \otimes \delta_{P_{\Sigma^0_\gamma}},
\end{equation}\label{glidounnome}
\begin{equation}\label{imposing}
V_\gamma^\infty:=0.
\end{equation}
\end{Definition}
Notice that $\Sigma_\gamma^0$ is $2$-rectifiable; moreover, it is 
nonspacelike.
Notice also that the multiplicity $\theta_\gamma^0$ 
takes values in $\mathbb{N} \cup \{+\infty\}$.

\begin{Remark}\label{questoaiuta}\rm
If $\gamma$ is an $L$-periodic relativistic string, we have
\begin{equation}\label{chepag}
\mu_{V_\gamma} = 
{\Phi_\gamma}_\# \Big(\mathcal L^2\rest \left(\R \times[0,L)\right)\Big).
\end{equation}
Indeed, from the area formula
 we have, for any Borel set $F \subseteq \spacetime$,
\begin{equation}\label{stira}
\int_{F \cap \Sigma^0_\gamma} \theta^0_\gamma 
~d\mathcal H^2
 = \int_{\Phi_\gamma^{-1}(F)} 
\vert 
{\rm det}
(\grad \Phi_\gamma^T \grad \Phi_\gamma)
\vert~dt du,
\end{equation}
where $\grad$ denotes the gradient and ${}^T$ denotes tranposition.
Now, the equality
\begin{equation}\label{gammatiperp}
\gamma_t^\perp := \gamma_t - 
\left(\gamma_t, \frac{\gamma_u}{\vert \gamma_u\vert_{\rm e}}\right)_{\rm e} 
\frac{\gamma_u}{\vert \gamma_u\vert_{\rm e}}=
\gamma_t =\vel,
\end{equation}
and a direct computation
using  \eqref{para:armo} give
$$
\vert {\rm det}
(\grad \Phi_\gamma^T \grad \Phi_\gamma)\vert = \vert \gamma_\xx\vert_{\rm e}
\sqrt{1+\vert \gamma_t\vert_{\rm e}^2} 
= \sqrt{(1 - \vert \gamma_t\vert_{\rm e}^2)(1 + \vert \gamma_t\vert_{\rm e}^2)}
= \sqrt{1-\vert \vel\vert^4_{\rm e}}.
$$
Formula \eqref{stira} implies, using \eqref{eq:areavel} and \eqref{eq:poo}, 
$$
\mathcal L^2(\Phi_\gamma^{-1}(F))=
\int_{F \cap \Sigma^0_\gamma}
\frac{\theta_\gamma^0 }{\sqrt{1-\vert \vel\vert^4_{\rm e}}} ~d\mathcal H^2 
=
\int_{F\cap \Sigma^0_\gamma}
\frac{\theta_\gamma^0}{1-\vert \vel\vert^2_{\rm e}} ~d\sigma^2 
= \mu_{V_\gamma}(F).
$$
\end{Remark}

\begin{Remark}\label{rem:ststst}\rm
{}From \eqref{eq:conse} it follows that 
$L$ is the energy of the varifold $V_\gamma$, even  when $\gamma$ is not injective.
Indeed, recalling \eqref{eq:conse}, 
the equality $\vel(t, \gamma(t,\xx)) = \gamma_t(t,\xx)$ for 
$\mathcal L^2$-almost every $(t,\xx) \in \R^2$, and
using  \eqref{para:armo}, we have
\begin{equation}\label{servesempre}
\mathcal E(t) = \int_{\Sigma^0_\gamma \cap \{x^0=t\}} \frac{\theta^0_\gamma}{\sqrt{1 - 
\vert \vel\vert_{\rm e}^2}}~d\mathcal H^1 = 
\int_{[0,L)} \frac{1}{\vert \gamma_\xx\vert_{\rm e}} 
\vert \gamma_\xx\vert_{\rm e} ~d\xx = L, \qquad t \in \R.
\end{equation}
\end{Remark}

One could think (observing
for instance what happens in Example \ref{exa:kink} below, where 
the set $C_\gamma$ consists of a discrete set of points)
that a stationary $V_\gamma$ may have some concentrated 
measures at ``null-like points'', 
the presence of which\footnote{We need to use 
the quotations: indeed in Example \ref{exa:kink}
the tangent space to $\Sigma^0_\gamma$ does not exist
at the points of $C_\gamma$.} should contradict \eqref{imposing}.
Actually, this is not the case, due essentially
to condition \eqref{eq:ipocons}. As shown in the 
next theorem, if we do not impose 
\eqref{imposing}, then $V_\gamma$ is not  stationary.

\begin{Theorem}[{\bf Stationarity of relativistic strings}]\label{senzalabel}
Let $\gamma$ be an $L$-periodic relativistic string. 
Then the rectifiable varifold $V_\gamma$ associated with $\gamma$ is stationary.
\end{Theorem}
\begin{proof}
Let $Y\in \left(\mathcal C_c^1(\spacetime)\right)^{N+1}$. 
We have to prove that
\begin{equation}\label{rina}
\int_{\Sigma_0^\gamma}
\theta_\gamma^0~
{\rm tr}\left(P_{\Sigma_0^\gamma} d Y\right)~ d\sigma^2=0.
\end{equation}
Recalling 
\eqref{eq:lordivtang}, and 
since $\mathcal H^2(C_\gamma)=0$, we have
\begin{equation}\label{eq:lavorone}
 \int_{\Sigma^0_\gamma} \theta^0_\gamma ~{\rm tr}(P_{\Sigma^0_\gamma}d Y)~d\sigma^2 =
\int_{\Sigma^0_\gamma\setminus C_\gamma} \theta^0_\gamma~ {\rm div}_\tau Y~d\sigma^2.
\end{equation}
Write $Y = (Y^0, \widehat Y) \in \R \times \Rn$,  and set
$(\psi,\Psi)=Y(\Phi_\gamma)$, that is,
\begin{equation}\label{phipsi}
\psi(t,\xx)
:= Y^0(t,\gamma(t,\xx)), \qquad
\Psi(t,\xx):=\widehat Y(t,\gamma(t,\xx)), \qquad (t,\xx) \in \R \times [0,L).
\end{equation}
Let us compute ${\rm div}_\tau Y$ in terms of $\gamma$, $\psi$ and 
$\Psi$. Define 
$$
\zeta := \frac{(1,\gamma_t)}{\vert \gamma_\xx\vert_{\rm e}},
\qquad
\xi := \frac{(0,\gamma_\xx)}{\vert \gamma_\xx\vert_{\rm e}}
\qquad {\rm on}~ \R \times [0,L) \setminus \{\gamma_\xx=0\}.
$$
Then $\zeta$ (resp. $\xi$) is a timelike (resp. 
spacelike) vector, $(\zeta, \zeta) = -1$ (resp. $(\xi,\xi) = 1$),
and $(\zeta,\xi)_{\rm e} = (\zeta,\xi)=0$.
We have\footnote{If $T = T_\alpha^\beta$ is a $(1,1)$-tensor,
we have ${\rm tr}(T) = T_\alpha^\alpha = 
\langle T (e_0), e^0\rangle + 
\langle T (e_\indicespaziale), e^\indicespaziale\rangle
= - (T (e_0), e_0) + 
(T (e_\indicespaziale),  
e_\indicespaziale)$. Moreover 
${\rm tr}(L^{-1}TL) = {\rm tr}(T)$ for any Lorentz transformation 
$L$.}, at the point $(t, \gamma(t,\xx))$, and supposing $\gamma_\xx(t,\xx)
\neq 0$,
\begin{equation}\label{divtau}
{\rm div}_\tau Y = 
-(\langle d Y, \zeta\rangle , \zeta)+
(\langle d Y, \xi\rangle , \xi).
\end{equation}
Now, differentiating \eqref{phipsi} we obtain
$$
\left(\begin{tabular}{lr}
  $\psi_t$ & $\psi_u$ \\
  $\Psi_t$ & $\Psi_u$ \\
\end{tabular}\right)
= d\left(Y(\Phi_\gamma)\right) = dY(\Phi_\gamma)\,d\Phi_\gamma
= dY(\Phi_\gamma)
\left(\begin{tabular}{lr}
  $1$ & $0$ \\
  $\gamma_t$ & $\gamma_u$ \\
\end{tabular}\right).
$$
In particular, 
$$
\frac{1}{\vert \gamma_u\vert_{\rm e}}
\left(\begin{tabular}{lr}
 $\psi_t$ & $\psi_u$ \\
  $\Psi_t$ & $\Psi_u$ \\
\end{tabular}\right) 
\left(\begin{tabular}{c}
  $1$ \\
  $\underline 0$ \\
\end{tabular}\right)
= 
\frac{1}{\vert \gamma_u\vert_{\rm e}}
dY(\Phi_\gamma)
\left(\begin{tabular}{c}
  $1$ \\
  $\gamma_t$ \\
\end{tabular}\right)
= dY(\Phi_\gamma)(\zeta),
$$
and
$$
\frac{1}{\vert \gamma_u\vert_{\rm e}}
\left(\begin{tabular}{lr}
 $\psi_t$ & $\psi_u$ \\
  $\Psi_t$ & $\Psi_u$ \\
\end{tabular}\right) 
\left(\begin{tabular}{c}
  $0$ \\
  $\underline 1$ \\
\end{tabular}\right)
= 
\frac{1}{\vert \gamma_u\vert_{\rm e}}
dY(\Phi_\gamma) 
\left(\begin{tabular}{c}
  $0$ \\
  $\gamma_u$ \\
\end{tabular}\right)
= dY(\Phi_\gamma)(\xi),
$$
where $\underline 0 = (0,\dots,0) \in \R^N$, $\underline 1 = 
(1,\dots,1) \in \R^N$.
Hence, recalling \eqref{divtau},
\begin{eqnarray}
\nonumber
&& {\rm div}_\tau Y = 
-(\langle d Y, \zeta\rangle , \zeta)+
(\langle d Y, \xi\rangle , \xi)
\\
\nonumber 
&=& -\left( \frac{1}{\vert \gamma_\xx\vert_{\rm e}}\left(\begin{tabular}{lr}
  $\psi_t$ & $\psi_u$ \\
  $\Psi_t$ & $\Psi_u$ \\
\end{tabular}\right)
\left(\begin{tabular}{c}
  $1$ \\
  $0$ \\
\end{tabular}\right),
\zeta\right)
+
\left( \frac{1}{\vert \gamma_\xx\vert_{\rm e}}\left(\begin{tabular}{lr}
  $\psi_t$ & $\psi_u$ \\
  $\Psi_t$ & $\Psi_u$ \\
\end{tabular}\right)
\left(\begin{tabular}{c}
  $0$ \\
  $1$ \\
\end{tabular}\right),
\xi\right)
\\ \label{meno}
&=& 
\frac{(\gamma_\xx,\Psi_\xx)_{\rm e} - 
(\gamma_t,\Psi_t)_{\rm e} + \psi_t}{\vert \gamma_\xx\vert^2_{\rm e}}.
\end{eqnarray}

Recall from \eqref{gammatiperp} and \eqref{para:armo} that 
\begin{equation}\label{caz}
\vel (t, \gamma(t,\xx))= \gamma_t(t,\xx),\qquad
\sqrt{1-\vert \vel(t, \gamma(t,\xx))\vert^2_{\rm e}} = \vert \gamma_\xx(t,\xx)\vert_{\rm e}.
\end{equation}
{}From the area formula and \eqref{eq:areavel} we have
$$
\int_B \theta^0_\gamma ~d\sigma^2 = \int_{\Phi_\gamma^{-1}(B)} 
\vert \gamma_\xx\vert^2_{\rm e} ~dtd\xx, \qquad B \subset \Sigma^0_\gamma.
$$
We choose $T>0$  large enough so that $Y$ has support contained in
$(-T,T) \times \Rn$.
Since 
$S_\gamma$ has zero 
Lebesgue measure,
from \eqref{meno} it follows 
\begin{eqnarray*}
&& \
\int_{\Sigma^0_\gamma\setminus C_\gamma} \theta^0_\gamma~ {\rm div}_\tau Y~d\sigma^2
\\
&=& \int_{(-T,T) \times [0,L) \setminus S_\gamma}  
\left[
(\gamma_\xx, \Psi_\xx)_{\rm e}
- (\gamma_t, \Psi_t)_{\rm e} + \psi_t\right]~d\xx dt
\\
&=& 
\int_{(-T,T) \times [0,L)}  
\left[(\gamma_\xx, \Psi_\xx)_{\rm e}
- (\gamma_t, \Psi_t)_{\rm e} + \psi_t\right]~d\xx dt.
\end{eqnarray*}
On the other hand, recalling our assumptions on the 
support of $\psi$, the validity of the 
linear wave system \eqref{linearwave} and integrating
by parts, we get
$$
\begin{aligned}
& \int_{(-T,T) \times [0,L)}  
\left[(\gamma_\xx, \Psi_\xx)_{\rm e}
- (\gamma_t, \Psi_t)_{\rm e} + \psi_t\right]~d\xx dt
\\
=&
\int_{(-T,T) \times [0,L)}  
\left[(\gamma_\xx, \Psi_\xx)_{\rm e}
- (\gamma_t, \Psi_t)_{\rm e} \right]~d\xx dt=0.
\end{aligned}
$$
Hence $V_\gamma$ is stationary.
\end{proof}

\subsection{Subrelativistic strings}
As mentioned in the Introduction, the uniform closure of relativistic
strings has been characterized, see \cite{Br:05,BHNO}. In this section 
we want to show that to any subrelativistic string we can associate
a weakly rectifiable (not rectifiable in general) {\it stationary} 
varifold.

\begin{Definition}[{\bf Closed subrelativistic strings}]\label{def:subrel}
We say that $\gamma: \R^2 \to \R^N$ is an $L$-periodic (or closed)
subrelativistic
string if there exist $L$-periodic maps 
$a,\, b\in {\rm Lip}(\R;\R^N)$ 
such that 
$$
|a'|_{\rm e}\le 1,\quad |b'|_{\rm e}\le 1, \qquad 
{\rm a.e.~in}~\R,
$$
and 
\begin{equation}\label{forsub}
\gamma(t,\xx):=\frac{a(\xx+t)+b(\xx-t)}{2}, \qquad (t,\xx) \in \R^2.
\end{equation}
\end{Definition}

If $\gamma$ is an $L$-periodic subrelativistic string, then
\begin{equation}\label{ahh}
\vert \gamma_t^\perp\vert^2_{\rm e}+
\vert \gamma_u\vert^2_{\rm e} \leq 
\vert \gamma_t\vert_{\rm e}^2  +\vert \gamma_\xx\vert_{\rm e}^2 \leq 1
\qquad {\rm a.e.~in~} \R^2.
\end{equation}

\begin{Remark}\rm
It is proven in \cite{ViSh:94,BHNO} that any $L$-periodic subrelativistic 
string is a uniform limit of a sequence of $L$-periodic relativistic strings. 
\end{Remark}

\begin{Remark}\label{remtl}\rm
For any $L$-periodic subrelativistic string $\gamma$ we can 
still define the Lipschitz map 
\begin{equation}\label{bis}
\Phi_\gamma(t,\xx) := (t, \gamma(t,\xx)), \qquad
(t,\xx) \in \R^2,
\end{equation}
and the sets
$\Sigma_\gamma^0:=\Phi_\gamma(\R \times [0,L))$ (which is $2$-rectifiable),
$C_\gamma := \Phi_\gamma(S_\gamma)$,
where\footnote{Note that, differently with respect to a relativistic
string, now $S_\gamma$ is not
necessarily closed.} $S_\gamma := \{(t,\xx) \in \R \times [0,L): 
\gamma {\rm ~is~differentiable~w.r.t.~} \xx {\rm ~at~} (t,u), \gamma_u(t,u)=0\}$.
{}From \eqref{ahh} it follows that 
$$
\Sigma^0_\gamma\setminus C_\gamma \qquad {\rm is~ timelike~}
$$
 and
\begin{equation}\label{nonhacap}
\mathcal H^2(C_\gamma)=0.
\end{equation}
Indeed, 
$$
\{\vert \gamma_t\vert_{\rm e} =1\} \subseteq
S_\gamma \quad {\rm up~to~a~set~of~zero~Lebesgue~measure},
$$
and therefore \eqref{nonhacap} follows
from the area formula.
\end{Remark}

Also in view of the examples considered in the sequel of the paper,
we need
to extend Theorem \ref{senzalabel} to subrelativistic strings.
This is the content of the next theorem: we point out that
one of the tools needed in the proof is
the compactness
of $\overline{B_{h, N+1}}$. To better understand the meaning of the result,
it is useful to keep in mind formula \eqref{chepag}.

\begin{Theorem}[{\bf 
Stationarity of subrelativistic strings}]\label{senzanome}
Let $\{\gamma_j\} \subset \mathcal C^2(\R^2;\Rn)$ be a sequence of $L$-periodic
relativistic strings uniformly converging to an 
$L$-periodic subrelativistic string
$\gamma \in {\rm Lip}(\R^2; \Rn)$.
Then there exist a subsequence $\{\gamma_{j_k}\}$ 
of $\{\gamma_j\}$ and a 
stationary weakly rectifiable varifold $V_{\gamma}
\in \LVdue$ such that
\[
V_{\gamma_{j_k}}\rightharpoonup V_{\gamma} \qquad{\rm as}~ k \to 
+\infty,
\]
\begin{equation}\label{lanumero}
\mu^{}_{V_\gamma} = {\Phi_{\gamma}}_\#\big( 
\mathcal L^2 \rest \left(\R \times [0,L)\right)\big),
\end{equation}
and
\begin{equation}\label{vgamma}
\begin{aligned}
\mu^{ac}_{\widetilde V^0_\gamma} =&~ \Theta^0_\gamma~\sigma^2 
\rest \Sigma^0_\gamma,
\\
\mu^{ac}_{V^\infty_\gamma} =& ~0,
\end{aligned}
\end{equation}
where $\Theta_\gamma^0$ is a real-valued multiplicity function
satisfying
\begin{equation}\label{Thetatheta}
\Theta^0_\gamma\geq \theta_\gamma^0 \geq 1.
\end{equation}
\end{Theorem}

\begin{proof}
{}From Theorem \ref{senzalabel} it follows that $V_{\gamma_j}$ 
is stationary for any $j \in \mathbb N$. Using 
Remark \ref{rem:ststst}
and formula \eqref{ambro}
we deduce that, if 
$T>0$,
$$
\mu^{}_{V_{\gamma_j}}((-T,T) \times \Rn) 
= 2T L .
$$
Then, by Proposition \ref{procida} 
there exist a subsequence $\{\gamma_{j_k}\}$ of 
$\{\gamma_{j}\}$ and a 
varifold $V_\gamma \in \LVdue$ such that 
\[
V_{\gamma_j}\rightharpoonup V_{\gamma} 
\]
as $k\to +\infty$. In particular, from Remark 
\ref{rem:passano},  we have 
$$
\mu_{V_{\gamma_{j_k}}} \rightharpoonup
\mu_{V_\gamma} \qquad {\rm as}~ k \to +\infty.
$$
Furthermore, 
from Remark \ref{remfirstvar} it follows that $V_\gamma$ is
stationary. 

{}From \eqref{chepag} we have
$$
\mu^{}_{V_{\gamma_{j_k}}} = 
{\Phi_{\gamma_{j_k}}}_\# (
\mathcal L^2\rest (\R \times [0,L))
).
$$
Using the uniform convergence of $\{\gamma_{j_k}\}$ to $\gamma$, 
it follows
that 
$$
\mu^{}_{V_{\gamma_{j_k}}} =
{\Phi_{\gamma_{j_k}}}_\# (\mathcal L^2\rest \R \times [0,L))
\rightharpoonup {\Phi_{\gamma}}_\# (\mathcal L^2\rest (\R \times [0,L)))
\qquad
{\rm as}~ k \to +\infty, 
$$
and 
\eqref{lanumero} follows.

Then   ${\rm spt}(\mu^{}_{V_\gamma}) = \Sigma^0_\gamma$, 
where $\Sigma^0_\gamma
\setminus C_\gamma$ is $2$-rectifiable and 
timelike and $\mathcal H^2(C_\gamma)=0$ 
by Remark \ref{remtl}. 

We now claim that 
there exists $\Theta^0_\gamma$ satisfying \eqref{Thetatheta} such that
\begin{equation}\label{piccolo}
\mu^{ac}_{V^0_\gamma} = \Theta^0_\gamma~ (P_{\Sigma^0_\gamma})^0_0~ \sigma^2 \rest 
\Sigma^0_\gamma.
\end{equation}
Indeed, 
let
$\theta_\gamma^0$ be as in \eqref{defteta}, and set
for notational simplicity 
$$
p := \frac{
\gamma_\xx(\Phi_\gamma^{-1})}
                {
\vert \gamma_\xx(
\Phi_\gamma^{-1})\vert_{\rm e}}.
$$
Reasoning as in Remark \ref{questoaiuta},  
using \eqref{lanumero} and the area formula, 
we have, for all Borel sets $F \subseteq \spacetime$,
\begin{equation}\label{dimiuno}
\begin{aligned} 
\mu^{ac}_{V_\gamma}(F)=&
\mathcal L^2(\Phi_\gamma^{-1}(F\setminus C_\gamma))
\\
 =& 
\int_{F \cap \Sigma^0_\gamma}
\frac{\theta_\gamma^0 }{
\vert \gamma_u\vert_{\rm e}
\sqrt{1+(
\sum_{\indicespaziale, {\rm b}=1}^N 
\gamma_t^\indicespaziale p^{\rm b} - \gamma_t^{\rm b} p^\indicespaziale)^2}
} 
~d\mathcal H^2
\\
 =& 
\int_{F \cap (\Sigma^0_\gamma \setminus C_\gamma)}
\frac{\theta_\gamma^0 }{
\vert \gamma_u\vert_{\rm e}
\sqrt{1+(
\sum_{\indicespaziale, {\rm b}=1}^N 
\gamma_t^\indicespaziale p^{\rm b} - \gamma_t^{\rm b} p^\indicespaziale)^2}
} 
~d\mathcal H^2 = 
\mu^{ac}_{V_\gamma^0}(F).
\end{aligned}
\end{equation}
In particular, \eqref{dimiuno} implies the second equality
in \eqref{vgamma}.

Recall now from \eqref{gammatiperp} that 
$\vel = \gamma_t^\perp$. 
Notice that 
$$
\left(\sum_{\indicespaziale, {\rm b}=1}^N 
\gamma_t^\indicespaziale p^{\rm b} - \gamma_t^{\rm b} p^\indicespaziale\right)^2
=
\left(\sum_{\indicespaziale, {\rm b}=1}^N 
{\gamma^\perp_t}^\indicespaziale p^{\rm b} - {\gamma_t^\perp}^{\rm b} 
p^\indicespaziale\right)^2 = \vert \gamma_t^\perp\vert_{\rm e}^2
=\vert \vel\vert_{\rm e}^2.
$$
Hence, using \eqref{eq:areavel}, 

\begin{equation}\label{dimi}
\begin{aligned} 
\mu^{ac}_{V_\gamma^0}(F)
= &
\int_{F\cap \Sigma^0_\gamma}
\frac{\theta_\gamma^0}{\vert \gamma_u\vert_{\rm e}} 
\frac{\sqrt{1 + \vert \vel\vert^2_{\rm e}}}{
\sqrt{1-
\vert \vel\vert_{\rm e}^2}
\sqrt{1+
\vert \vel\vert_{\rm e}^2
}
} ~d\sigma^2
\\
= &
\int_{F\cap \Sigma^0_\gamma}
\frac{\theta_\gamma^0}{\vert \gamma_u\vert_{\rm e}} 
\frac{
1}{
\sqrt{1-
\vert \vel\vert_{\rm e}^2}
} ~d\sigma^2
\\
=&  \int_{F\cap \Sigma^0_\gamma}
\frac{\Theta_\gamma^0}{1-\vert \vel\vert^2_{\rm e}} ~d\sigma^2, 
\end{aligned}
\end{equation}
where 
$$
\Theta_\gamma^0 := 
\theta^0_\gamma
~ \frac{
 \sqrt{1 - \vert \vel \vert_{\rm e}^2}}{\vert 
\gamma_u\vert_{\rm e}}.
$$
Recalling  \eqref{ahh} we obtain \eqref{Thetatheta}, and this concludes
the proof of the claim. It follows from \eqref{piccolo} that 
the first equality in \eqref{vgamma} holds. 

{}From \eqref{lanumero} and the 
definition of diffuse part of a measure (Section \ref{sub:ass}) it follows
that 
\begin{equation}\label{nodiffuse}
\mu_{V_\gamma}^d=0,
\end{equation}
 since $\mu_{V_\gamma}$ is a Radon 
measure concentrated on a $\dimension$-rectifiable set. This concludes 
the proof that $V_\gamma$ is weakly rectifiable.
\end{proof}

\begin{Remark}\label{nome}\rm
{}From \eqref{lanumero}, \eqref{vgamma} and \eqref{nodiffuse}
 it follows that 
$$
\mu_{V_\gamma}^s = 
\mu_{V_\gamma^0}^s + \mu_{V_\gamma^\infty} = 
\mu_{V_\gamma^0}^s + \mu^s_{V_\gamma^\infty} 
={\Phi_\gamma}_\# \left(
\mathcal L^2 \rest S_\gamma\right).
$$
\end{Remark}

\begin{Remark}\rm
The sequence $\{\mu_{V_{\gamma_j}}\}$ converges to $\mu_{V_\gamma}$,
without passing to a subsequence $\{j_k\}$.
Moreover $\mu^{}_{V_\gamma}$ is independent of $\{\gamma_j\}$ (see
\eqref{lanumero}) and depends
only on $\gamma$, while 
a priori $V_\gamma$ could depend on $\{\gamma_{j_k}\}$.
\end{Remark}

%%%%%%%%%%%%%%%%%%%%%%%%%%%%%%%%%%%%%%%%%%%%%%%%%%%%%%%%%%%%%%%%%%%%%%%
\subsection{Examples of varifolds associated 
with relativistic and subrelativistic strings}\label{subkink}
%%%%%%%%%%%%%%%%%%%%%%%%%%%%%%%%%%%%%%%%%%%%%%%%%%%%%%%%%%%%%%%%%%%%%%%
A first example of stationary rectifiable varifold is the so-called kink.
This example is not completely trivial, since the set $C_\gamma$ is not
empty. 

\begin{Example}[{\bf Kink}]\label{exa:kink}\rm 
Let $N=2$, $h=2$,  $R>0$, and 
\begin{equation}\label{exkink}
\gamma(t,\xx) := R\big(\cos(\xx/R), \sin(\xx/R)\big) \cos(t/R), 
\qquad (t,\xx) \in \R^2,
\end{equation}
be the kink. 
Note that $S_\gamma$ is nonempty, precisely
$$
S_\gamma = \left\{ 
\left(\frac{\pi R}{2} + k \pi R, 0\right): k \in \mathbb{Z}\right\},
$$
which, being a discrete set of points, satisfies condition \eqref{ifweassume}.
Since it is immediate to check the validity of the system 
\eqref{linearwave}, 
$\gamma$ is therefore a $2\pi$-periodic relativistic string. 
The associated varifold $V_\gamma \in \LVdue$ defined by
$V_\gamma = V_\gamma^0 + V_\gamma^\infty$, where
$$
\widetilde V^0_{\gamma} = \left(\sigma^2 \rest \Sigma^0_\gamma\right) \otimes
\delta_{P_{\Sigma^0_\gamma}},  \qquad V_\gamma^\infty=0
$$
is proper, rectifiable and stationary by Theorem \ref{senzalabel}. 
\end{Example}

\begin{figure}[H]
\begin{center}
\includegraphics[height=5cm]{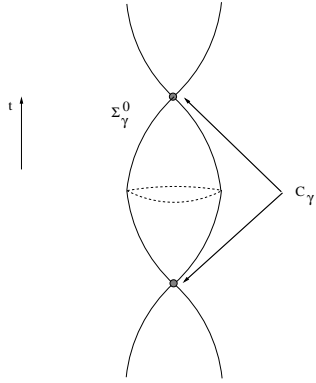}
\smallskip
\caption{\small The set $\Sigma^0_\gamma$ in Example \ref{exa:kink}.}
\label{fig:kink}
\end{center}
\end{figure}

Cylinders over any closed curve support a stationary weakly rectifiable
varifold, as shown in the following example.

\begin{Example}[{\bf Cylindrical strings}]\label{exa:sub}\rm
Let $N=2$ and $\dimension = 2$. We consider a particular class of 
subrelativistic strings.
Precisely, let $a\in \C^2(\R;\R^2)$ be a $1$-periodic map, 
with $|a'(s)|=1$ for
all $s\in\R$,
and let
$$
\Sigma^0=\Sigma := 
\R \times a(\R) \subset \R^{1+2}
$$
be the cylinder over $a(\R)$.
{}From Definition \ref{def:subrel} we have that the function 
$$
\gamma(t,\xx):=\frac{a(\xx+t)}{2}, \qquad (t,u) \in \R^2,
$$
is a subrelativistic string 
with the corresponding $\Phi_\gamma$ parametrizing $\Sigma$. 

Observe that $\gamma$ is the 
uniform limit of the sequence $\{\gamma_n\}$ of $1$-periodic relativistic strings
$$
\gamma_n(t,\xx) := \frac{a(\xx+t)
+\frac{1}{n} a\left(n(\xx-t)\right)}{2},\qquad
(t,\xx)\in\R^2.
$$

By Theorem \ref{senzanome}
there exists a stationary weakly rectifiable varifold $V \in \LVdue$ such that
$$
\mu_{\widetilde V^0} = \sigma^2\rest \Sigma 
$$
and $V= \lim_{k\to +\infty}V_{n_k}$,
where 
$$
\widetilde V_{n}^0 := \left(
\sigma^2\rest \Sigma_n\right) \otimes \delta_{P_{\Sigma_n}},
\qquad 
\Sigma_n^0 = \Sigma_n := \R \times \gamma_n(\R \times [0,1)).
$$

\newpage

Since $\gamma_\xx(t,\xx)\ne 0$ for all $(t,\xx)$ we have 
$S_\gamma=\emptyset$, hence 
by Remark \ref{nome} it follows 
that  $V$ is a proper varifold. Moreover,
$${\rm Range}\left(\overline P_\Sigma(z)\right)\subset T_z \Sigma$$
(note the strict inclusion)
and,
recalling 
\eqref{eqtr}, also
\begin{equation}\label{eqtrim}
{\rm tr}\left(\overline P_\Sigma(z)\right)=2.
\end{equation}
Furthermore,
from the necessary stationarity condition \eqref{eqHgen}, 
and the fact that $\Theta^0_\gamma=1$, we have
\begin{equation}\label{eqdiv}
{\rm div}\left(\overline P_\Sigma(z)\right)=0, \qquad {\rm\ for\ all\ }z\in \Sigma.
\end{equation}
These conditions necessarily imply that 
\begin{equation}\label{eqp2}
\overline P_\Sigma(z) = {\rm diag}(2,0,0).
\end{equation} 
Indeed, assuming without loss of generality
 $\n_\Sigma=e_2$, from \eqref{eqtrim} we get 
$$
\overline P_\Sigma(z) = {\rm diag}(1+\alpha, 1-\alpha,0)
$$
for some $\alpha\in\R$.
As a consequence, we have ${\rm div}(\overline P_\Sigma(z))=(1-\alpha)(0,0,H_{\Sigma}(z))$.
Recalling \eqref{eqdiv} we then get $\alpha=1$, which implies \eqref{eqp2}.

\smallskip

Notice that, given a multiplicity function $\theta> 0$ depending only on $x$,
by \eqref{eqCweak} the varifold $\theta V$ is still stationary. This is a peculiar phenomenon of cylindrical strings 
and may be not true for stationary rectifiable varifolds, 
that is, in general for relativistic strings.

Notice also that the lorentzian mean curvature of $\Sigma$ is not
zero everywhere, and therefore $V$ is not rectifiable.
\end{Example}

The following example was originally considered in \cite{BHNO} 
in a classical parametrized setting. 

\begin{Example}[\bf Polyhedral string]\label{exa:square}\rm
Assume $N=2$, $h=2$, let $L>0$ and let $a  : \R \to \R^2$
be a Lipschitz continuous
 $\energy:=4L$-periodic map, such that $a_{\vert [0,4\rightextremum]}$
is the counterclock-wise arc-length parametrization of the boundary of
the square $Q_0 = [-L/2,L/2]^2$.
Obviously $a_{\vert [0,4\rightextremum]}\in (\C^2([0,4\rightextremum]
\setminus \{0,
\rightextremum,2 \rightextremum,3 \rightextremum\}))^2$.
We define the
map
$$
\spaceimm(t,\xx):= 
\frac{a(\xx+t) + a(\xx-t)}{2},
\qquad (t,\xx) \in \R^2,
$$
and we let $\Phi_\gamma$ be as in \eqref{bis}.
Recalling Definition \ref{def:subrel}, we have that $\gamma$ 
is a
 $4 \rightextremum$-periodic  subrelativistic string.
Notice that $\spaceimm(t,\cdot)$ is a Lipschitz parametrization of
$\partial Q(t)$, where $Q(t)$ is the square defined as
\[
Q(t) := Q_0\cap \left\{ (x^1,x^2)\in\R^2:\,|x^1|+|x^2|\le L-|t| \right\},
\qquad t\in [-\rightextremum,\rightextremum].
\]
By \cite[Theorem 3.1]{BHNO} 
$\gamma$ is the uniform limit of a sequence $\{\gamma_j\}
\subset \mathcal C^2(\R^2; \R^N)$
of $4 \rightextremum$-periodic relativistic strings with 
zero initial velocity
(in particular, the strings $\gamma_j$ have equibounded energy).
 Let $V_\gamma \in \LVdue$ be the corresponding 
stationary weakly rectifiable varifold given by
Theorem \ref{senzanome}. 

Referring also to Figure \ref{fig:square3D}, observe that: 
 $\Sigma_\gamma^0=\Phi_\gamma(\R^2)$ is the
support of $\mu_{V_\gamma}$; moreover
$$
\mu_{\widetilde V^0_\gamma}=\sigma_2 \rest \Sigma_\gamma^0.
$$
Indeed, 
as a consequence of the equality $(\gamma_t, \gamma_\xx)_{\rm e}=0$ valid 
almost everywhere in $\R^2$, we have, from
 \eqref{dimi}, 
$$
\Theta_\gamma^0=1.
$$
Note also that by Remark \ref{nome}
$$\mu^{}_{V^s_\gamma}={\Phi_\gamma}_\#(\mathcal L^2\rest S_\gamma).
$$
For $|t|\in [0,L/2)$ the set
$Q(t)$ is an octagon and $\mu^{s}_{V_\gamma}\rest \left((-L/2,L/2)\times
\R^2\right)=0$, that is,
the restriction of $V_\gamma$ to $(-L/2,L/2)\times \R^2
\times \overline{B_{2,3}}$ is a proper varifold.

For $t\in [L/2,L)$, $Q(t)$ is a square of sidelength $\sqrt
2(L-t)$, rotated of an angle $\pi/2$
with respect to the initial square $Q_0$, which shrinks to the point $(0,0)$
as $t\uparrow L$.
%Considering the disintegration $\mu^{s}_{V_\gamma}=\mathcal
%L^1 \otimes \mu^s_t$,
%we get
%$$
%\mu^s_t=0 \qquad {\rm for}~ |t|< L/2,
%$$
%so that, for times $|t|\le L/2$, there is neither singular part nor part at
%infinity of the varifold $V_\gamma$.
%On the other hand, 
For $t \in (L/2,L)$ the four vertices of the rotated
square
$Q(t)$ move at speed 1, and the edges move with 
normal velocity equal to $\frac{1}{\sqrt{2}}$.
% for $|t|\in [L/2,L)$
%$\mu^s_t$ is supported on the four vertices of $Q(t)$,
Moreover we have
$$
\mu^s_{V_\gamma} = \alpha(t) \mathcal H^1 \rest \ell,
$$
where the singular set $\ell\subset\R^{1+2}$ is the union of four line 
segments $\ell_1,\ldots,\ell_4$, see Figure \ref{fig:square3D}.
To find $\alpha(t)$, which turns
out to be a linearly increasing function,  we use the conservation of energy
given by Theorem \ref{procons}.
{}From \eqref{eq:conse} we have 
$$
\mathcal E(t) = 8(L-t) + 4 \alpha(t) = 4L, 
\qquad t \in (L/2,L).
$$
We get 
$$
\alpha(t) = 2t -L,\qquad t \in (L/2,L).
$$
We conclude by observing that we expect
$V_\gamma^\infty$ to be concentrated on the singular set $\ell$: 
$$
V_\gamma^\infty=\alpha(t) \mathcal H^1 \rest \ell\otimes \delta_{Q_\gamma},
$$
where $Q_\gamma(z)=-(1,v_i(z))\otimes \eta (1,v_i(z))$ 
for $z\in \ell_i$, $i\in\{1,\ldots,4\}$,
and $v_i(z)\in\R^2$ is such that $|v_i(z)|_{\rm e}=1$ and $(1,v_i)$ is parallel to $\ell_i$.
%\end{enumerate}
\end{Example}

%%%%%%%%%%%%%%%%%%%%%%%%%%%%%%%%%%%%%%%%%%%%%%%%%
\section{Further examples: null hyperplanes and collisions}\label{secexa}
%%%%%%%%%%%%%%%%%%%%%%%%%%%%%%%%%%%%%%%%%%%%%%%%%
A null hyperplane $\rect \subset \spacetime$
 is not a minimal hypersurface in the classical sense,
since the vector $\n_1$ is not well defined
(heuristically, it should be parallel to $\rect$
and should have infinite euclidean norm). 
Therefore it becomes meaningless computing the classical
mean curvature $H_\rect$.
However, we can interpret these planes as stationary varifolds.  

\begin{Example}[{\bf Null $\dimension$-spaces as stationary varifolds}]\label{exa:subspace}\rm
Given $\indicesuccessione
 \in \mathbb N$, 
let $\manist_\indicesuccessione
=\manist_\indicesuccessione^0$ be an $h$-dimensional
 timelike subspace of $\spacetime$,
let $\theta_\indicesuccessione
=\theta_\indicesuccessione^0 \in (0,+\infty)$, 
and set
$$
\widetilde V^0_\indicesuccessione=\theta_\indicesuccessione~
(\H^\dimension\rest \manist_\indicesuccessione) \otimes \delta_{P_\indicesuccessione}
\in \LV
$$
be the proper rectifiable $h$-varifold associated with $\manist_\indicesuccessione$ and $\theta_\indicesuccessione$, where $P_\indicesuccessione := P_{\manist_\indicesuccessione}$.
Recall from \eqref{eq:poo} and \eqref{eq:areavel} that 
\begin{equation}\label{spiderman}
\lormeas \rest \manist_\indicesuccessione = \Big(({P_\indicesuccessione})^0_0\Big)^{-1/2} (\mathcal L^1 \otimes {\mathcal H}^{\dimension-1}) \rest \manist_\indicesuccessione.
\end{equation}
Hence
$$
\mu_{\widetilde V^0_\indicesuccessione}  = \theta_\indicesuccessione~ \H^\dimension \rest \manist_\indicesuccessione 
= \theta_\indicesuccessione\Big( (P_\indicesuccessione)^0_0\Big)^{-1/2}(\mathcal L^1 \otimes {\mathcal H}^{\dimension-1}) \rest \manist_\indicesuccessione.
$$
Note that from \eqref{eq:Vzerotilde}
and \eqref{spiderman} we have
\begin{equation}\label{tours}
V_\indicesuccessione = V_\indicesuccessione^0 = \theta_\indicesuccessione~ (P_\indicesuccessione)^0_0 ~ \H^\dimension\rest \manist_\indicesuccessione\otimes \delta_{q(P_\indicesuccessione)}
= \theta_\indicesuccessione~\sqrt{(P_\indicesuccessione)^0_0}~(
\mathcal L^1 \otimes {\mathcal H}^{\dimension-1}) \rest \manist_\indicesuccessione \otimes \delta_{q(P_\indicesuccessione)}.
\end{equation}
Suppose now that
\begin{itemize}
\item[-]
 $\manist_\indicesuccessione$ converge to a null $\dimension$-plane $\manist^\infty$ as
$\indicesuccessione \to +\infty$,
 \item[-]
 there exists the limit 
\begin{equation}\label{eq:piano}
 \lim_{\indicesuccessione \to +\infty} \theta_\indicesuccessione \sqrt{(P_\indicesuccessione)^0_0} =: C \in (0,+\infty).
\end{equation}
\end{itemize}
In particular
$$
\lim_{\indicesuccessione \to +\infty}\theta_\indicesuccessione =0,
$$
so that $\widetilde V^0_\indicesuccessione \rightharpoonup 0$. Recalling \eqref{tours} we have
$$
V_\indicesuccessione \rightharpoonup 
C\,\left(\mathcal L^1 \otimes {\mathcal H}^{\dimension-1}\right) \rest \Sigma^\infty
\otimes \delta_{Q_{\Sigma^\infty}}
= \frac{C}{\sqrt{2}} (\mathcal H^\dimension \rest \Sigma^\infty)
\otimes \delta_{Q_{\Sigma^\infty}} =: V.
$$
Hence 
$$
V=V^\infty,
$$
and $V\in \LV$ is 
rectifiable.
Finally, it follows from Remark \ref{remfirstvar} that 
$$
V {\rm ~ is~ stationary}.
$$
\end{Example}
      
\begin{Example}[\bf Collisions and splittings]\label{exa:collsplit}\rm
Let 
$$
N=1,  \qquad \dimension=1,
$$
let be given angles $\alpha,\,\beta\in (\pi/4,\pi/2)$, and
real multiplicities $\theta_i\in (0,+\infty)$ for $i\in \{ 1,2,3\}$.
We consider the proper rectifiable varifold $V=V^0 \in \LVuno$, where
\begin{equation}\label{puntotriplo}
\widetilde V^0 = \sum_{i=1}^3 \theta_i\,\left(\H^1\rest\Sigma_i \right)
\otimes \delta_{P_{\Sigma_i}}.
\end{equation}
We have indicated here by $\Sigma_i = \Sigma_i^0$ 
the relatively open three half-lines depicted in Figure 
\ref{fig:creation} meeting at the point $p$ of the plane
$\mathbb R^{1+1}$. The condition $\alpha, \beta > \pi/4$ yields
that $\Sigma_2$ and $\Sigma_3$ are timelike.
We are interested in the following problem:
{\it given the multiplicity $\theta_1$ on the incoming half-line 
$\Sigma_1$, find conditions on $\theta_2$, $\theta_3$, 
$\alpha$ and $\beta$ which ensure that 
$\widetilde V^0$ in \eqref{puntotriplo} is 
stationary.}

We regard each $\Sigma_i$ as a one-dimensional manifold with boundary (the point $p$).
We indicate by $\tau_i$  the euclidean unit conormal vector at $p$ pointing
out of  $\Sigma_i$:
$$
\tau_1 = (1,0), \qquad \tau_2 = (-\sin\alpha,
\cos \alpha), \qquad \tau_3 =(-\sin \beta, -\cos \beta).
$$
The stationarity requirement (Definition \ref{defsta}) 
reads as
\begin{equation}\label{stazzz}
0 = \sum_{i=1}^3 \theta_i \int_{\Sigma_i} {\rm div}_{\tau}   Y ~d \H^1,
\qquad 
Y\in (\mathcal C^1_c(\R^{1+1}))^2.
\end{equation}
Following  \eqref{perpm}, and since the lorentzian (mean) curvature of each 
$\Sigma_i$ vanishes,  we have
\begin{equation}\label{nientehacapi}
{\rm div}_{\tau}   Y  =  {\rm div}_{\tau}  \left(
P_{\Sigma_i} Y\right) \qquad {\rm on}~ \Sigma_i.
\end{equation}
Moreover if  $Z \in (\C^1(\R^{1+1}))^2$  is tangent to
$\Sigma_i$ we have 
\begin{equation}\label{verodritto}
 {\rm div}_{\tau}   Z  =  {\rm div}_{\tau}^{\rm e}  Z \qquad {\rm on}~\Sigma_i,
 \end{equation}
where ${\rm div}^{\rm e}_\tau$ is the euclidean tangential divergence.
Indeed, if $\mathcal Z$ is an extension of $Z_{\vert \Sigma_i}$ in an 
open neighbourhood of $\Sigma_i$, constant
along $\n_i$, we have ${\rm div}_\tau Z = {\rm tr}(d \mathcal Z)
= {\rm div}_\tau^{\rm e} Z$. 

Applying this observation to $Z = P_{\Sigma_i} Y$ and using 
\eqref{nientehacapi}, we get 
$$
{\rm div}_\tau Y = {\rm div}_\tau^{\rm e} (P_{\Sigma_i} Y)
\qquad {\rm on}~ \Sigma_i.
$$
Set for notational simplicity
$$
\n_i := {\n_1}^{}_{\Sigma_i}, \qquad 
\nu_i:= \nu^{}_{\Sigma_i} = \frac{\eta \n_i}{\vert \eta \n_i\vert_{\rm e}},
\qquad i \in \{1,2,3\},
$$
see \eqref{def:nufi}. Observe that $\n_1 \in \{(0,1),(0,-1)\}$
(in particular $\n_1$ has the same direction of a euclidean
normal to $\Sigma_1$), and we
choose 
\begin{equation}\label{eq:nuno}
\n_1 = (0,1).
\end{equation}
Moreover, using \eqref{mattex} and choosing $\nu_2 = (-\cos \alpha, -\sin \alpha)$,
$\nu_3 = (-\cos \beta, \sin \beta)$, we find
\begin{equation}\label{eq:ndue}
\n_2=\frac{1}{\sqrt{-\cos^2 \alpha+  \sin^2\alpha}} (\cos\alpha, -\sin\alpha), 
\quad \n_3=
\frac{1}{\sqrt{-\cos^2\beta+\sin^2\beta}}
(\cos\beta, \sin\beta)
\end{equation}
(recall that by definition the time component of $\n_i$ are required to be nonnegative).

{}From \eqref{stazzz} and \eqref{verodritto}  we obtain, using 
also \eqref{sigman} and integrating by parts,  
\begin{equation}\label{jj}
\begin{aligned}
0 &= \sum_{i=1}^3 \theta_i \int_{\Sigma_i} {\rm div}_{\tau}^{\rm e}  \left(P_{\Sigma_i} Y\right) ~d \H^1 = 
\sum_{i=1}^3 \theta_i \vert \nu_i\vert 
\int_{\Sigma_i} {\rm div}_{\tau}^{\rm e}  \left(P_{\Sigma_i} Y\right) ~d \mathcal H^1
\\ 
&= \sum_{i=1}^3 \theta_i \vert \nu_i\vert (P_{\Sigma_i} Y(p), \tau_i)_{\rm e} 
=  \sum_{i=1}^3 \theta_i \vert \nu_i\vert (Y(p), P_{\Sigma_i} \tau_i)_{\rm e} 
= (Y(p), \sum_{i=1}^3 \theta_i \vert \nu_i\vert  P_{\Sigma_i} \tau_i)_{\rm e}.
\end{aligned}
\end{equation}
Denote by $R_i$ the matrix representing 
the euclidean rotation of angle $\pi/2$ such that $R_i \nu_i = \tau_i$
($R_1,R_2$ are counterclockwise, and $R_3$ is clockwise). 
Then 
$$
R_1 \n_1 = (1,0), \qquad
R_2 \n_2 = \frac{(-\sin \alpha, \cos \alpha)}{\sqrt{\sin^2\alpha-\cos^2 \alpha}},
\qquad R_3 \n_3 = \frac{(-\sin \beta, -\cos \beta)}{\sqrt{\sin^2\beta-\cos^2 \beta}}.
$$
We claim that 
\begin{equation}\label{eq:claim}
\vert \nu_i\vert  P_{\Sigma_i} \tau_i = R_i ~\n^{}_{i}, \qquad
i \in \{1,2,3\}.
\end{equation}
Indeed, 
$$
\eta \n_i \otimes \n_i~ \tau_i = 
\eta \n_i (\n_i,\tau_i)_{\rm e} = 
\eta \n_i \left(\frac{\nu_i}{\vert \eta^{-1} \nu_i\vert},
\eta^{-1} \tau_i\right)_{\rm e} 
= 
\frac{\nu_i}{\vert \nu_i\vert}
\left(\frac{\nu_i}{\vert \nu_i\vert} ,\eta^{-1} \tau_i\right)_{\rm e},
$$
 hence
$$
\vert \nu_i\vert  P_{\Sigma_i} \tau_i =
\frac{\vert \nu_i\vert^2  \tau_i - 
\left(\nu_i, \eta^{-1} \tau_i\right)_{\rm e}~ \nu_i}{\vert \nu_i\vert}
= -\frac{\eta^{-1}\tau_i}{\vert \nu_i\vert} = R_i \n_i,
$$
where the last equality follows 
from the fact that $\eta^{-1} R_i=-R_i\eta^{-1}$. Therefore, 
\eqref{eq:claim} is proven, and
from \eqref{jj} it the follows
\[
(Y(p), \sum_{i=1}^3 \theta_i R_i \n_i)_{\rm e} =0,
\]
which in turn implies
\begin{equation}\label{tripunti}
\sum_{i=1}^3 \theta_i R_i \n_i =0
\end{equation}
by the arbitrariness of $Y$.
Equality \eqref{tripunti}, recalling \eqref{eq:nuno}, \eqref{eq:ndue}, is the solution to the problem
posed at the beginning of the example.

\begin{figure}
\begin{center}
\includegraphics[height=4cm]{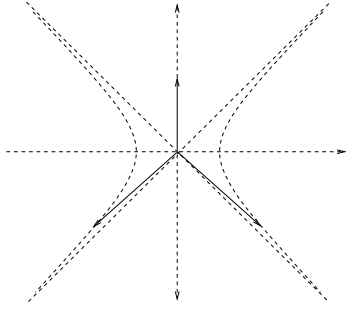}
\smallskip
\caption{\small 
We take 
$\theta_1 = 4$, and $\theta_2=\theta_3=1$, and $\alpha=\beta$ 
as in Figure \ref{fig:creation}, which are 
determined by equations \eqref{eq:conscons}: $\alpha=  {\rm arctg}(\frac{2}{\sqrt{3}})$. 
The vertical vector is $R_1 \n_1$, the lower left one is $R_3 \n_3$ and 
the lower right one is $R_2 \n_2$. Recall that $\n_1$, $\n_2$ and 
$\n_3$ 
have unit lorentzian length. For any $i=1,2,3$ the vector
$R_i \n_i$ is timelike, and
is obtained from $\n_i$ through a
$\pi/2$-rotation. 
 The vectors $R_i\n_i$ 
satisfy the weighted balance condition \eqref{tripunti} at the 
triple junction.}
\label{fig:vectors}
\end{center}
\end{figure}

It is interesting to observe that \eqref{tripunti} is in this example equivalent to
the conservation of energy and momentum. Indeed
\eqref{tripunti} becomes 
\begin{equation}\label{eq:conscons}
\left\{
\begin{aligned}
& \theta_1 =  \frac{\sin\alpha}{\sqrt{\sin^2\alpha-\cos^2\alpha}}\,\theta_2
+ \frac{\sin\beta}{\sqrt{\sin^2\beta-\cos^2\beta}}\,\theta_3, 
\\
\\
& \frac{\cos\alpha}{\sqrt{\sin^2\alpha-\cos^2\alpha}}\,\theta_2 =
\frac{\cos\beta}{\sqrt{\sin^2\beta-\cos^2\beta}}\,\theta_3.
\end{aligned}
\right.
\end{equation}

Recalling Remark \ref{rem:consrett}
let us check that
the first equation 
in \eqref{eq:conscons} is equivalent to the conservation of energy, 
and that the second one  is equivalent to the conservation of 
momentum. Indeed, from \eqref{eq:enemome} we have 
$$
\mathcal E(t) = 
\int_{\cup_{i=1}^3 \Sigma_i(t)} 
{P_{\Sigma_i}}^0_0 ~\theta_i ~d\mathcal H^0
= 
\sum_{i=1}^3  
{P_{\Sigma_i}}^0_0~ \theta_i.
$$

\newpage

Hence for $t<0$ 
$$
\mathcal E(t)=  \theta_1 {P_{\Sigma_1}}^0_0  = \theta_1,
$$
and for $t>0$ 
$$
\mathcal E(t) = 
{P_{\Sigma_2}}^0_0~ \theta_2 + {P_{\Sigma_3}}^0_0~ \theta_3 = 
\frac{\theta_2}{\sqrt{1-\vert\vel_2\vert^2_{\rm e}}}
+
\frac{\theta_3}{\sqrt{1-\vert\vel_3\vert^2_{\rm e}}}
=
\theta_2\sqrt{1+(\n^0_2)^2}
+\theta_2\sqrt{1+(\n^0_3)^2}.
$$
The conservation of energy then becomes
$$
-\theta_1  + 
\theta_2 \sqrt{1+(\n_2^0)^2} + 
\theta_3 \sqrt{1+(\n_3^0)^2}=0,
$$
which is equivalent to 
 the first equation in \eqref{eq:conscons}.

Furthermore, since $\vel_1=0$,
$$
\mathcal P^1(t) = 
\int_{\cup_{i=1}^3 \Sigma_i(t)} \frac{\theta_i \vel_i}{\sqrt{1-\vert 
\vel_i\vert^2_{\rm e}}} ~d\mathcal H^0= 
-\theta_2 \n_0^2 + \theta_3 \n_0^3.
$$
The conservation of momentun reads therefore as
$$
-\theta_2 \n_0^2 + \theta_3 \n_0^3 =0,
$$
which is equivalent to 
 the second equation in \eqref{eq:conscons}.

We conclude this example with some remarks.
Concerning  the solvability of 
\eqref{eq:conscons}: 
given $\theta_i$ for $i=1,2,3$, \eqref{eq:conscons} has a unique solution in the variables
$\alpha$ and $\beta$. Given only $\theta_1$, there are infinitely
many solutions $\theta_2, \theta_3, \alpha, \beta$ of \eqref{eq:conscons}. 
Given $\theta_1\in \mathbb N \setminus \{0\}$, there are only a finite number
of solutions $\alpha, \beta$, and 
$\theta_2, \theta_3 \in \mathbb N \setminus \{0\}$.

Notice that the image of this varifold through the time-reversing map $t \mapsto -t$
is also a stationary rectifiable varifold.  

\smallskip
We now let the two exiting directions to converge to the null directions,
i.e., we let 
$$
\alpha,\,\beta\to \pi/4,
$$
keeping $\theta_1>0$ and $\Sigma_1$ fixed.
We get $\theta_2,\,\theta_3\to 0$, and
the corresponding varifolds in \eqref{puntotriplo} tend to the limit varifold
\[
 \widetilde V^0  
=
\theta_1 (\H^1\rest\Sigma_1) \otimes \delta_{P_{\Sigma_1}},
\qquad 
V^\infty = \frac{\theta_1}{2\sqrt{2}}
~\sum_{i=2}^3 (\mathcal H^1\rest\Sigma_i)\otimes \delta_{Q_{\Sigma_i}},
\]
where $Q_{\Sigma_i}=(1,(-1)^i)\otimes (1,(-1)^{i+1})\in \partial B_{1,2}$
are projections on null lines.
Notice that $V$ is rectifiable and stationary, but no longer proper.

We conclude this example by observing that it refers to a local
situation: indeed, other stationary varifolds
can be obtained by 
furtherly (and properly) 
splitting $\Sigma_2$ and $\Sigma_3$).
\end{Example}
%
%%%%%%%%%%%%%%%%%%%%%%%%%%%%%%%%%%%%%%%%%%%%%%%%%%%%%%%%%%%%
\section{Some pathological examples}\label{sec:pathological}
%%%%%%%%%%%%%%%%%%%%%%%%%%%%%%%%%%%%%%%%%%%%%%%%%%%%%%%%%%%%
In the following elementary example we exhibit
a non weakly rectifiable $1$-varifold obtained as a
 limit of rectifiable
varifolds.
\begin{Example}[{\bf
Limits of zig-zag null curves}]\label{exa:zigzag}\rm
Let $N=1$, $h=1$ and let $\Sigma_\indicesuccessione=
\Sigma_\indicesuccessione^\infty\subset \R^{1+1}$ be the null Lipschitz curve defined as
\[
\manist_\indicesuccessione := \left\{(t,x)\in\R^{1+1}:\ 
x=\frac{1}{\indicesuccessione}\left| \indicesuccessione t-[\indicesuccessione t]-\frac 1 2\right|\right\}, \qquad \indicesuccessione \in \mathbb N
\setminus \{0\},
\]
where $[\alpha]$ denotes the integer part of $\alpha \in \R$, see
Figure \ref{fig:scalette}. 
\begin{figure}
\begin{center}
\includegraphics[height=5cm]{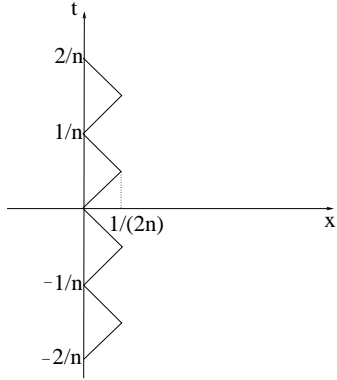}
\smallskip
\caption{\small 
The Lipschitz curve $\Sigma_\indicesuccessione$; the slopes are $\pm 1$, i.e.,
null slopes.}
\label{fig:scalette}
\end{center}
\end{figure}
We can associate with $\manist_\indicesuccessione$ the
rectifiable $1$-varifold 
\[
V_\indicesuccessione = V_\indicesuccessione^\infty := \left(\mathcal H^1\rest\manist_\indicesuccessione\right) \otimes \delta_{Q_{\manist_\indicesuccessione}},
\]
consisting of a null part only. 
Notice that the varifolds $V_\indicesuccessione$ are not stationary. 
Indeed, the lorentzian curvature 
vector 
of $\Sigma_\indicesuccessione$ is concentrated on the vertices of $\Sigma_\indicesuccessione$, where it is a
Dirac delta multiplied by $(0, 2)$ or $(0,-2)$ (depending 
on whether the vertex belongs to the vertical axis or not).

As $\indicesuccessione\to +\infty$ we have
\[
V_\indicesuccessione\rightharpoonup V 
= \sqrt{2} \left(\mathcal H^1\rest \Sigma \right)\otimes V_z,
\]
where 
$$
\Sigma := \left\{z=(t,x)\in\R^{1+1}:\ x=0\right\} = 
\Sigma^\infty
$$
is the vertical 
axis (see \eqref{decodeco}) and \eqref{eq:bdrySn}),
and
\[
V_z = V^\infty_z = \frac{\delta_{(1,1)\otimes (1,-1)}+\delta_{(1,-1)\otimes
(1,1)}}{2}.
\]
In particular, $V
\in \LVuno$ is not weakly rectifiable, since condition 
2c in Definition \ref{def:weaklyrectifiable} is not satisfied.
Notice also that $V$ has multiplicity $\sqrt{2}$, even if 
all approximating vaifolds $V_\indicesuccessione$ have multiplicity $1$. 
\end{Example}

The next example shows a sequence of rectifiable stationary $2$-varifolds
the limit of which is a proper $2$-varifold having only singular part,
and which is not rectifiable.

\begin{Example}[{\bf 
Limits of superpositions of kinks}]\label{exa:patella}\rm
Let $N=2$ and $h=2$. Given $R>0$, let $\gamma=\gamma_R:\R^2\to\R^{1+2}$ be the 
kink solution \eqref{exkink} 
considered in Example \ref{exa:kink}, and let $V_{\gamma^{}_R}\in
\LVdue$ 
be the stationary rectifiable varifold 
associated with $\gamma_R$
in the sense of Definition  \ref{anchelui}. Recall from
\eqref{servesempre}
that the energy of $V_{\gamma^{}_R}$ 
is equal to $2\pi R$ for any $t \in \R\setminus 
\displaystyle \bigcup_{k\in \mathbb Z}
\left\{\frac{\pi R}{2}+ k\pi R\right\}$.
For any $\indicesuccessione\in\mathbb N \setminus \{0\}$ let 
\[
V_\indicesuccessione := \indicesuccessione V_{\gamma^{}_\frac{1}{\indicesuccessione}}. 
\]
The support of $\mu_{V_\indicesuccessione}^{}$ is the superposition of $\indicesuccessione$ kinks of radius $1/\indicesuccessione$,
centered at the origin. Note that the uniform bound \eqref{eqmass}
on $\mu_{V_\indicesuccessione}$ is satisfied.

Recalling also the kink Example \ref{exa:kink}, we have that $V_\indicesuccessione$ 
is a stationary rectifiable $2$-varifold 
with multiplicity $\indicesuccessione$ and energy $2\pi$ at any time 
$t\in\R\setminus 
\displaystyle \bigcup_{k\in \mathbb Z, n \in \mathbb N \setminus \{0\}}
\left\{\frac{\pi}{2n}+ \frac{k\pi}{n}\right\}$.
As $\indicesuccessione\to +\infty$, we have $V_\indicesuccessione\rightharpoonup V$, where 
$V\in \LVdue$ is a stationary weakly rectifiable varifold such that 
\[
\mu^{}_{V} = 2\pi~ \sigma^1\rest\{(t,0,0):\,t\in\R\},
\]
where we have used Remark \ref{rem:ststst}. 
In particular $\mu_{V}$ has only singular part, i.e.,
$$
\mu^{}_V = \mu^s_V,
$$
and therefore $V$ is not rectifiable. Note  that $V$ is a $2$-varifold,
despite the fact that the support of $\mu_V$ is one-dimensional.

Let us now show that the varifolds $V_\indicesuccessione$ satisfy 
the uniform bound \eqref{eqLp} guaranteeing that $V$ is proper.
It is enough to check that there exists $p >1$ such that 
\begin{equation}\label{enou}
\sup_{\indicesuccessione \in \mathbb N} ~ \indicesuccessione \int_{[-T,T]}\int_
{\Sigma_\indicesuccessione(t)}
\left(\frac{1}{\sqrt{1-\vert \vel\vert^2_{\rm e}}}\right)^p
~d\mathcal H^1 dt < +\infty,
\end{equation}
where $\Sigma_\indicesuccessione := \Sigma_{\gamma_{1/\indicesuccessione}}
=\Phi_{\gamma_n}(\R \times [0,2\pi R))$ and 
$\Sigma_\indicesuccessione(t)=
\Sigma_\indicesuccessione\cap \{x^0=t\}$ is the $t$-time slice of $\Sigma_\indicesuccessione$.
{}From \eqref{exkink} it follows $\vert \vel\vert_{\rm e} = \vert 
\sin (\indicesuccessione t)\vert$ on $\Sigma_\indicesuccessione(t)$, hence
$$
\indicesuccessione \int_{[-T,T]}\int_
{\Sigma_\indicesuccessione(t)}
\left(\frac{1}{\sqrt{1-\vert \vel\vert^2_{\rm e}}}\right)^p
~d\mathcal H^1 dt = 
2\pi  \int_{-T}^T \vert \cos(\indicesuccessione t)\vert^{1-p} dt = 
\frac{2\pi}{\indicesuccessione} \int_{-\indicesuccessione T}^{\indicesuccessione T} \vert\cos \tau \vert^{1-p} d\tau
$$
If we choose $p \in (1,2)$ we have 
$$
\sup_{\indicesuccessione \in \mathbb N} 
\frac{2\pi}{\indicesuccessione} \int_{-\indicesuccessione T}^{\indicesuccessione T} 1/\vert\cos \tau \vert^{p-1} d\tau < 
+\infty,
$$
where we use also the periodicity of $\cos \tau$. Hence 
\eqref{enou} holds, and
we conclude from Remark \ref{rem:equi} that  $V$
is a proper varifold. 
\end{Example}

We have seen in Example \ref{exa:patella} that 
limits of stationary rectifiable
varifolds can have only singular part; 
in addition, in the next example
we show that limits of stationary rectifiable
varifolds 
can have only diffuse part.

\begin{Example}[\bf A diffuse limit varifold]\label{exa:patologico}\rm
Let $\gamma = \gamma_R^{}$ be the kink solution
\eqref{exkink} considered in Example \ref{exa:kink}, and define
\[
V_n:= \sum_{i,j\in \{0,\ldots, n-1\}}
V_{\left(\frac{i}{n},\frac{j}{n}\right)+\gamma^{}_{1/n^2}}, 
\qquad n\in\mathbb N \setminus \{0\}.
\]
The support of $\mu^{}_{V_n}$ consists of $n^2$ disjoint kinks of radius $1/n^2$, 
uniformly distributed in  the unit square $[0,1]^2$.
Then $V_n$ is a stationary rectifiable $2$-varifold with multiplicity one
and energy $2\pi$ at any time  $t\in\R$.
As $n\to +\infty$, we have $V_n\rightharpoonup V$, where 
$V$ is a stationary $2$-varifold such that 
\[
\mu_{V}^{} = 2\pi~ \mathcal L^3\rest \left(\R\times [0,1]^2\right).
\]
In particular $\mu_V$ has only diffuse part and $V$ is not weakly 
rectifiable. With a similar computation as in Example \ref{exa:patella}, 
one can show that $V$ is proper.
\end{Example} 

%%%%%%%%%%%%%%%%%%%%%%%%%%%%%%%%%%%%%%%%%%%%%%%%%%
\section{Appendix: measure theory}\label{sec:mea}
%%%%%%%%%%%%%%%%%%%%%%%%%%%%%%%%%%%%%%%%%%%%%%%%%%
Let $\mu$ be a positive measure on $\R^m$ defined on Borel sets. 
We recall \cite{AmFuPa:00} that:
\begin{itemize}
\item[-]  the support of $\mu$ is 
the closure of the set of all points 
$x \in X$ such that $\mu(U) >0$ for any neighbourhood $U$ of $x$; 
\item[-]
$\mu$ is said to be concentrated on $S$ if $S$ is $\mu$-measurable and
$\mu(X \setminus S) =0$;
\item[-] $\mu$ is called a Radon measure if $\mu$ is finite 
on compact sets;

\item[-] 
if $A$ is $\mu$-measurable,
$\mu \rest A$ denotes the restriction of $\mu$ to $A$, 
defined as $\mu(E) := \mu(E \cap A)$. If $\mu$ is a Radon measure, 
then $\mu \rest A$ is a Radon measure;
\item[-] if $p \geq 1$ and $A$ is measurable, $L^p(A,\mu)$ (resp.
$L^p_{\rm loc}(A,\mu)$)
is the space of $p$-integrable (resp. locally $p$-integrable) functions with respect to $\mu$;
\item[-] if $u: \R^m \to \R^k$ is Borel-measurable, 
the push-forward measure
(or image measure)
$u_{\#} \mu$ 
is the Borel measure on $\R^k$
defined by $u_\#\mu (B) := \mu(u^{-1}(B))$. We have 
$\int_{\R^k} f ~d u_\#\mu = \int_{\R^m}  f \circ u d\mu$
for any $f$ summable with respect to $u_\#\mu$. 
\end{itemize}

We recall the following definition \cite{FoLe:07}.
\begin{Definition}
Let $\mu,\nu$ be two measures on $\R^d$ defined on the Borel
subsets of $\R^d$. 
\begin{itemize}
\item[-] $\mu,\nu$ are said to be mutually singular, and we write
$\nu \perp \mu$, if there exists two disjoint Borel sets $X_\mu, X_\nu
\subseteq \R^d$ such that $\R^d = X_\mu \cup X_\nu$ and for every 
Borel set $E \subseteq\R^d$ we have
$$
\mu(E) = \mu(E \cap X_\mu), 
\qquad
\nu(E) = \nu(E \cap X_\nu).
$$
\item[-] 
$\nu$  is  said to be absolutely continuous with respect to $\mu$, and 
we write $\nu << \mu$ if for every Borel set $E\subseteq \R^d$ with 
$\mu(E)=0$ we have $\nu(E) =0$.
\item[-]
$\nu$ is said to be diffuse with respect to $\mu$ if 
for every Borel set $E \subseteq \R^d$ with 
$\mu(E) < +\infty$ we have  $\nu(E)=0$.
\end{itemize}
\end{Definition}

%%%%%%%%%%%%%%%%%%%%%%%%%%%%%%%%%%%%%%%%%%%%%%%%%%%%%%%%%%%%%%%%%%%%%%%
\subsubsection{Absolutely continuous, singular and diffuse parts}\label{sub:ass}
%%%%%%%%%%%%%%%%%%%%%%%%%%%%%%%%%%%%%%%%%%%%%%%%%%%%%%%%%%%%%%%%%%%%%%%
Since we need to 
split a measure with respect to a Hausdorff measure $\mathcal H^\dimension$,
which is not $\sigma$-finite, we need the 
Radon-Nikod\'ym 
theorem 
in a generalized form
\cite{FoLe:07}, where a diffuse part is present.

Let $\mu, \nu$ be positive  measures defined on the Borel 
subsets of $\R^d$. Define,
for every Borel set $E \subseteq \R^d$,  
$$
\begin{aligned}
\nu^{ac}(E) := &
\sup \Big\{\int_E u d\mu : ~ u : \R^d \to [0,+\infty] {\rm ~measurable},
\\
&\qquad\qquad \int_{E'} u d\mu \leq \nu(E') {\rm ~for~all~Borel~} E' \subset E
\Big\},
\\
\nu^s(E) := & \sup \Big\{
\nu(E') :  E' \subset E, E' {\rm ~Borel}, \mu(E')=0
\Big\},
\\
\nu^d(E) := & \sup\Big\{
\nu(E') : E' \subset E, E' {\rm ~Borel~such ~that}
\\
& \qquad \qquad  {\rm for~all~Borel}~ E'' \subset E' {\rm ~with~} \nu(E'')>0
{\rm ~we~have~} \mu(E'') = +\infty
\Big\}.
\end{aligned}
$$
Then the following result holds, see \cite[Theorem 1.114]{FoLe:07}.

\begin{Theorem}[{\bf Generalized Radon-Nikod\'ym Theorem}]\label{th:gRN}
Let $\mu,\nu$ be two positive measures defined on 
the Borel subsets of $\R^d$. Then 
$\nu^{ac}$, $\nu^s$, $\nu^d$ are measures, 
$$
\nu = \nu^{ac} + \nu^s + \nu^d,
$$
with $\nu^{ac} << \mu$ and $\nu^d$ diffuse with respect to $\mu$. 
Moreover, if $\nu$ is $\sigma$-finite,
then $\nu^{ac}$, $\nu^s$, $\nu^d$ are mutually singular, and $\nu^s \perp \mu$;
\end{Theorem}
If $\mu$ is $\sigma$-finite (which corresponds
to the classical Radon-Nyko\'ym Theorem)
 we have $\nu^d=0$.

The density of $\nu$ with respect to $\mu$ will be denoted by $\frac{d\nu}{d\mu}$.

%%%%%%%%%%%%%%%%%%%%%%%%%%%%%%%%%%%%%%%%%%%%%%%%%%%%%%%%%%%%%%%%%%%%%%%
\subsubsection{Disintegration of Radon measures}
%%%%%%%%%%%%%%%%%%%%%%%%%%%%%%%%%%%%%%%%%%%%%%%%%%%%%%%%%%%%%%%%%%%%%%%
Let $\mu$ be a positive Radon measure on $\R^d$, and $z \to 
\nu_z$ be a map which assigns to each $z \in \R^d$ a finite Radon
measure $\nu_z$ on $\R^m$, such that the function $z \to
\nu_z(B)$ is $\mu$-measurable for any Borel set $B \subseteq \R^d$. 

We denote by
$$
\nu = \mu \otimes \nu_z
$$
the Radon measure on $\R^d \times \R^m$ defined by
$$
\mu \otimes \nu_z(B) := \int_{\R^d} \left(
\int_{\R^m} \chi_B(z,y)~d\nu_z(y)\right)
~d\mu(z)
$$
for any Borel set $B \subseteq K \times \R^m$, where $K \subset \R^d$
is any compact set.

The following result is proven for instance in 
 \cite[Th. 2.28]{AmFuPa:00}.

\begin{Theorem}[{\bf Disintegration}]\label{th:disint}
Let $\nu$ be a positive Radon measure on $\R^d \times \Rm$,
let  $\pi :  \R^d \times \R^m \to \R^d$ be the 
projection on the first factor, and set
$$
\mu := \pi_\#\nu.
$$
Assume that $\mu$ is a Radon measure,
namely that 
\begin{equation}\label{eq:disint}
\nu(K \times \R^m) <+\infty
\qquad {\rm for~ any~ compact~ set~} K \subset \Rn.
\end{equation}
Then there exist positive Radon measures $\nu_z$ in $\Rm$ such that
\begin{itemize}
\item[-] for any Borel set
$B \subseteq \Rm$ the function $z \to \nu_z(B)$ is $\mu$-measurable, and 
$$
\nu_z(\Rm) =1 \qquad {\rm for}~ \mu-{\rm a.e.~ in}~ z \in \R^d,
$$
\item[-] for any $f \in L^1(\R^d \times \Rm, \nu)$ we have 
$$
f(z,\cdot) \in L^1(F, \nu_z) \qquad {\rm for~} \mu {\rm ~a.e.~} z \in \R^d,
$$
\begin{equation}\label{dueediciannove}
z \to \int_{\R^d} f(z,y) ~d\nu_z(y) \in L^1(\R^d, \mu),
\end{equation}
and 
\begin{equation}\label{dueeventi}
\int_{\R^d \times \Rm}
f(z,y)~d\nu(z,y) = 
\int_{\R^d}
\left(
\int_{\Rm}
f(z,y)~ d\nu_z(y)
\right)~d\mu(z).
\end{equation}
\end{itemize}
Hence we have the following disintegration of $\nu$:
\begin{equation}\label{eq:disintnu}
\nu = \mu \otimes \nu_z.
\end{equation}
Moreover, if $z \to \nu'_z$ is any other  Radon measures-valued map
such that the function $z \to \nu_z'(B)$ is $\mu$-measurable
for any Borel set $B \subseteq \R^d$, and 
 satisfying 
\eqref{dueediciannove}, \eqref{dueeventi} for every
bounded Borel function with compact support and such that 
$\nu'_z(F) \in L^1_{\rm loc}(\R^d, \mu)$, then 
$\nu_z = \nu_z'$ for $\mu$-almost every $z \in \R^d$.
\end{Theorem}

%%%%%%%%%%%%%%%%%%%%%%%%%%%%%%%%%%%%%%%%%%%%%%%%%%%%%%%%%%%%%%%%%%%%%%%%%%%%%%%%%%%%%
\bibliographystyle{plain} 

%%%%%%%%%%%%%%%%%%%%%%%%%%%%%%%%%%%%%%%%%%%%%%%%%%%%%%%%%%%%%%%%%%%%
\end{document}